\documentclass[11pt,preprint]{aastex}

\def\Ldiss{L_{\rm diss}}
\def\Bpole{B_{\rm pole}}
\def\BQED{B_{\rm QED}}

\def\>{$>$}
\def\<{$<$}

\def\simlt{\lower.5ex\hbox{$\; \buildrel < \over \sim \;$}}
\def\simgt{\lower.5ex\hbox{$\; \buildrel > \over \sim \;$}}
\def\ch2{$\chi^{2}$}

\def\ee{\'{e}}

\def\sT{\sigma_{\rm T}}
\def\dd{{\rm d}}

\def\be{\begin{equation}}
\def\ee{\end{equation}}

\def\bB{{\,\mathbf B}}
\def\bE{{\,\mathbf E}}
\def\bj{{\,\mathbf j}}

\def\bxi{{\,\mathbf \xi}}

\def\jB{j_B}
\def\RNS{R_{\rm NS}}
\def\gres{\gamma_{\rm res}}

\def\tT{\tau_{\rm T}}
\def\natm{n_{\rm atm}}
\def\lres{l_{\rm res}}
\def\Lres{a_{\rm res}}

\def\lD{\lambda_{\rm D}}
\def\lDe{\lambda_{{\rm D}e}}

\def\Ethr{E_{\rm thr}}
\def\fthr{f_{\rm thr}}
\def\fpair{f_B}
\def\lCoul{\ell_{\rm Coul}}
\def\Uturb{U_{\rm turb}}

\def\kB{k_{\rm B}}
\def\lb{l_b}
\def\nb{n_b}
\def\gb{\gamma_b}
\def\ND{{\cal N}_{\rm D}}
\def\calN{{\cal N}}

\def\th{\theta_{\rm em}}
\def\them{\theta_{\rm em}}
\def\tdyn{t_{\rm dyn}}

\newbox\grsign \setbox\grsign=\hbox{$>$} \newdimen\grdimen \grdimen=\ht\grsign
\newbox\simlessbox \newbox\simgreatbox \newbox\simpropbox
\setbox\simgreatbox=\hbox{\raise.5ex\hbox{$>$}\llap
     {\lower.5ex\hbox{$\sim$}}}\ht1=\grdimen\dp1=0pt
\setbox\simlessbox=\hbox{\raise.5ex\hbox{$<$}\llap
     {\lower.5ex\hbox{$\sim$}}}\ht2=\grdimen\dp2=0pt
\setbox\simpropbox=\hbox{\raise.5ex\hbox{$\propto$}\llap
     {\lower.5ex\hbox{$\sim$}}}\ht2=\grdimen\dp2=0pt
\def\simgt{\mathrel{\copy\simgreatbox}}
\def\simlt{\mathrel{\copy\simlessbox}}

\begin{document}

 \title{CORONA OF MAGNETARS}

\author{Andrei M. Beloborodov\altaffilmark{1}}
\affil{Physics Department and Columbia Astrophysics Laboratory, 
Columbia University, 538  West 120th Street New York, NY 10027;
amb@phys.columbia.edu}
\author{Christopher Thompson} 
\affil{Canadian Institute for Theoretical Astrophysics,
University of Toronto, 60 St. George Street, Toronto,
ON M5S 3H8, Canada; thompson@cita.utoronto.ca}
                                                                                
\altaffiltext{1}{Also at Astro-Space Center of Lebedev Physical
Institute, Profsojuznaja 84/32, Moscow 117810, Russia}

\begin{abstract}
We develop a theoretical model that explains the formation of hot coronae
around strongly magnetized neutron stars --- magnetars. The starquakes of 
a magnetar shear its external magnetic field, which becomes non-potential 
and is threaded by an electric current. Once twisted, the magnetosphere 
cannot untwist immediately because of its self-induction. The self-induction
electric field lifts particles from the stellar surface, accelerates 
them, and initiates avalanches of pair creation in the magnetosphere. 
The created plasma corona maintains the electric current demanded by 
${\rm curl}\bB$ and regulates the self-induction e.m.f. by screening.
This corona persists in dynamic equilibrium: it is continually lost to 
the stellar surface on the light-crossing time $\sim 10^{-4}$~s and
replenished with new particles. In essence, the twisted magnetosphere 
acts as an accelerator that converts the toroidal field energy to 
particle kinetic energy. Using a direct numerical experiment, we show 
that the corona self-organizes quickly (on a millisecond timescale)
into a quasi-steady state, with voltage $e\Phi_e\sim 1$~GeV along the 
magnetic lines. The voltage is maintained near threshold for ignition 
of pair production, and pair production occurs at a rate just enough 
to feed the electric current. The heating rate of the corona is 
$\sim 10^{36}$~erg/s, in agreement with the observed persistent, 
high-energy output of magnetars. We deduce that a static twist that is 
suddenly implanted into the magnetosphere will decay on a timescale of 
1-10 yrs. The particles accelerated in the corona impact the solid 
crust, knock out protons, and regulate the column density of the 
hydrostatic atmosphere of the star. The transition layer between the 
atmosphere and the corona may be hot enough to create additional $e^\pm$ 
pairs. This layer is the likely source of the observed 100~keV emission 
from magnetars. The corona emits curvature radiation up to $10^{14}$~Hz 
and can supply the observed IR-optical luminosity. We outline the
implications of our results for the non-thermal radiation in other bands,
and for the post-burst evolution of magnetars.
\end{abstract}



\section{INTRODUCTION}

At least 10\% of all neutron stars are born as magnetars, with 
magnetic fields $B>10^{14}$~G. 
Their activity is powered by the decay of the 
ultrastrong field and lasts about $10^4$~years. They are observed at this 
active stage as either Soft Gamma Repeaters (SGRs) or Anomalous X-ray 
Pulsars (AXPs) (see Woods \& Thompson 2006 for a recent review).
Twelve active magnetars are currently known, and all of them have similar
parameters: spin period $P\sim 10$~s, persistent X-ray luminosity
$L\sim 10^{35}-10^{36}$~erg/s, and surface temperature 
$k_{\rm B}T\sim 0.5$~keV.
The large-scale (dipole) component of the magnetic field 
$B_{\rm dipole}>10^{14}$~G has been inferred from the measurement
of rapid spindown in several AXPs, and more recently in two SGRs
(Kouveliotou et al. 1998, 1999).
However, there is evidence for even stronger magnetic field inside
the star, as was originally deduced from considerations of magnetic
field transport (Thompson \& Duncan 1996).   In particular,
the occurence of enormously energetic X-ray flares in SGRs
points to internal magnetic fields approaching $\sim 10^{16}$~G
(Hurley et al. 2005).
Bursting activity has also been detected in two AXPs (Kaspi et al. 2003),
which strongly supports the hypothesis that
active rearrangement of the magnetic field takes place in
AXPs as well as SGRs, although at differing rates.

Besides the sporadic X-ray outbursts, a second, persistent, form of 
activity has been discovered by studying the emission spectra of 
magnetars. Until recently, the spectrum was known to have a thermal 
component with temperature $\kB T\sim 0.5$~keV, interpreted as blackbody 
emission from the star's surface. The soft X-ray spectrum also showed a 
tail at $2-10$~keV with photon index $\Gamma=2-4$. Intriguingly, the 
hardness of the $2-10$ keV spectrum correlates with long-term activity
as a transient burst source (Marsden \& White 2001). This deviation from 
pure surface emission already suggested that energy is partially released 
above the star's surface.  
Recently, observations by RXTE and INTEGRAL have revealed even more 
intriguing feature: magnetars are bright, persistent sources of 100 keV 
X-rays (Kuiper et al. 2004; Molkov et al. 2005; Mereghetti et al. 2005;
den Hartog et al. 2006; Kuiper et al. 2006).
This high-energy emission forms a separate 
component of the magnetar spectrum, distinct from the soft X-ray component 
that was known before. The new component becomes dominant above 10~keV, 
has a very hard spectrum, $\Gamma\simeq 1$, and peaks above 100~keV
where the spectrum is unknown.
The luminosity in this band, $L\sim 10^{36}$~erg~s$^{-1}$, even exceeds
the thermal luminosity from the star's surface. These hard X-rays
can be emitted only in the exterior of the star and demonstrate
the presence of an active plasma corona.
                                                                                
Our goal in this paper is to understand the origin and composition of the 
magnetar corona. The starting point of our analysis is the fact that
an evolving magnetar experiences strong deformations of its crust, due to 
a shifting configuration of the internal magnetic field.
These ``starquakes'' shear the magnetic field that is anchored in the crust,
thereby injecting an electric current into the stellar magnetosphere.
This process bears some resemblence to the formation of
current-carrying structures in the Solar corona,
although there are important differences which shall be discussed below.
We investigate how the
starquakes and twisting of the magnetosphere 
can lead to the formation of a plasma corona
around a magnetar.

The persistence of magnetospheric twists can explain several observed 
properties of the AXPs and SGRs (Thompson, Lyutikov, \& Kulkarni 2002; 
hereafter TLK).  TLK investigated the observational consequences using 
a static, force-free model, idealizing the magnetosphere as a globally 
twisted dipole. They showed that a twist affects the spindown rate of 
the neutron star: it causes the magnetosphere
to flare out slightly from a pure dipole configuration, thereby 
increasing the braking torque acting on the star. 
The force-free configuration is independent of the plasma behavior in 
the magnetosphere. It does rely, however, on the presence of plasma
that can conduct the required current $\bj_B=(c/4\pi)\nabla\times\bB$. 
This requires a minimum density $n_c=j_B/ec$. TLK showed that even this
minimum density significantly modifies the stellar radiation 
by multiple resonant scattering, which can explain the 
tail of the surface emission at $2-10$~keV.  

In this paper, we formulate the problem of plasma dynamics in the 
closed twisted magnetosphere and find a simple solution to this problem,
which allows one to understand the observed energy output of the magnetar
corona and the evolution of magnetic twists. Plasma behavior around neutron
stars has been studied for decades in the context of radio pulsars, and that
problem remains unsolved. The principle difference with radio pulsars is that 
their activity is caused by {\it rotation}, and 
dissipation occurs on {\it open magnetic lines} that connect the star 
to its light cylinder. Electron-positron pairs are then created on open 
field lines (e.g. Ruderman \& Sutherland 1975; Arons \& Scharlemann 1979;
Harding \& Muslimov 2002). By contrast, the formation of a corona around 
a  magnetar does not depend on rotation of the neutron star. All observed 
magnetars are slow rotators ($\Omega=2\pi/P\sim 1$~Hz), and their 
rotational energy cannot feed the observed coronal emission. 
(The bolometric output of some magnetars
is $\sim 10^4$ times larger than the spindown luminosity, as deduced from 
the rotational period derivative!) The open 
field lines form a tiny fraction $\RNS\Omega/c\sim 3\times 10^{-5}$ of 
the magnetic flux threading the neutron star. Practically the entire
plasma corona is immersed in the {\it closed} magnetosphere and its 
heating must be caused by some form of dissipation on the closed field 
lines. In contrast with canonical radio pulsars, the magnetar corona
admits a simple solution. 

Plasma near a neutron star is strongly constrained compared with many
other forms of plasma encountered in laboratory and space, and therefore
easier to model.  
The relative simplicity is due to the fact that the magnetic field is
strong and fixed --- its evolution timescale is much longer
than the residence time of particles in the corona. Furthermore,
particles are at the lowest Landau level and move strictly along
the magnetic field lines, which makes the problem effectively one-dimensional.
Finally, the magnetic lines under consideration are closed, with both
ends anchored in the highly conducting stellar material.
The basic questions that one would like to answer are as follows.
How is the magnetosphere populated with plasma? The neutron star 
surface has a temperature $k_{\rm B}T\simlt 1$~keV and the scale-height 
of its atmospheric layer (if it exists) 
is only a few cm.  How is the plasma supplied
above this layer and what type of particles populate the corona?
How is the corona heated, and what are the typical energies of the particles?
If the corona conducts a current associated with a magnetic
twist $\nabla\times\bB\neq 0$, how rapid is the dissipation of this 
current, i.e. what is the lifetime of the twist? Does its decay imply the
disappearance of the corona?
                                                                                
An outline of the model proposed in this paper is as follows.
The key agent in corona formation is an electric field $E_\parallel$
parallel to the magnetic field. $E_\parallel$ is generated by the 
self-induction of the gradually decaying current and in essence
measures the rate of the decay. It determines the heating rate of
the corona via Joule dissipation. If $E_\parallel=0$ then the corona is
not heated and, being in contact with the cool stellar surface, it will
have to condense to a thin surface layer with $k_{\rm B}T\simlt 1$~keV.
The current-carrying particles cannot flow upward against
gravity unless a force $eE_\parallel$ drives it. On the other hand, when
$E_\parallel$ exceeds a critical value, $e^\pm$ avalanches
are triggered in the magnetosphere, and the created pairs screen the electric
field. This leads to a ``bottleneck'' for the decay of a magnetic twist,
which implies a slow decay.

The maintenance of the corona and the slow decay of the magnetic twist
are intimately related because both are governed by $E_\parallel$.
In order to find $E_\parallel$, one can use Gauss' law 
$\nabla\cdot\bE=4\pi\rho$, where $\rho$ is the net charge density of 
the coronal plasma. This constraint implies that 
$\bE$ and $\rho$ must be found self-consistently. 
We formulate the problem of plasma dynamics in a twisted
magnetic field and the self-consistent electric
field, and investigate its basic properties.
The problem turns out to be
similar to the classical double-layer problem of plasma physics
(Langmuir 1929) with a new essential ingredient: $e^\pm$ pair creation.

We design a numerical experiment that simulates the behavior of the 1-D
plasma in the magnetosphere. The experiment shows how the plasma and
electric field self-organize to maintain the time-average magnetospheric
current $\bar{\bj}=\bj_B$ demanded by the magnetic field,
$(4\pi/c)\bar{\bj}=\nabla\times\bB$. The electric current admits no steady
state on short timescales and keeps fluctuating, producing $e^\pm$ avalanches.
This state may be described as a self-organized criticality.
Pair creation is found to provide a robust mechanism for limiting the voltage
along the magnetic lines to $e\Phi_e \la 1$~GeV
and regulate the observed luminosity of the corona.

The paper is organized as follows.
We start with a qualitative description of the mechanism of corona formation, 
and show that it can be thought of as an 
electric-circuit problem
(\S~2). A careful formulation of the circuit problem
and the set up of numerical experiment are described in \S~3. 
In \S~4 we find that the circuit without $e^\pm$ creation
does not provide a self-consistent model. Pair creation is 
found to be inevitable and described in detail in \S~5.
In \S~6 we discuss the transition layer
between the corona and the relatively cold surface 
of the star. The maintenance of a dense plasma layer on the 
surface is addressed in \S~7. 
Observational implications and further developments of the 
model are discussed in \S~8.
                                                                            

\section{MECHANISM OF CORONA FORMATION}

\subsection{Ejection of Magnetic Helicity from a Neutron Star}

A tightly wound-up magnetic field is assumed to exist inside magnetars
(e.g. Thompson \& Duncan 2001). The internal toroidal field can be 
much stronger than the external large-scale dipole 
component.\footnote{Existing calculations of the relaxation to magnetostatic 
equilibrium in a fluid star (e.g., a nascent magnetar) assume that
the initial toroidal and poloidal fluxes are comparable (Braithwaite \& Spruit 
2004). The initial toroidal flux can, in fact, be substantially stronger
due to the presence of rapid differential rotation in the newly formed star.}
The essence of magnetar activity is the transfer of magnetic helicity 
from the interior of the star to its exterior, where it dissipates. 
This involves rotational motions of the crust, which inevitably twist 
the external magnetosphere that is anchored to the stellar surface.
The magnetosphere is probably twisted in a catastrophic way, when the 
internal magnetic field breaks the crust and rotates it. Such starquakes
are associated with observed X-ray bursts (Thompson \& Duncan 1995).
The most interesting effect of a starquake for us here is the partial 
release of the winding of the internal magnetic field into the exterior, 
i.e., the injection of magnetic helicity into the magnetosphere.

Since the magnetic field is energetically dominant in the magnetosphere,
it must relax there to a force-free configuration with $\bj\times\bB=0$. 
The effective footpoints of the force-free field lines sit in the inner
crust of the star. At this depth
currents may flow across the magnetic lines 
because the deep crust can sustain a significant Amp\`ere force 
$\bj\times\bB/c\neq 0$: this force is balanced by elastic forces of the 
deformed crust. 
The ability of the crust to sustain $\bj\nparallel\bB$ quickly decreases 
toward the surface, and it is nearly force-free 
where\footnote{Throughout this paper we use the notation 
$X = X_n\times 10^n$, where quantity $X$ is measured in c.g.s. units.}  
$\rho\simlt 10^{13}B_{15}^{2.5}$~g/cm$^3$ (Thompson \& Duncan 2001).
The build-up of stresses in the crust from an initial magnetostatic
equilibrium has been studied recently by Braithwaite \& Spruit (2006).

The ejection of helicity may be thought of as 
directing an interior current (which previously flowed across the field 
and applied a ${\bf j}\times {\bf B}/c$ force) into the force-free
magnetosphere. The ejected current flows along the magnetic lines 
to the exterior of the star, reaches the top of the line and comes back
to the star at the other footpoint. 
The currents emerging during the starquake may percolate through a 
network of fractures in the crust. Then large gradients in the field are 
initially present which can be quickly erased.
A relatively smooth magnetospheric twist of scale $\sim\RNS$ 
will have the longest lifetime.

Globally twisted magnetic fields are observed to persist in the Solar
corona over many Alfv\'en crossing times.
Twisted magnetic configurations have also been studied in 
laboratory experiments. In particular, the toroidal pinch configuration
has been studied in detail and is
well explained as a relaxed plasma state in which the imparted
magnetic helicity has been nearly conserved (Taylor 1986).
The toroidal pinch relaxes to the minimum-energy state which has
a uniform helicity density: $\nabla\times\bB=\mu\bB$ with $\mu={\rm const}$.
This state is achieved because the experiment has toroidal geometry
and the current is isolated from an external conductor. The relaxation 
process is then free to redistribute the current between neighboring 
magnetic flux tubes via small-scale reconnection, which leads to a 
uniform $\mu$. 
The geometry is different in the case of a magnetar (or the sun) --- the
field lines are anchored to the star and their configuration depends on 
the footpoint motion (for related calculations of the relaxed state of a 
driven plasma see, e.g., Tang \& Boozer [2005]). 
The long lifetime ($\simgt 1$~year) of a twisted configuration around 
magnetars is supported by observations of long-lived changes in 
the spindown rate and X-ray pulse profiles (Woods \& Thompson 2006; TLK) 
as well as persistent nonthermal X-ray emission.

The shape of a twisted force-free magnetosphere may be calculated 
given the boundary conditions at the footpoints.
An illustrative model of a globally twisted dipole field was
considered by TLK. It has one parameter --- twist angle $\Delta\phi$ 
--- and allows one to see the qualitative effects of increasing twist
on the magnetic configuration. 
The current density in this configuration is 
\be\label{eq:j}
{\bj}(r,\theta) = {c\over 4\pi}{\bf\nabla}\times{\bf B} \simeq
{c {\bf B}\over 4\pi r}\sin^2\theta\,\Delta\phi\;\;\;\;\;\;
(r > \RNS),
\ee
where $r,\theta,\phi$ are spherical coordinates 
with the polar axis chosen along the axis of the dipole.
Even for strong twists, $\Delta\phi\sim 1$, 
the poloidal components $B_r$ and $B_\theta$ are found to be close to 
the normal dipole configuration, although the magnetosphere is somewhat 
inflated.  The main difference is the appearance of a toroidal field 
$B_\phi$.   

The global structure of a real twisted magnetosphere may be complex.
The displacement ${\bxi}$ of the magnetosphere's footpoints during 
a starquake may be described as a 2D incompressible motion with
$\nabla\cdot\bxi =0$ and $\nabla\times\bxi\neq 0$. Suppose that the yielding 
behavior of the crust produces a motion $\bxi\neq 0$ that is localized 
within a plate $S$, and this plate is connected by magnetospheric field 
lines to another plate on the opposite side of the star. 
Then $\int\nabla\times\bxi\,\dd S = \int\bxi\cdot\dd {\mathbf L}=0$ 
where $L$ is the boundary of $S$. Hence $\nabla\times\xi$ must change 
sign within $S$. For example, for an axially symmetric rotational motion, the 
ejection of current in the center of the plate is accompanied by initiation 
of an opposite current on its periphery that screens the ejected current
from the unmoved footpoints with $\xi=0$. 

In this paper, we study the plasma dynamics in a flux tube $dS$ with a 
given twist current $\jB=(c/4\pi)|\nabla\times\bB|$. 
The magnetosphere may be thought of as a collection
of such elementary flux tubes, whose currents $j_B$ vary from footpoint to
footpoint. The variation of $\jB$ is expected to be gradual, so that the 
characteristic thickness of a flux tube with a given $\jB$ is comparable 
to $\RNS$. 

The emerging currents are easily maintained during the X-ray outburst 
accompanying a starquake. A dense, thermalized plasma is then present in 
the magnetosphere, which easily conducts the current. Plasma remains 
suspended for some time after the starquake because of the transient 
thermal afterglow with a high blackbody temperature, up to $\sim 4$~keV.
The cyclotron resonance of the ions is at the energy
$\hbar eB/m_pc = 6.3 B_{15}$~keV and, during the afterglow, 
the radiative flux at the resonance is high enough to lift ions 
from the surface against gravity (Ibrahim et al. 2001).

Eventually the afterglow extinguishes and the radiative flux becomes unable
to support the plasma outside the star. The decreasing density then threats 
the capability of the magnetosphere to conduct the current of the
magnetic
twist. A minimal ``corotation'' charge density 
$\rho_{\rm co} = -{\bf\Omega}\cdot\bB/2\pi c$ 
is always maintained because of rotation of the star with 
frequency $\Omega$ (Goldreich \& Julian 1969), but it is far smaller than 
the minimal density that is needed to supply the current $\bj$ 
(eq.~\ref{eq:j}):
\be\label{rhoco}
{|\rho_{\rm co}|\over |{\bj}|/c} \sim {1\over\Delta\phi}\,
\left({\Omega R_{\rm max}\over c}\right)\ll 1.
\ee
Here $R_{\rm max}(r,\theta) = r/\sin^2\theta$ is the maximum radius of
a dipolar field line that passes through coordinates $(r,\theta)$.
The current, however, will be forced to continue to flow by its
self-induction, and the magnetosphere finds a way to re-generate the
plasma that can carry the current (see \S~2.2).
So, the twisted force-free configuration persists.

The stored
energy of non-potential (toroidal) magnetic field associated with 
the ejected current is subject to gradual dissipation.
In our model, this dissipation
feeds the observed activity of the corona. 
The stored energy is concentrated near the star and carried
by closed magnetic lines with a maximum extension radius 
$R_{\max}\sim 2R_{NS}$.
Thus, most of the twist energy will be released if these lines
untwist, and we focus in this paper on the near magnetosphere 
$R_{\max}\sim 2R_{NS}$.

\subsection{Bottleneck for the Twist Decay and Plasma Supply to the Corona}

Consider a magnetic flux tube\footnote{In an axisymmetric magnetosphere,
we could  equally well focus on a flux surface of revolution which sits
within a range   $\Delta\theta$
of polar angle $\theta$ and has a cross section 
  $S = 2\pi r^2\sin\theta \Delta\theta$.}
with cross section 
$S\simlt \RNS^2$ and length $L\sim\RNS$ which carries a current $I=Sj$. 
The stored magnetic energy of the current per unit length of the tube is 
\be
   {{\cal E}_{twist}\over L} \sim \frac{I^2}{c^2}\, S.
\ee
The decay of this energy is associated with an electric field parallel to 
the magnetic lines $\bE_\parallel$: this field can accelerate particles 
and convert the magnetic energy into plasma energy. Conservation of energy
can be expressed as
\be
  \frac{\partial}{\partial t}\left(\frac{B^2}{8\pi}\right)=-\bE\cdot\bj
         -\nabla\cdot\left(c\frac{\mathbf E\times \mathbf B}{4\pi}\right),
\ee
as follows from Maxwell's equations with $E\ll B$ (e.g.
Landau \& Lifshitz 1975).
The first term on the right-hand side is the Joule dissipation caused by 
$\bE_\parallel$. The second term --- the divergence of the Poynting 
flux --- describes the redistribution of magnetic energy in space.

\begin{figure}
\begin{center}
\epsscale{.80}
\plotone{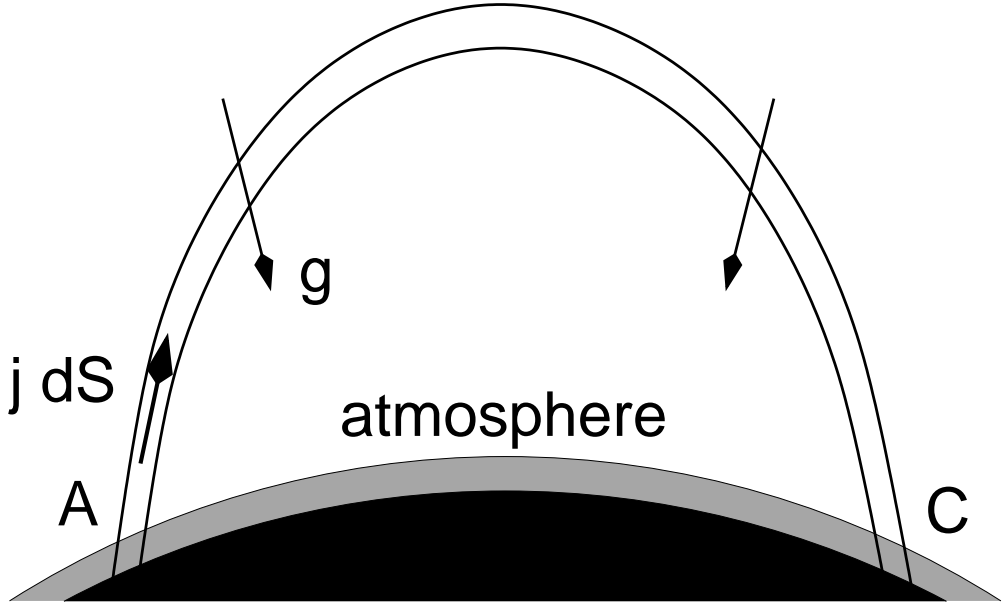}
\label{fig1}
\caption
{Schematic picture of a current-carrying closed magnetic flux tube,
anchored to the star's surface. The current is initiated by a starquake 
that twists one (or both) footpoints of the tube. The current flows along 
the tube into the magnetosphere and returns to the star at the other 
footpoint, where it enters the atmosphere and the crust, 
following the magnetic field lines ($\bj\parallel\bB$).
Sufficiently deep in the crust, the current can flow across the field lines
and change direction, so that it can flow back to the twisted footpoint 
through the magnetosphere (or directly through the star, depending on the
geometry of the twist). A self-induction voltage is created along the tube 
between its footpoints, which accompanies the gradual decay of the current. 
The voltage is generated because the current has a tendency to become 
charge-starved above the atmospheric layer whose scale-height 
$h=k_{\rm B}T/gm_p$ is a few cm. 
}
\end{center}
\end{figure}

The decay of the twist is related to the 
voltage between the footpoints of a magnetic line, 
\be
\label{eq:Phi_e}
   \Phi_e=\int_A^C {\mathbf E}\cdot \dd{\mathbf l}.
\ee
Here A and C are the anode and cathode footpoints and $\dd{\mathbf l}$
is the line element; the integral is taken along the magnetic line outside
the star (Fig.~1).  The product $\Phi_e I$ approximately represents
the dissipation rate $\dot{\cal E}$ in the tube.  (The two quantities are 
not exactly equal in a time-dependent circuit.)
The instantaneous dissipation rate is given by
\be
  \dot{\cal E}=\int_A^C {\mathbf E}\cdot {\mathbf I}(l) \dd{\mathbf l},
\ee
where ${\mathbf I}(l)=S{\mathbf j}$ is the instantaneous current at 
position $l$ in the tube. The current is fixed at the footpoints A and C, 
$I=I_0$;  it may, however, fluctuate between the footpoints.

Note that the true scalar potential $\Phi$ is different from $\Phi_e$
and not related to the dissipation rate.
$\Phi$ is the 0-component of the 4-potential
$\Phi^\mu=(\Phi,{\mathbf A})$ which is related to the electric field by
\be
  {\mathbf E}=-\nabla \Phi -\frac{1}{c}\frac{\partial{\mathbf A}}{\partial t}.
\ee
If anode and cathode are connected by an ideal conductor
then $\Phi(C)-\Phi(A)=0$ while the voltage $\Phi_e$ may not vanish.

The voltage $\Phi_e$ is entirely maintained by the self-induction that 
accompanies the 
gradual untwisting of the magnetic field. Thus, $\Phi_e$ reflects
the gradual decrease of the magnetic helicity in the tube,
\be
\dot{\cal H} = {\partial\over\partial t}\int_A^C {\bf A}\cdot{\bf B}\, \dd V 
= -\int_A^C {\bf E\cdot B}\, S(l)\, \dd l.
\ee
(Here $\dd V = S(l)\dd l$ is the volume element.)  It also 
determines the effective resistivity of the tube: ${\cal R}=\Phi_e/I$.
The self-induction voltage passes the released magnetic energy to charged
particles, and a higher rate of untwisting implies a higher energy $e\Phi_e$
gained per particle. A huge magnetic energy is stored in the twisted tube, 
and a quick untwisting would lead to extremely high Lorentz factors of the 
accelerated particles. 

There is, however, a bottleneck that prevents a fast decay of the twist:
the tube responds to high voltages through the copious production of 
$e^\pm$ pairs.  Runaway pair creation is ignited when electrons are 
accelerated to a certain Lorentz factor $\gamma_\pm\sim 10^3$.   
(We describe the microphysics of pair creation in \S~5.)  
The created $e^\pm$ plasma does not tolerate large $E_\parallel$ --- the 
plasma is immediately polarized, screening the electric field. 
This provides a negative feedback that limits the magnitude of $\Phi_e$ 
and buffers the decay of the twist.

A minimum $\Phi_e$ is needed for the formation of a corona. Two mechanisms 
can supply plasma: (1) Ions and electrons can be lifted into the 
magnetosphere from the stellar surface. This requires a minimum voltage,
\be
  e\Phi_e \sim gm_i R_{\rm NS} \sim 200\;\;{\rm MeV},
\ee
corresponding to $E_\parallel$ that is strong enough to lift ions 
(of mass $m_i$) from the anode footpoint and electrons from the
cathode footpoint. (2) Pairs can be created in the magnetosphere if
\be
  e\Phi_e \sim \gamma_\pm mc^2 = 0.5\gamma_\pm\;\;{\rm MeV},
\ee
which can accelerate particles to the 
Lorentz factor $\gamma_\pm$ sufficient to ignite $e^\pm$ creation.  
A third possible source --- thermal $e^\pm$ production in a heated 
surface layer --- will be discussed in \S~6; the corresponding minimum
voltage is currently unknown.

If $\Phi_e$ is too low and the plasma is not supplied, a flux tube is 
guaranteed to generate a stronger electric field. The current becomes 
slightly charge-starved: that is, the net density of free charges becomes 
smaller than $|\nabla\times\bB|/4\pi e$.
The ultrastrong magnetic field, whose twist carries an enormous energy 
compared with the energy of the plasma, does not change and responds 
to the decrease of $\bj$ by generating an electric field: a small 
reduction of the conduction current  $\bj$ induces a displacement current 
$(1/4\pi)\partial {\mathbf E}/\partial t = (c/4\pi)\nabla\times\bB -\bj$.
The longitudinal electric field
$E_\parallel$ then quickly grows until it can pull
particles away from the stellar surface and ignite pair creation.

The limiting cases $E_\parallel\rightarrow 0$ (no decay of the twist) 
and $E_\parallel\rightarrow\infty$ (fast decay) both imply a contradiction.
The electric field and the plasma content of the corona must regulate each 
other toward a self-consistent state, and the gradual decay of the twist 
proceeds through a delicate balance: $E_\parallel$ must be strong enough 
to supply plasma and maintain the current in the corona; 
however, if plasma is oversupplied $E_\parallel$ will be reduced by 
screening.  A cyclic behavior is possible, in which plasma is periodically 
oversupplied and $E_\parallel$ is screened. Our numerical experiment 
will show that such a cyclic behavior indeed takes place.


\section{ELECTRIC CIRCUIT}\label{sc:circuit}

\subsection{Basic Properties}\label{sc:basic}

Three facts facilitate our analysis of plasma dynamics in the electric 
circuit of magnetars:

1. The ultrastrong magnetic field makes the particle dynamics 1-D. 
The rest-energy density of the coronal 
plasma is $\sim mcj/e$, where $m$ is the mass of a plasma particle. 
It is much smaller than the magnetic energy density $B^2/8\pi$: their ratio is
$\sim m c^2/eB R_{\rm NS} \sim 10^{-15} (m/m_p)\,B_{15}^{-1}$. 
The magnetic field lines are not perturbed significantly
by the plasma inertia, and they can be thought of as fixed ``rails'' along 
which the particles move. The particle motion is confined to the
lowest Landau level and is strictly parallel to the field.

2. The particle motion is collisionless in the magnetosphere.
It is governed by two forces only: the component of gravity projected onto 
the magnetic field and a collective electric field $E_\parallel$ which is 
determined by the charge density distribution and must be found 
self-consistently.   

3. The star possesses a dense and thin atmospheric layer.\footnote{Such a 
layer initially exists on the surface of a young neutron star. Its 
maintenance is discussed in \S~\ref{spall}.} 
Near the base of the atmosphere, the required current
$\bj_B=(c/4\pi)\nabla\times {\mathbf B}$ is easily conducted,
with almost no electric field. Therefore the circuit
has simple boundary 
conditions: $E_\parallel=0$ and fixed current. The atmosphere 
is much thicker than the skin depth of the plasma and 
screens the magnetospheric electric field from the star. 

Our goal is to understand the plasma behavior above the screening 
layer, where the atmospheric density is exponentially reduced and
an electric field $E_\parallel$ must develop.
The induced electric field $\bE$ and conduction current $\bj$ satisfy the 
Maxwell equation,
\be
\label{eq:Maxw_}
  \nabla\times \bB=\frac{4\pi}{c}\bj
                   +\frac{1}{c}\frac{\partial \bE}{\partial t}.
\ee 
Here $\bj$ is parallel to the direction of the magnetic field. 
Projection of equation~(\ref{eq:Maxw_}) onto the magnetic field gives
\be
\label{eq:jB}
   \frac{\partial E_\parallel}{\partial t}=4\pi(\jB-j), \qquad
   \jB\equiv\frac{c}{4\pi}\,|\nabla\times \bB|,
\ee
where we have taken into account the force-free condition 
       $\nabla\times \bB\parallel\bB$.
If the conduction current $j$ is smaller than $\jB$, 
then $\partial E_\parallel/\partial t>0$ and an electric field appears
that tends to increase the current. Alternatively, if $j>\jB$ then 
an electric field of the opposite sign develops which tends to 
reduce the current. Thus, $j$ is always regulated toward $j_B$.
This is the standard self-induction effect.

The timescale of this regulation, $\tau$, is very short.
Consider a deviation $|j-\jB|\sim \jB$; then equation~(\ref{eq:jB}) gives
\be
\label{eq:tau1}
  \frac{E_\parallel}{\tau}\sim 4\pi\jB.
\ee 
The current $j=env$ (carried by particles with charge $e$
and mass $m$) responds to the induced $E_\parallel$ by changing $v$,
\be
\label{eq:tau2}
  \frac{v}{\tau}\sim \frac{eE_\parallel}{m}.
\ee
From equations~(\ref{eq:tau1}) and (\ref{eq:tau2}) we get
\be
\label{eq:omega_p}
   \frac{1}{\tau^2}\sim\frac{4\pi e^2 n}{m}=\omega_P^2.
\ee 
Thus the timescale of the current response to deviations of $j$ from $\jB$
is simply the Langmuir plasma timescale $\omega_P^{-1}$, which is very short.
Taking the characteristic density of the corona,
\be
   n\sim n_c\equiv\frac{\jB}{ec},
\ee
and $\jB\sim (c/4\pi)(B/r)$ (eq.~\ref{eq:j}), 
we estimate $n\sim 10^{17}B_{15}r_6^{-1}$cm$^{-3}$. 
The corresponding electron plasma frequency ($m=m_e$) is 
$\omega_P\sim 10^{13}$~Hz. The current relaxes to 
$\jB$ on the timescale $\omega_P^{-1}\sim 10^{-13}$~s, which is
tiny compared with the light crossing timescale of the circuit.

Local deviations from charge neutrality tend to be erased on the same 
plasma timescale.
When the corotation charge density is neglected, the net charge imbalance 
of the corona must be extremely small, i.e. the corona is neutral. 
Indeed, the total number of paricles in the corona is 
$N_c\sim n_c\RNS^3\sim 10^{35}$ and a tiny imbalance of positive and 
negative charges $\delta N_c/N_c\sim 10^{-14}$ would create an electric 
field $E\sim e\delta N_c/\RNS^2 >GMm_p/e\RNS^2$ that easily pulls out the 
missing charges from the surface. 

Adopting a characteristic velocity $v\sim c$ of the coronal particles,
the Debye length of the plasma $\lD=v/\omega_P$ is essentially
the plasma skin depth $c/\omega_P$. An important physical 
parameter of the corona is the ratio of the Debye length to the circuit 
size $L\sim\RNS$. 
\be
\label{eq:zeta}
  \zeta=\frac{\lD}{L}=\frac{c}{L\omega_P}.
\ee
In the case of the twisted dipolar magnetosphere (eq. [\ref{eq:j}]), this
becomes
\be\label{zetwist}
  \zeta= 3\times 10^{-9}\,B_{15}^{-1/2}
\left(\Delta\phi\,\sin^2\theta\right)^{-1/2}.
\ee
The same parameter may be expressed using a dimensionless current
$j_B/j_*\sim 10^{17}B_{15}\Delta\phi\sin^2\theta$.
Here $j_*$ corresponds to the plasma density $n_*=j_*/ec$ such that $\lD=L$,
\be
\label{eq:j*}
  \frac{\jB}{j_*}=\zeta^{-2},
  \qquad
  j_*=\frac{m_ec^3}{4\pi eL^2}.
\ee
Note that equation~(\ref{eq:omega_p}) for the plasma frequency is 
valid only for non-relativistic plasma. We will use the above definition
of $\omega_P$, however, one should keep in mind that plasma oscillations 
in the corona may have a smaller frequency because of relativistic effects.

\subsection{Quasi-Neutral Steady State?}

Suppose for now that no $e^\pm$ are created in the corona, so that the 
circuit must be fed by charges lifted from the stellar surface.
May one expect that a steady current $j=j_B$ is maintained 
in the neutral corona by a steady
electric force $eE_\parallel\sim gm_i$ or $eE_\parallel\sim gm_e$? 
It turns out that such a balance between gravitational and electric forces 
is impossible in the circuit for the following reason: $E_\parallel$ exerts 
opposite 
forces on positive and negative charges while 
the force of gravity has the same sign. 
This precludes a steady state where $E_\parallel$ lifts particles into the 
magnetosphere without a significant deviation from neutrality appearing 
somewhere in the circuit. The point is illustrated by the following toy model.

Suppose positive charges $e$ (mass $m_+$) flow with 
velocity $v_+>0$ and negative charges $-e$ (mass $m_-$) flow with 
velocity $v_-<0$.  The flow of each species is assumed to be cold in 
this illustration, with a vanishing velocity dispersion at each point, 
and the total current is $j=j_++j_-$ where $j_\pm=\pm en_\pm v_\pm$.
The continuity equation for $+/-$ charges reads
\be
 \frac{\partial (\pm en_\pm/B)}{\partial t}
   +\frac{\partial (j_\pm/B)}{\partial l}=0,
\ee
where $l$ is the coordinate along the field line.
Suppose the circuit is steady, $\partial/\partial t=0$, and no strong 
deviation from neutrality occurs, i.e. $n_+\simeq n_-$ everywhere. 
Then one finds $j_+/j_i=-v_+/v_-=const$, that is
the ratio of velocities must remain constant along the field line.
Hence $v_+$ and $v_-$ must {\it both} vanish at the footpoints.
This cannot be: for example, at the anode $E_\parallel$ must lift ions 
against gravity, and hence it is accelerating the electrons downward 
rather than stopping them\footnote{A static plasma
configuration is possible with the electric and gravitational forces 
balancing each other at each point of the circuit. However, $j=0$ in this 
configuration and it is not relevant to our circuit 
problem.  Besides, such a configuration would be unstable.}

So, if the current-carrying particles are supplied by the star, we expect 
either a significant charge separation at some places in the 
circuit, or a time-dependent behavior (or both). 
One must also allow for the creation of $e^\pm$ pairs
if the light charges are accelerated
to high enough energies, which will be discussed in \S~5.
Our approach to this problem is a direct numerical experiment that simulates
the time-dependent behavior of plasma particles in the circuit.


\subsection{Numerical Simulation of a One-Dimensional Circuit}\label{sc:sim}

We shall describe below a time-dependent simulation of a circuit operating 
along a thin magnetic tube with given $\jB=(c/4\pi)|\nabla\times{\bf B}|$ 
that is fixed in time.  
Strictly speaking, $\jB$ is {\it quasi}-steady: it will be found to 
decay on a long resistive timescale ($\sim $~yrs), much longer than the 
dynamic time of the current $\tdyn=L/c\sim 10^{-4}$~s. The purpose of 
the simulation is to understand how plasma is supplied above the star's
surface layer, and to find the electric field that develops along the tube.

The electric field may be decomposed as $\bE=\bE_\parallel+\bE_\perp$ 
where $\bE_\parallel$ is parallel to the magnetic field. It is 
$\bE_\parallel$ that governs the plasma motion and determines the release 
of magnetic energy in the circuit.
In the fixed magnetic configuration, $\nabla\times \bB$ does not vary 
with time and remains parallel to $\bB$ (the force-free condition).
Then equation~(\ref{eq:Maxw_}) implies that $\partial\bE/\partial t$ is 
parallel to $\bB$, i.e.
no perpendicular field $\bE_\perp$ is created by the induction effect,
and $\nabla\cdot\bE_\perp=0$ may be assumed. Then Gauss' law reads
\be
\label{eq:Gauss1}
   4\pi\rho=\nabla\cdot\bE_\parallel=\frac{\dd E_\parallel}{\dd l},
\ee
where $l$ is length measured along the magnetic tube. The condition
$\nabla\cdot\bE_\perp=0$ significantly simplifies the 
problem: it becomes strictly 1-D since $\bE_\perp$ has no 
relation to charge density and falls out from the problem. 
Hereafter we assume $\bE=\bE_\parallel$ and drop the subscript $\parallel$.

We note that the approximation of 1D circuit, $\jB(t)=const$,
excludes the excitation of transverse waves in the magnetosphere,
which in reality can exist.
These waves are described by the coupled fluctuations $\delta\jB$
and $\bE_\perp$ on scales much smaller than the circuit size $L$.
(Wavelengths $\lambda\sim c/\omega_P$ can be excited by
plasma oscillations.)  These fluctuations may be expected to have a 
small effect on the circuit solution 
unless $\delta\jB$ becomes comparable to $\jB$.

With $\jB(t)=const$, the particle motions 
on different magnetic lines are decoupled. Indeed, the particle dynamics 
is controlled by electric field $E=E_\parallel$ which 
is related to the instantaneous charge distribution by Gauss'
law (\ref{eq:Gauss1}).\footnote{The Maxwell equation~(\ref{eq:jB}) and 
Gauss' law (\ref{eq:Gauss1}) are equivalent when charge conservation 
$\partial j/\partial l=-\partial\rho/\partial t$ is taken into account.} 
Thus, $E(l)$ on a given magnetic line is fully determined by 
charge density $\rho(l)$ {\it on the same line}. The plasma and electric 
field evolve along the line as if the world were 1-D. 

The force-free condition $\bj_B\times\bB$ together with $\nabla\cdot\bj_B=0$
requires that $j_B(l)\propto B(l)$ along the magnetic 
line.   We can scale 
out this variation simply dividing all local quantities (charge density, 
current density, and electric field) by $B$. This reduces the problem to an 
equivalent problem where $\jB(l)=const$. Furthermore, only
forces along magnetic lines control the plasma dynamics,
and the curvature of magnetic lines falls out from the problem.
Therefore, we can set up the experiment so that plasma 
particles move along a straight line connecting anode and cathode (Fig.~2). 
We designated this line as the $z$-axis, so that $l=z$.

\begin{figure}
\begin{center}
\epsscale{.90}
\plotone{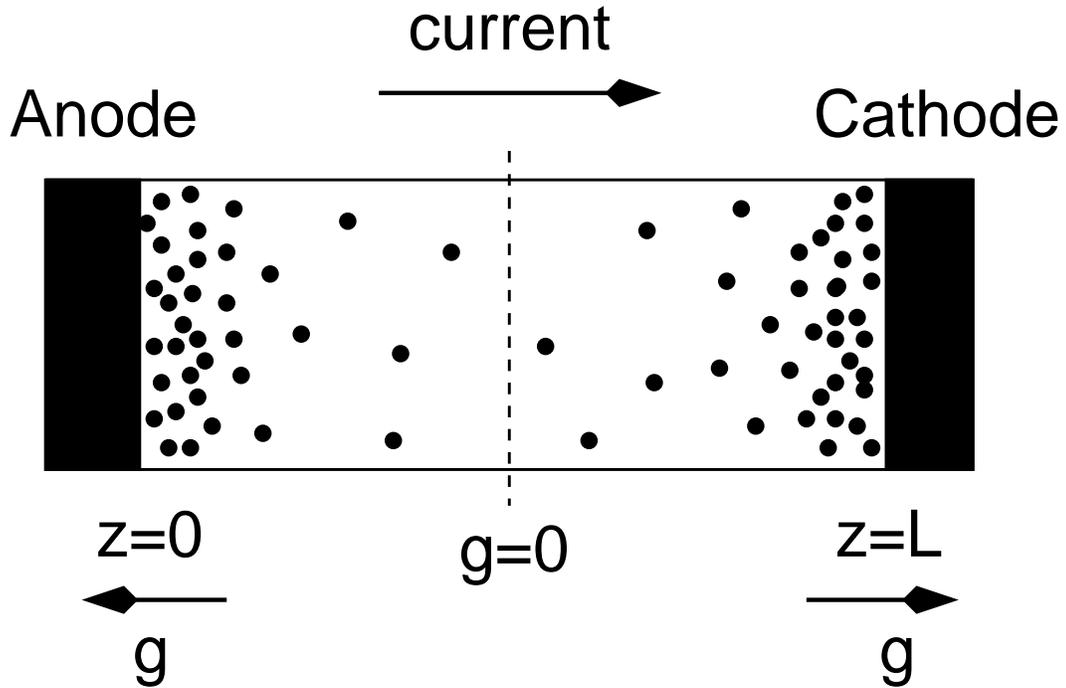}
\label{fig2}
\caption
{Set up of numerical experiment. Thin and dense plasma layers
are maintained near the cathode and anode by injecting cold particles 
through the boundaries of the tube (the footpoints of a magnetic flux tube). 
The electric current is kept constant at the boundaries and the system 
is allowed to evolve in time until a quasi-steady state is reached. 
Without voltage between anode and cathode, the current cannot flow 
because gravity $g$ traps the particles near the boundaries.
The constant current at the boundaries, however, implies that 
voltage is immediately generated should the flow of charge stop 
in the tube. The experiment aims to find the induced voltage
that keeps the current flowing.
}
\end{center}
\end{figure}

This one-dimensional system may be simulated numerically. 
Consider $N$ particles moving along the $z$-axis ($N\sim 10^6$ in our
simulations). Their positions $z_i$ and velocities $v_i$ evolve in 
time according to
\be
\label{eq:pev}
  \frac{\dd z_i}{\dd t}=v_i, \qquad 
  \frac{\dd p_i}{\dd t}=g \gamma_i m_i+e_iE,
  \qquad i=1,..., N,
\ee 
where $e_i$ and $m_i$ are charge and mass of the $i$-th particle,
$p_i=\gamma_im_iv_i$, and $\gamma_i=(1-v_i^2/c^2)^{-1/2}$.  Also,
$g(z)$ is the projection of the gravitational acceleration onto the 
magnetic line. In the simulations, we assume $g=g_0(2z/L-1)$ where $L$ 
is the distance between the anode and cathode; then $g=\pm g_0$ at the 
cathode/anode surfaces and $g=0$ in the middle $z=L/2$.
The true gravitational acceleration depends on $\gamma$ and could be 
refined at large $\gamma$ by a factor $\sim 2$ using the exact geodesic 
equation. However, the crude Newtonian model is sufficient for our purposes. 
The electric field appearing in equation~(\ref{eq:pev}) is given by 
\be
\label{eq:Gauss}
   E(z)=4\pi\int_0^z\rho\,\dd z,
\ee
where $E(0)=E(L)=0$ are the boundary condition discussed above.
Since the current $j=\jB$ is fixed at both ends, the net charge between
anode and cathode does not change with time 
and remains zero (in agreement with 
Given the instantaneous positions $z_i$ of charges $e_i$ we immediately find 
$E(z)$ from Gauss's law (eq.~[\ref{eq:Gauss}]).  Thus, we have a 
well-defined 1-D problem of particle motion in a self-consistent electric 
field and the fixed gravitational field.
To find a quasi-steady state of such a system we start with an initial 
state and let it relax by following the particle motion.

The evolution of the system 
may be followed without calculating $\rho$ 
or $j$: the equations contain only integrated charge $Q(z)=\int_0^z\rho dz$. 
In contrast with usual particle-in-cell simulations, no grid or cells 
are needed. We find
the {\it exact} $E$ at any point $z$ simply by 
counting the net charge of particles between $z$ and anode. The only 
source of numerical error is caused by the finite timestep of 
particle dynamics, and we keep it small,
$\Delta t=10^{-5}-10^{-4}(L/c)$.

The main dimensionless parameter of the problem is
$\zeta=c/\omega_PL\sim 3\times 10^{-9}$ (see \S~\ref{sc:basic}). 
Although it is not possible to simulate circuits with such a low value of 
$\zeta$, we can experiment with a similar system with 
$\zeta\simeq 10^{-2}$. This allows one to understand the mechanism of
the circuit and find plasma parameters that will be easy to scale
to a real magnetosphere.
A large number of particles $N\simgt 10^6$ may be followed with modern 
computers in a reasonable time, and $\zeta\ll 1$ may be achieved by 
choosing an appropriate charge per particle. Two other basic 
requirements must be satisfied in the simulations:
(1) the number of particles within the Debye length is much larger than one;
and (2) the timestep is much smaller than $\omega_P^{-1}$ and small 
enough to follow accurately the particle dynamics.
Finally, we require the particle charge $e$ to be sufficiently small, so 
that its electric field in our 1-D problem satisfies $eE=4\pi e^2\ll m_e g$.
Given a number of particles $N$, this last condition constrains $\zeta$ 
from below. The minimum $\zeta$ achieved in our numerical experiment is
$\sim 5\times 10^{-3}$.

The ions and electrons in the simulated circuit have different masses
$m_i$ and $m_e$. The ratio $m_i/m_e$ is smaller than it would be in a 
real magnetosphere. Below we will show models with $m_i/m_e=10$ and 30.  
Hereafter subscripts $i$ and $e$ refer to ions and electrons, 
respectively, and subscript $p$ will refer to positrons. 

The neutral atmospheric layers near the boundaries of the computational 
box $z=0,L$ are created by injecting a flux of positive and negative 
charges with small temperatures $T_e$ and $T_i$.
Although most charges are bound to the star, some of them must be
lifted by the electric field into the magnetosphere. The numerical 
simulation will show precisely how this happens.
The boundary conditions $E=0$ and $j=\jB$ are imposed {\it beneath}
the atmospheric layers. The condition $j=\jB$ implies keeping track
of the flux of leaving charges, and injecting new particles with
the correct imbalance between the positive and negative charges.
Thus the required current is fed at a constant rate at the conducting
boundaries. If net charge escaping through the cathode boundary during 
timestep $dt$ is smaller than $\jB dt$, we inject electrons to maintain 
the required $\jB$, otherwise we inject the needed number of ions. 
Similarly, $j=\jB$ is enforced at the anode boundary. 
This maintains the current, but does not determine the
contributions to $\jB$ from different types of particles. The plasma 
self-organizes during the experiment and decides itself what the 
contributions must be.

The forced injection of net positive charge at anode and 
net negative charge at cathode would immediately create a long range 
electric field if the current were not flowing in the system. 
A deviation of $j$ from $\jB$ at any point of the circuit leads to 
appearance of $E$ at that point. Thus, the boundary 
conditions $j(0)=j(L)=\jB$ are equivalent to equation~(\ref{eq:jB}).

The electric field does a net positive work on the current-carrying 
particles. In our experiment, the deposited kinetic energy sinks through 
the boundaries and never comes back.
We envision dense radiatively efficient layers just outside the simulation 
box and assume that the lost energy is emitted there.
The surface layers maintained in the experiment have density only 
$\sim 20-100$ times $n_0=\jB/ce$, which is just sufficient to screen the 
electric field and avoid interference of the boundaries in the 
circuit solution. A future full simulation may include radiative losses 
{\it in situ} in the tube and allow one to track the fate of released 
energy and the spectrum of produced radiation. 

Our final choice regards the initial plasma state in the tube.
We tried different initial states and got same results. 
In the runs shown below we take as initial conditions $j(z)=j_B$, 
a mildly relativistic bulk velocity $v_e=v_i=0.4c$ and Maxwellian 
momentum distribution with temperature equal to the bulk kinetic energy.

The problem has four scales: tube length $L$, electron plasma scale 
$\lambda_{{\rm D}e}=c/\omega_{Pe}$, ion plasma scale
$\lambda_{{\rm D}i}=(m_i/m_e)^{1/2}\lambda_{{\rm D}e}$, 
and scale-height of the atmospheric layer $h=k(T_e+T_i)/g(m_i+m_e)$. 
For a real magnetosphere, 
$\lambda_{{\rm D}e}<\lambda_{{\rm D}i}< h < L$. We choose the parameters 
of the experiment so that these relations are satisfied. The inclusion of 
pair production in the simulation brings a new scale: mean free path $l$
(see \S~3.5 below).

A numerical code implementing the described method was developed and 
first tested on simple plasma problems without gravity, $g=0$. 
For example, we calculated the two-stream instability starting with two 
oppositely directed cold beams of positive and negative charges,
and found the correct development of Langmuir oscillations on the plasma
timescale. Various technical tests have been done, e.g. independence  
of the results of timestep and charge of individual particles $e$. 
The main parameter of the problem is $\zeta$, and circuits
with equal $\zeta$ and different $N$, $e$, $dt$ gave the same result. 
Another test passed is the agreement of the linear-accelerator state 
obtained by the code (\S~5) with the solution of Carlqvist (1982).

As we shall see below, $e^\pm$ creation is inevitable in the magnetospheric
circuit. However, first we investigate what happens in the circuit with if
pair creation ``switched-off,''
when the current-carrying charges must be supplied by the atmospheric layers.


\section{CIRCUIT WITHOUT PAIR CREATION}

The circuit model has the following parameters:
(1) dimensionless current $j_B/j_*=\zeta^{-2}$ (see \S~\ref{sc:basic}),
(2) gravitational potential barrier $\Phi_g$ 
(we will assume $\Phi_g=c^2/4$ in numerical examples),
(3) density of the hydrostatic
boundary layer $\natm$ (which is regulated in the 
numerical experiment by the particle injection rate at the boundaries), 
and (4) injection temperatures of ions, $T_i$, and electrons, $T_e$.

Note that $T_i$ and $T_e$ may be well above
the surface temperature of the star $\kB T\sim 0.5$~keV
because the atmosphere is heated from above by the corona. 
The vertical temperature profile that is established below the corona
is a separate problem which we do not attempt to solve in this paper.
$T_i$ and $T_e$ serve as parameters of our model, and we explore 
first how the mechanism of the circuit depends on these parameters.

\subsection{Thermally-Fed Circuit}

If the hydrostatic
atmosphere is sufficiently hot, its Boltzmann tail at large heights 
may be able to conduct the current without the need to lift particles by an 
electric field. The plasma must remain quasi-neutral in the Boltzmann tail. 
One may think of the neutral atmosphere as a gas of composite particles 
$m=m_i+m_e$ with effective temperature $T=T_i+T_e$. Its scale-height is 
given by
\be
\label{eq:h}
  h=\frac{k_{\rm B}T}{g_0m}=\frac{\kB(T_i+T_e)}{g_0(m_i+m_e)}.
\ee
The plasma density is reduced exponentially on scale $h$. The reduced 
density remains high enough to conduct the current $\jB$ all the way to 
the top of the gravitational barrier if
\be
\label{eq:thermal}
  \natm \exp\left(-\frac{m\Phi_g}{k_{\rm B}T}\right) > n_c = \frac{j_B}{ec}.
\ee

One example of a circuit that satisfies condition~(\ref{eq:thermal})
is shown in Figure~3. The electrons and ions are
injected at the boundaries with $k_{\rm B}T_i=k_{\rm B}T_e=0.5m_ec^2$
and $m_i=10m_e$; this gives a scale-height $h=0.09L$.
After several light-crossing times $L/c$, the circuit develops
the Boltzmann atmosphere that conducts the required current $\jB=10^3j_*$.
The electric field remains small and dynamically unimportant, $eE<g_0m_i$.

\begin{figure}
\begin{center}
\epsscale{1.0}
\plotone{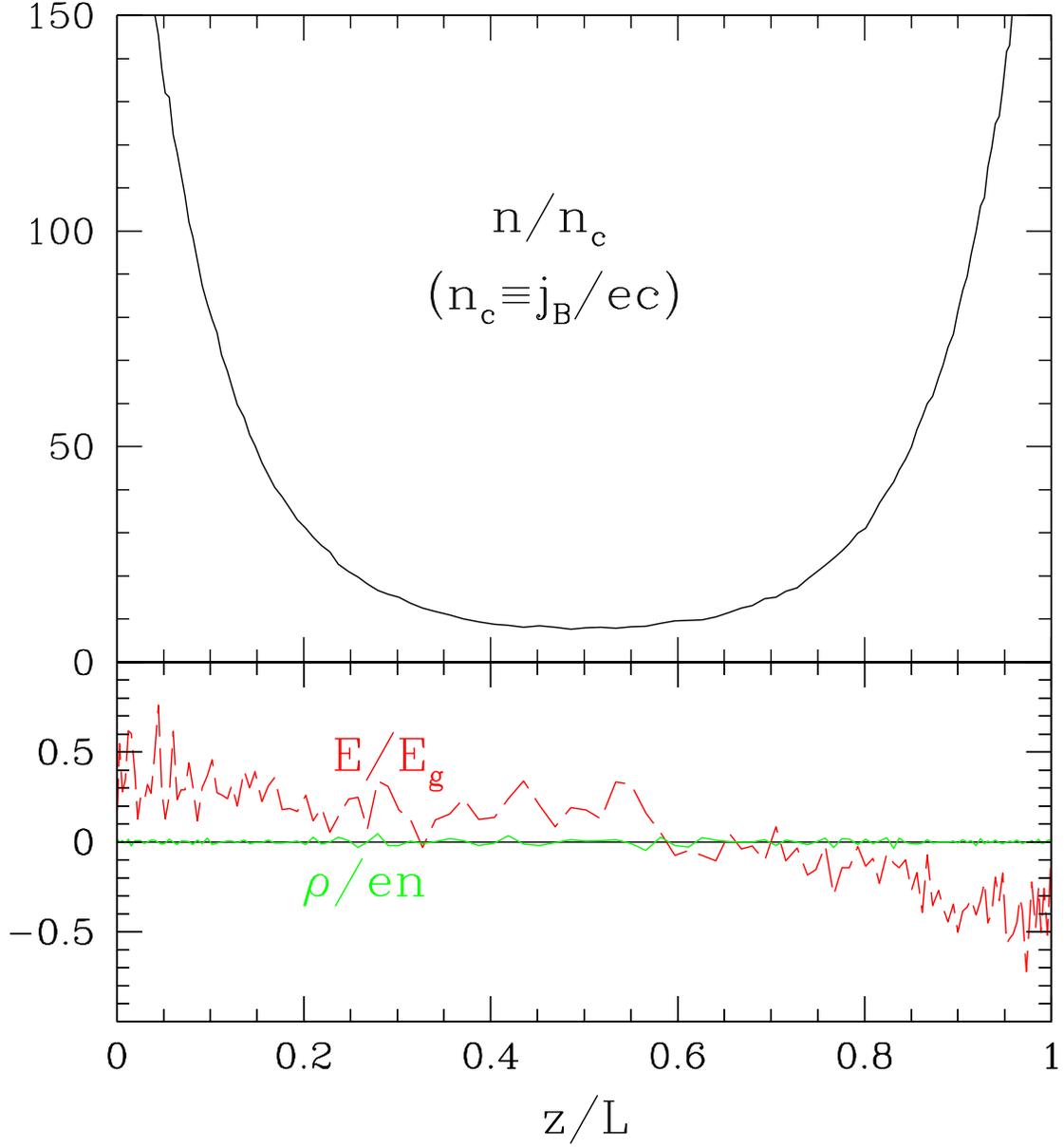}
\label{fig3}
\caption
{Thermally-fed circuit without $e^\pm$ production. 
Upper panel: the plasma density $n=n_e+n_i$ in units of $n_c=\jB/ec$.
Lower panel: deviation from neutrality $\rho/en$, and 
electric field in units of $E_g$ defined by $eE_g=m_ig_0$. 
In this simulation, $m_i=10m_e$ and $\zeta=c/\omega_PL\simeq 0.03$. 
The snapshot is taken at $t=20L/c$, after the system has reached the
quasi-steady state.
}
\end{center}
\end{figure}

The thermally-fed regime requires too high a temperature for the pure
ion-electron circuit. Indeed, $mc^2=(m_i+m_e)c^2\simeq 1$~GeV and 
then $k_{\rm B}T> 1{\rm~GeV}/\log(\natm/n_c)$ would be required. 
Pair creation would take place at much lower temperatures
(see \S~\ref{pairatm} below). Therefore,
the thermally-fed pair-free circuit solution cannot apply to a real
magnetosphere. The rest of this section focuses on circuits that do not
satisfy condition~(\ref{eq:thermal}).

\subsection{Linear Accelerator}
\label{LA}

When the condition~(\ref{eq:thermal}) is not satisfied, the circuit 
becomes plasma-starved and an electric field develops that lifts 
particles from the dense surface layer.
We find that such a circuit quickly relaxes to a state where it
acts as an ultra-relativistic linear accelerator. 
One example is shown Figure~4. The parameters of
this particular model are $\jB/j_*=10^4$, $k_{\rm B}T_e=
0.1k_{\rm B}T_i=0.04m_ec^2$, and 
$\natm/n_c\simeq 20$. We obtained similar solutions for various combinations
of $T_e$, $T_i$, and $\natm/n_0$ that do not satisfy 
condition~(\ref{eq:thermal}) of a thermally-fed regime.
In all cases, the circuit reached the quasi-steady state at $t\simlt 5L/c$.
In this state, 
oscillations of electric field (on the plasma timescale $\omega_P^{-1}$) 
are confined to the thin atmospheric layers. A static accelerating electric 
field is created above the layers where the atmosphere density is 
exponentially reduced. The striking result is that on top of each
layer, a large charge builds up, which is positive ($+Q$) at the anode 
and negative ($-Q$) at the cathode. The two charges create a long-range 
electric field $E=4\pi Q$.

\begin{figure}
\begin{center}
\plotone{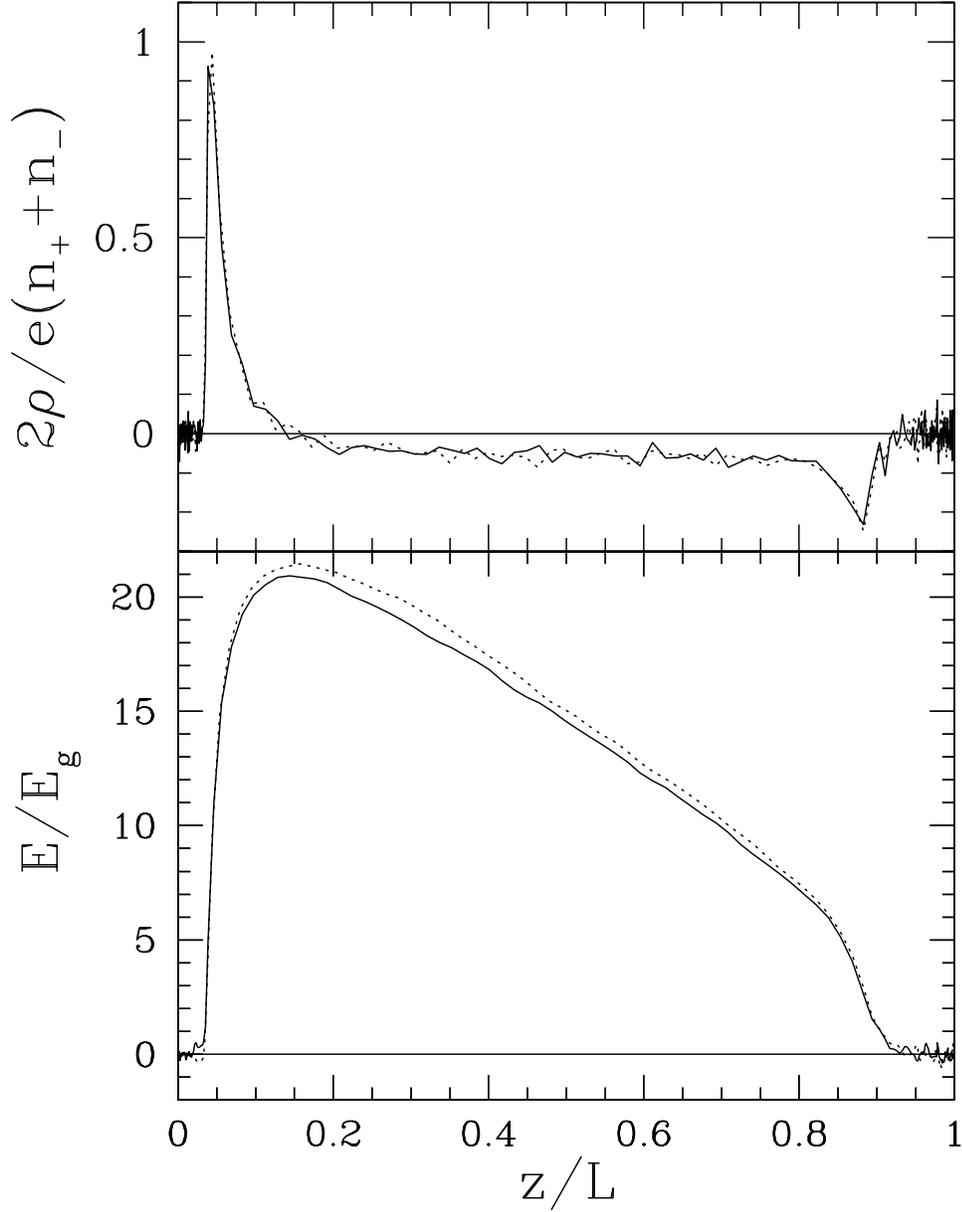}
\label{fig4}
\caption
{Circuit without $e^\pm$ production. Upper panel shows the normalized 
charge density and lower panel shows the electric field in units of 
$E_g$ defined by $eE_g=m_ig_0$. Solid and dotted curves correspond to two 
different moments of time, demonstrating that a quasi-steady state has 
been reached. In this simulation, $m_i=10m_e$ and 
$\zeta=c/\omega_PL\simeq 0.01$. The final state is the relativistic 
double layer described analytically by Carlqvist (1982). 
}
\end{center}
\end{figure}

This configuration is a relativistic double layer (Carlqvist 1982).
It is well described by Carlqvist's solution, which has no gravity 
in the circuit and assumes zero temperature at the boundaries, so that 
particles are injected with zero velocity. According to this solution,
the potential drop between anode and cathode, in the limit $e\Phi_e\gg m_ic^2$,
is given by 
\be
\label{eq:PhiC}
   e\Phi_e = \frac{1}{2}\left[\left(\frac{m_i}{Zm_e}\right)^{1/2}+1\right]
   \,L\omega_P\, m_ec, \qquad 
   \omega_P\equiv\left(\frac{4\pi j e}{m_ec}\right)^{1/2},
\ee
where $Z$ is the charge number of the ions (in our experiment $Z=1$ was
assumed). In the limit $m_i\gg Zm_e$, 
this expression depends only on the ion and not the electron mass.

The established voltage is much larger than is needed to 
overcome the gravitational barrier $\Phi_g$.  It does not even depend on 
$\Phi_g$ as long as $\Phi_g$ 
is large enough to prohibit the thermally-fed regime.
Gravity causes the transition to the
linear-accelerator state, but the state itself does not depend on $\Phi_g$.
The asymmetry of the double-layer solution (Fig.~4) is caused by the 
difference between the electron and ion masses. A similar but symmetric 
configuration is obtained for $m_i=m_e$. 

The symmetric double layer
with $m_e=m_i$ is especially simple and can be 
qualitatively understood as follows. The currents carried by ions and 
electrons are equal by symmetry, $j_i=j_e=\frac{1}{2}\,j$, and hence 
plasma is approximately neutral between the anode and cathode where 
$v_i\simeq v_e\simeq c$.
A significant deviation from neutrality $|n_e-n_i|\sim n_i+n_e$ takes place 
in the acceleration regions $\Delta z$ near anode and cathode
because here $v_i$ differs from $v_e$. 
For instance, near the 
anode the ions are not accelerated yet to $c$, and their 
charge density $\rho_i=j_i/v_i$ is larger than that of the incoming electron 
beam, $\rho_e=j_e/v_e\simeq j_e/c$, so the net charge density is positive. 
One can show that the anode charge $Q(z)=\int_0^z (j_i/v_i-j_e/v_e)dz$ 
peaks at height $z\sim \Delta z$ 
where ions move with a mildly relativistic velocity $v_i\sim c/2$.
Similarly, at cathode end, where the emitted electron flow 
is still slow, the net charge density is negative, and the charge $-Q$ peaks
at the distance $\Delta z$ from the cathode. The thickness $\Delta z$ 
and the characteristic electric field $E$ in the mildly relativistic region 
are related by $eE\Delta z\sim m_ec^2$. Combining this
with $E=4\pi Q\sim 4\pi en_c\Delta z$, one finds $\Delta z\sim\lD$ and 
$Q\sim en_c\lD$. This is the only possible self-consistent static solution 
for a double layer (besides the trivial solution $E=0$ which is prohibited 
in our case by gravity).
$E=4\pi Q$ must be this large in a static double layer because
a smaller $E$ would accelerate the particles more slowly, which would 
imply larger non-neutral regions $\Delta z$ near the ends of the circuit
and larger $Q$.  Gauss' law would then imply a larger 
$E$, leading to inconsistency with the assumed smaller $E$. 

Similar estimates may be made for the asymmetric double layer with $m_i>m_e$.
In the limit $e\Phi_e\gg m_ic^2$, one again finds 
$j_i\simeq j_e$ from the Langmuir condition (see Carlqvist 1982),
\be
 \frac{j_i}{j_e}=\left(\frac{e\Phi_e+2m_ec^2}{e\Phi_e+2m_ic^2/Z}\right)^{1/2} 
                 \simeq 1.
\ee
There is only one qualitative difference from the symmetric case.  Now the
mildly relativistic region near the anode ($v_i\sim c/2$) has a thickness 
$\Delta z_A\sim \lambda_{{\rm D}i}$, and near the cathode ($v_e\sim c/2$) 
--- $\Delta z_C\sim \lambda_{{\rm D}e}$.
The spikes of charge near anode and cathode are now different: the cathode
spike is smaller by a factor of 
$\lambda_{{\rm D}e}/\lambda_{{\rm D}i}=Z(m_e/m_i)^{1/2}$.
The net charge of a double layer, however,
must vanish to satisfy the boundary 
conditions $E=0$. The ``missing'' negative charge is distributed 
between the two spikes, throughout the circuit. This behavior is 
observed in Fig.~4 and derived analytically by Carlqvist (1982).

The static double-layer solution has been much studied previously, 
but its possible astrophysical applications remained unclear.
Our numerical experiment shows that an ion-electron circuit in a 
gravitational field relaxes to the double layer of the macroscopic size 
$L$ and huge voltage $\Phi_e$. 
The system does not find any state with a lower $\Phi_e$, even though it 
is allowed to be time-dependent. The experiment shows that, regardless
the initial state, the charges 
$\pm Q$ near the anode/cathode build up with time in spite of the strong
fluctuations that persist in the atmospheric layers (reversing the sign 
of $E$), and $Q$ grows until the Carlqvist solution is reached.

The effective boundaries of the double layer in our experiment are 
not exactly at $z=0,L$ because the atmosphere has a finite scale-height.
Height $z_*$ where the circuit becomes plasma-starved
may be estimated from the condition $n_{\rm atm}\exp(-z_*/h)\sim n_c$.

If the linear accelerator were maintained in the twisted magnetosphere,
the twist would be immediately killed off. The huge voltage implies a large
untwisting rate and a fast dissipation of the toroidal magnetic energy (\S~2). 
The electron Lorentz factor developed in the linear 
accelerator is (taking the real $m_i/m_e=1836$ and $\zeta\sim 3\times 10^{-9}$),
\be
\label{eq:gamma}
    \gamma_e=\frac{e\Phi_e}{m_ec^2}\simeq \frac{20}{\zeta}
    \sim 6\times 10^9.
\ee
However, new processes will become important before the particles could
acquire such high energies: production of $e^\pm$ pairs will take place.
Therefore, the linear-accelerator solution cannot describe a real 
magnetosphere. We conclude that pair creation is a key ingredient of the 
circuit that will regulate the voltage to a smaller value.


\section{PAIR CREATION IN THE CORONA}\label{sc:pairs}

In the magnetospheres of canonical radio pulsars with $B\sim 10^{12}$~G, 
$e^\pm$ pairs are created when seed electrons are 
accelerated to large Lorentz factors $\gamma_e\sim 10^7$. Such electrons 
emit curvature gamma rays that can convert to $e^\pm$ off the magnetic field.

In stronger fields, another channel of $e^\pm$ creation appears. 
It is also two-step: an accelerated particle resonantly upscatters a 
thermal X-ray photon, which subsequently converts to a pair.
This channel is dominant in radio pulsars with relatively strong magnetic 
fields (e.g. Hibschman \& Arons 2001). 
A similar channel of $e^\pm$ creation operates in the super-QED field
of a magnetar. Because of the large magnetic field an accelerated electron 
can resonantly upscatter
the ambient X-rays and produces $e^\pm$ pairs when its Lorentz factor is
$\gamma_e\sim 10^3$, much smaller than in a normal pulsar.

\subsection{Resonant Channel of Pair Creation}

Consider an electron moving along a magnetic line with Lorentz factor 
$\gamma_e \gg 1$ and an ambient X-ray photon $\hbar \omega_X$ propagating 
at an angle $\theta_{kB}$ with respect to the electron velocity. In the 
electron frame, the photon is strongly aberrated and moves nearly parallel 
to the magnetic field with an energy 
$\hbar\omega_X' = \gamma_e(1-\cos\theta_{kB})\hbar\omega_X$. The resonant
scattering may be viewed as a two-step process: the excitation of the 
electron into the first Landau level with energy 
\be 
\label{eq:E_B}
E_B = \left(\frac{2B}{B_{\rm QED}}+1\right)^{1/2} m_ec^2,
\ee
followed rapidly by a de-excitation.
Here $\BQED=m_e^2c^3/e\hbar\approx 4.4\times 10^{13}$~G.

The resonant condition on photon energy $\hbar\omega^\prime_X$ is obtained
from energy and momentum conservation, 
\be
  \hbar\omega^\prime_X+m_ec^2=\gamma E_B, \qquad
  \frac{\hbar\omega^\prime_X}{c}=\gamma\beta \frac{E_B}{c}. 
   \qquad 
\ee
Here $\beta$ is the velocity along $\bB$ acquired by the excited electron 
(relative to its initial rest frame) and $\gamma=(1-\beta^2)^{-1/2}$.
The photon $\hbar\omega^\prime_X$ is assumed to propagate parallel to 
$\bB$, which is a very good approximation since we consider ultrarelativistic 
electrons and the ambient radiation is strongly beamed in the electron frame.
From these equations one finds
$\gamma=\frac{1}{2}(E_B/m_ec^2+m_ec^2/E_B)$ and 
\be
  \hbar\omega^\prime_X=\frac{1}{2}\left[
        \left(\frac{E_B}{m_ec^2}\right)^2-1\right]
    =\frac{B}{\BQED}\,m_ec^2=\hbar\,\frac{eB}{m_ec}. 
\ee
This resonant condition may also be written as a condition on $\gamma_e$
for given $\hbar\omega_X$ and $\theta_{kB}$,
\be
\label{eq:gres}
  \gres = {B/B_{\rm QED}\over1-\cos\theta_{kB}}\,
            \left({\hbar\omega_X\over m_ec^2}\right)^{-1}
          \approx 10^3\,B_{\rm 15}\,
            \left({\hbar\omega_X\over{\rm 10 keV}}\right)^{-1}.
\ee

De-excitation is examined conveniently in the frame 
where the parallel momentum of the excited electron vanishes
(Herold, Ruder, \& Wunner 1982). The energy of 
de-excitation photon $E_\gamma$
depends on its emission angle $\th$ with respect to the magnetic field.
This angle determines how the released energy $E_B$
is shared between the emitted photon and the electron recoil.
The relation
$E_\gamma(\th)$ is found from energy and momentum conservation,
\be 
  E_\gamma+(c^2p_e^2+m_e^2c^4)^{1/2}=E_B, \qquad E_\gamma\cos\th= -cp_e.
\ee
Here $p_e$ is recoil momentum of the electron, which is directed 
along $B$. Eliminating $p_e$, one finds
\be
  E_\gamma(\th)=\frac{E_B}{\sin^2\th}
     \left[1-\left(\cos^2\th+\frac{m_e^2c^4}{E_B^2}
        \,\sin^2\th\right)^{1/2} \right].
\ee  
In the super-critical field, $B\gg B_{\rm QED}$, this simplifies to
\be
\label{eq:E_gamma}
    E_\gamma(\th)\approx\frac{E_B}{1+|\cos\th|}.
\ee 
The de-excitation photon has the maximum energy $E_\gamma=E_B$ when it is 
emitted at $\th=\pi/2$ (no recoil) and the minimum energy 
$E_\gamma=E_B/2$ when $\th=0$ (maximum recoil). 

Photons have two polarization states in the strongly magnetized
background: O-mode and E-mode. These eigenmodes are determined by the effect
of vaccuum polarization in response to an electromagnetic wave. Both modes
are linearly polarized: the O-mode has its electric vector in the plane of 
the wave vector ${\mathbf k}$ and the background magnetic field $\bB$.
The electric vector of the E-mode is perpendicular to this plane.
De-excitation photons are emitted in either the O or E mode: the gyrational 
motion of the excited electron overlaps both linear polarization states.

O-mode photons can convert to a pair with both particles
in the lowest Landau state (Daugherty \& Harding 1983).
The threshold condition for this process is
\be
\label{thr_O}
    E_\gamma>E_{\rm thr}=\frac{2m_ec^2}{\sin\th}.
\ee
The threshold energy becomes $2m_ec^2$ when transformed
to the frame where the photon propagates perpendicular to $\bB$.
Using equation~(\ref{eq:E_gamma}), we find the interval of emission angles
$\th$ for which the O-mode photon will immediately convert to a pair,
\be
\label{eq:thrcond}
  1-|\cos\th|< 8\left(\frac{m_ec^2}{E_B}\right)^2, \qquad B\gg\BQED.
\ee 
O-mode photons emitted below the threshold still have huge energy in 
the lab frame, $\sim \gres E_B$, and will convert to $e^\pm$ after 
propagating a small distance $\Delta l\sim \RNS/\gres$. Indeed, even 
a photon emitted with $\th=0$ will quickly build up a sufficient pitch 
angle for conversion because of the curvature of magnetic lines and/or 
gravitational bending of the photon trajectory.

The threshold energy for an E-mode photon is $E_B$ in the frame where
the photon propagates normally to the field. Therefore, an E-mode 
de-excitation photon cannot immediately convert to $e^\pm$. It will, 
however, quickly split in two daughter photons, at least one of which 
will be in O-mode (Adler et al. 1970) and will convert to a pair.\footnote{
Splitting of the O-mode is kinematically 
forbidden. All three components of the momentum are conserved in the 
splitting process. This allows splitting only for the mode with the smaller 
index of refraction, which is the E-mode in a super-QED magnetic field. }  

Thus, resonant scattering in the ultrastrong field always leads to $e^\pm$ 
creation, sometimes multiple because E-mode may split repeatedly and
give multiple O-mode photons that convert to $e^\pm$. The average 
multiplicity ${\cal M}$ is not much larger than unity, and
${\cal M}=1$ may be taken as a first approximation.
The rate of pair creation by an accelerated electron is then
determined by
the cross section of resonant scattering and the number density of the 
target photons. The cross section of resonant absorption for a photon 
propagating parallel to $\bB$ in the rest frame of the electron 
is given by (Daugherty \& Ventura 1978),
\be
\label{eq:sigres}
  \sigma_{\rm res}(\omega_X^\prime) = 
    2\pi^2r_ec\,\delta\left(\omega_X^\prime - {eB\over m_ec}\right)
    =\sigma_{\rm res\,0}\, \omega_X^\prime\,
    \delta\left(\omega_X^\prime - {eB\over m_ec}\right),
\ee
where $r_e=e^2/m_ec^2$ and $\sigma_{\rm res\,0}=2\pi^2 e/B$.
This expression is valid for any $B/\BQED$. 

Magnetic Compton scattering 
has also been calculated for arbitrary $B/\BQED$  by Herold (1979)
and Melrose \& Parle (1983). The calculations
must be modified at the resonance to take into account the finite 
lifetime $\Gamma^{-1}$ of the intermediate resonant state 
(e.g. Ventura 1979; Harding \& Daugherty 1991). Using the results 
of Herold (1979) and Herold et al. (1982), we find the resonant Compton
cross section that agrees with equation~(\ref{eq:sigres}) within 
a factor of 2 (details will be given elsewhere). 
Below we use the cross section~(\ref{eq:sigres}).
Note also that at
$B\gg \BQED$, there are equal probabilities for the photon to end up 
in the E-mode or O-mode in the final state, independent of $\them$.

The number density of target photons for resonant scattering is 
\be
     n_X \approx \frac{\dd L_X/\dd\ln\omega_X}{4\pi\RNS^2\hbar\omega_X c} 
   \approx 2\times 10^{20} 
   \left(\frac{\dd L_X/\dd\ln\omega_X}{10^{35}{\rm ~erg~s}^{-1}}\right)
   \left({\hbar\omega_X\over {\rm keV}}\right)^{-1}\;{\rm ~cm}^{-3}.
\ee
The mean free path of an electron to scattering a photon is then
\be\label{mfp}
  l_{\rm res} 
    = {1\over n_X \sigma_{\rm res\,0}}\,\left|{d\ln B\over d\ln l}\right|
    \approx {3\over n_X \sigma_{\rm res\,0}} 
     \sim 10^3\,B_{15}
  \left(\frac{\dd L_X/\dd\ln\omega_X}{10^{35}{\rm ~erg~s}^{-1}}\right)^{-1}
    \left({\hbar\omega_X\over {\rm  keV}}\right)\;\;{\rm cm}.
\ee
The upscattered photons produce pairs at a small
distance from the scattering site compared with $l_{\rm res}$.  
So, the distance over which an accelerated electron creates a pair is
approximately equal to the free path to scattering.
It depends only on $B$ and the spectrum of target photons. 
While being accelerated, an electron with initially low $\gamma_e$ may 
first experience a resonance with a high-energy photon. Therefore
it is important to know
the radiation spectrum at high energies, above the blackbody peak.
It is possible that the transition region between the corona and the star 
emits 100~keV photons (Thompson \& Beloborodov 2005; see \S\S~6 and 8.5 
below), which may be targets for resonant scattering by electrons with 
relatively low $\gamma_e\sim 100$. Photons with energy much above 
$\sim 100$~keV are not good targets for resonant scattering in the 
strong-$B$ region near the star because then $\lres$ becomes comparable to
$\RNS$.

\subsection{Inconsistency of the Linear Accelerator}

Now we can show that the linear accelerator discussed in \S~4.2 
becomes inconsistent in the presence of pair creation and cannot
be established in the magnetosphere. 
Suppose it is established. Electrons in the accelerator will be in 
resonance with photons at the peak of the stellar blackbody spectrum,
$\hbar\omega_X\sim 1$~keV, when they have Lorentz factors 
$\gres\sim 10^4\,B_{15}$.
Using the Carlqvist solution for the 
electron Lorentz factor $\gamma_e(z)$, we find the characteristic distance 
from cathode where electrons reach $\gres$. This distance is given by
\be
  \Lres \simeq \gres\lambda_{{\rm D}e}.
\ee
Here we have used $\lDe\ll \Lres\ll (m_i/m_e)^{1/2}L$;
the electric field of the double layer
varies little across this region and equals
$E=4\pi(j/c)\lDe$, where $j$ is the current flowing 
through the circuit.  The resonance region is at the distance 
$\Lres\sim 30$~cm from cathode and its thickness is $\sim\Lres$. The mean 
free path of an electron in this region is $\lres\sim 10^3$~cm 
(see eq.~[\ref{mfp}]), which is much larger than $\Lres$. This means that 
only a small fraction $f=\Lres/\lres$ of the flowing electrons will 
upscatter a photon
and the bulk of them pass through the resonance region without any 
interaction. 

In sum, $f\ll 1$ 
positrons are created per unit electron flowing through the circuit.
The created positrons will flow to the cathode. 
This backflow carries a current $fj$ and charge density $\rho_+=fj/c$.
The charge density $\rho_+$ creates a potential drop 
between the resonance region and cathode,
\be
      \Delta\Phi=2\pi\,\frac{\Lres^3}{\lres}\,\frac{j}{c}.
\ee 
The linear-accelerator configuration becomes inconsistent if 
$e\Delta\Phi>\gres m_ec^2$ --- then the created positrons will screen 
the assumed electric field. This condition may be written as
\be
       \lres<\frac{1}{2}\,\gres^2\zeta L.
\ee
We can evaluate it by substituting $L = R_{\rm NS} = 10$~km and
equation~(\ref{mfp}) for $\lres$. Using the 
example of a twisted dipole magnetosphere with 
the corresponding expression~(\ref{zetwist}) for $\zeta$, one finds
\be
 {\dd L_X\over \dd\ln\omega_X} >  7\times  10^{32}
 \,B_{15}^{-1/2}\, (\Delta\phi)^{1/2}\sin\theta
 \,\left({\hbar\omega_X\over 1~{\rm keV}}\right)^3\;\;\;\;{\rm erg~s^{-1}}.
\ee
The linear accelerator configuration is inconsistent if this condition is 
satisfied at some $\hbar\omega_X$. It is easily satisfied at 
$\hbar\omega_X\sim 1$~keV 
(the observed luminosity in this range is $\sim 10^{35}$~erg/s).

Even if the linear-accelerator state were formally allowed as a 
self-consistent solution, the coronal circuit 
would not have to evolve to this state. Indeed, if the voltage in
the circuit evolves gradually to higher and higher values, an $e^\pm$
discharge occurs
well before the linear accelerator could be established. We now consider 
the circuit that operates on $e^\pm$ discharge, at a much lower voltage 
compared to the linear accelerator.

\subsection{Discharge via Pair Breakdown}

The development of pair breakdown may be illustrated with the following
toy model. Suppose a uniform electric field $E$ is fixed in the tube
of size $L$. To start, we introduce one seed electron at the cathode with 
zero velocity. The electron accelerates toward the anode and its Lorentz 
factor $\gamma_e$ grows.
The toy model assumes that the electron creates a pair when 
it reaches a certain threshold $\gres$. At this point, the electron 
energy is shared
between three particles, so the energy of the original electron is 
reduced. In this illustration we use a random 
energy distribution between the three particles with the mean values
of 1/2, 1/4, and 1/4 of $\gres m_ec^2$. This introduces stochasticity 
in the pair 
breakdown.\footnote{In reality there are other sources of stochasticity, 
in particular, the random free-path to resonant scattering 
with a mean value $\lres(\gamma_e)$. It is taken into account in the 
numerical experiment described below (\S~5.3). For the illustrative 
purposes of this section, it is sufficient to introduce any
random element in the pair creation process to make it stochastic.}

The key parameter is $\Lres/L$ where 
$$
  \Lres=\frac{(\gres-1)m_ec^2}{eE}
$$
is the length of electron (or positron) acceleration to the energy 
$\gres m_ec^2$, starting from rest. The necessary condition for breakdown 
is $\Lres/L<1$ --- otherwise the seed electron reaches the anode before 
gaining enough energy for pair creation.

\begin{figure}[h]
\begin{center}
\plotone{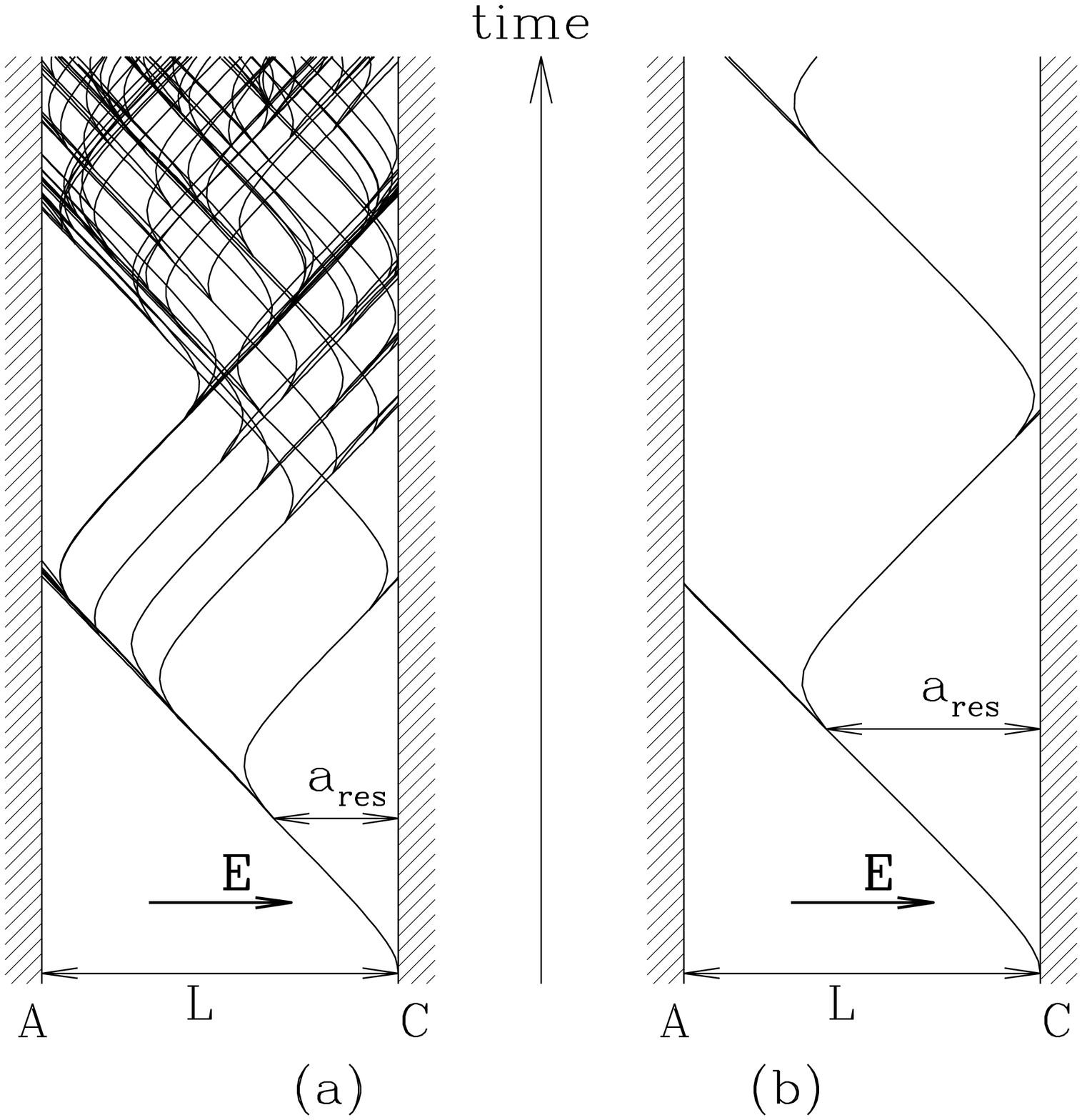}
\label{fig5}
\caption
{Spacetime diagrams illustrating the critical character of the 
$e^\pm$ breakdown and formation of avalanches. The toy model used
in this illustration is described in the text; $e\Phi_e=eEL=11m_ec^2$ 
is assumed. Worldlines of particles are shown, starting with one seed 
electron at the cathode. 
(a) $\Lres/L=0.35$ (supercritical case). 
(b) $\Lres/L=0.6$ (critical case).
The magnetar corona is expected to evolve close to the threshold for 
triggering $e^\pm$ avalanches, when $\Lres$ is comparable to but somewhat 
smaller than $L$ (or, equivalently, when the voltage $\Phi_e$ is just
above the minimum needed to trigger $e^\pm$ creation).
The $e^\pm$ avalanches are then initiated stochastically in the regime
of self-organized criticality. This expectation is confirmed by the 
numerical experiment described in \S~5.4.
}
\end{center}
\end{figure}

The development of the pair breakdown is shown in a spacetime diagram in 
Figure~5. 
Each pair-creation event gives two new particles of opposite charge, 
which initially move in the same direction. One of them is accelerated by 
the electric field and lost at the boundary after a time $<L/c$, 
while the other is decelerated and can reverse direction before reaching
the boundary. After the reversal, the particle is accelerated toward the 
opposite boundary and creates at least two new particles, one of which
can reverse direction etc. This reversal of particles in the tube 
and repeated pair creation allows the $e^\pm$ plasma to be continually 
replenished. In the super-critical regime
(left panel in Fig.~5) more than one reversing 
particle is created per passage time $L/c$ and an avalanche develops 
exponentially on a timescale $\sim \Lres/c$.
In the near-critical regime shown in the right panel of Fig.~5, 
just one reversing particle is created per passage time $L/c$. 
This critical state is unstable: sooner or later the avalanche will 
be extinguished.

This toy model is very idealized because it assumes a fixed electric
field $E$.  In reality, the behavior of $E$ 
is coupled to the charge density in the tube, which is determined
by the particle dynamics and the boundary conditions $j=\jB$.
$E$ changes on a timescale much shorter than $L/c$ and the behavior
of the coupled plasma and electric field is complicated.

The toy model, however, shows an essential property of the $e^\pm$ 
breakdown: it is a critical stochastic phenomenon. 
Above a critical voltage pair creation proceeds in a runway manner, 
and the current and the dissipation rate would run away if the voltage 
were fixed. Below the critical voltage pair creation does not ignite.
The criticality 
parameter is $L/\Lres=e\Phi_e/(\gres-1)m_ec^2$. The tube with enforced
current at the boundaries must self-organize to create pairs in the 
near-critical regime $e\Phi_e\sim \gres m_ec^2$ and maintain the current.

If the current demanded by $\nabla\times\bB$ is not maintained, the 
self-induction voltage will grow until it exceeds the critical value
$e\Phi_e\sim \gres m_ec^2$ (corresponding to $\Lres\sim L$) and a pair 
breakdown develops. 
Avalanches of $e^\pm$ then supply plasma that tends to screen the 
electric field by conducting the necessary current, and the 
voltage in the tube is reduced.
The discharging tube is similar to other phenomena that show self-organized 
criticality, e.g., a pile of sand on a table (Bak, Tang, \& Weisenfeld 1987).
If sand is steadily added, a quasi-steady state is established with a 
characteristic mean slope of the pile. The sand is lost (falls from the 
table) intermittently, through avalanches --- a sort of ``sand discharge.''
In our case, charges of the opposite signs are added steadily instead of 
sand (fixed $j$ at the boundaries), and voltage $\Phi_e=EL$ plays the role 
of the mean slope of a pile. The behavior of the discharging system is 
expected to be time-dependent, with stochastic avalanches. 

A steady state is not expected for a pair-creating circuit for two reasons:
(1) The created pairs may not maintain a static electric field. Like the
Carlqvist double layer, a self-consistent static voltage must be huge, 
much above the critical value tolerated by pair discharges. No static 
solution exists at modest voltages because it
implies a slow particle acceleration and a large charge imbalance in a 
broad region, leading to inconsistency (see the end of \S~4.3).
(2) The near-critical behavior is unstable: either a runaway is ignited
or pair creation extinguishes soon after a seed particle is injected.

Discharges are well known in tubes with low-temperature 
plasma, which is weakly ionized. In that case, the critical voltage 
accelerates electrons to the ionization threshold and avalanches of 
ionization develop. 
Charge carriers are then supplied by ionization rather than pair 
creation. Another difference is that the fast electrons are lost to the 
cylindrical wall of the tube, so the problem is not 1-D, and coherent 
waves of ionization can develop.

  
\subsection{Numerical Simulations}

\subsubsection{Setup}

We implement the pair-creation process in our numerical experiment in a
simple way: electrons (or positrons) can create pairs if their Lorentz 
factors are in a specified range $\gamma_1<\gamma_e<\gamma_2$.
The mean free-path for $e^\pm$ emission by an accelerated electron 
in this window, $\lres$, is a parameter of the experiment.
It determines the probability of pair creation 
by the electron during a timestep $dt$: 
this probability is $cdt/l_{\rm res}$. 
Note that $\lambda_{{\rm D}e} \ll l_{\rm res} \ll \RNS$.

In a real magnetosphere, the resonance window 
$\gamma_1<\gamma_e<\gamma_2$ is shaped by the spectrum of target photons
$F_\omega$ and the local magnetic field $B$. In our simplified model,
this window is constant along the magnetic line. 
It is possible to use a more realistic
resonance condition that
takes into account variations in $B$ and $F_\omega$ along a magnetospheric
field line.
However, our major goal is to see whether $e^\pm$ creation leads to a 
qualitatively different state of the circuit. We shall see that
the new state weakly depends on the details of the 
resonance condition when $B > \BQED$ and $\gamma_{\rm res} \gg 1$
everywhere in the circuit.

We consider here only circuits with 
boundary temperatures too low for plasma to be supplied 
thermally (\S~4.1). Then the circuit solution weakly depends on the chosen
values of $T_i$ and $T_e$. In this section, we show a sequence of models 
with $k_{\rm B}T_e=0.04m_ec^2$ and $k_{\rm B}T_i=0.04m_ic^2$. 
The corresponding scale height of the surface layers is $h=0.04L$.  

To illustrate the weak dependence 
on details of the pair creation process we consider two extreme 
cases of very efficient and very inefficient $e^\pm$ creation.
Pair creation is less efficient if the resonant window $(\gamma_1,\gamma_2)$
is narrow and the mean free-path $\lres$ is large even in this window.
We therefore focus on two models:
Model~A: an infinitely wide resonance window $(\gres,\infty)$
         with free path $\lres=0$ and
Model~B: a narrow resonance window $(\gres,2\gres)$ with 
         a large mean free path $\lres=L/3$.
The inclusion of pair creation brings two new parameters into the 
experiment: $\gres$ and $\lres$. 
We are especially interested in values of $\gres$ that are comparable to 
or below $m_i/m_e$, which corresponds to a real magnetosphere.

\begin{figure}[h]
\begin{center}
\plotone{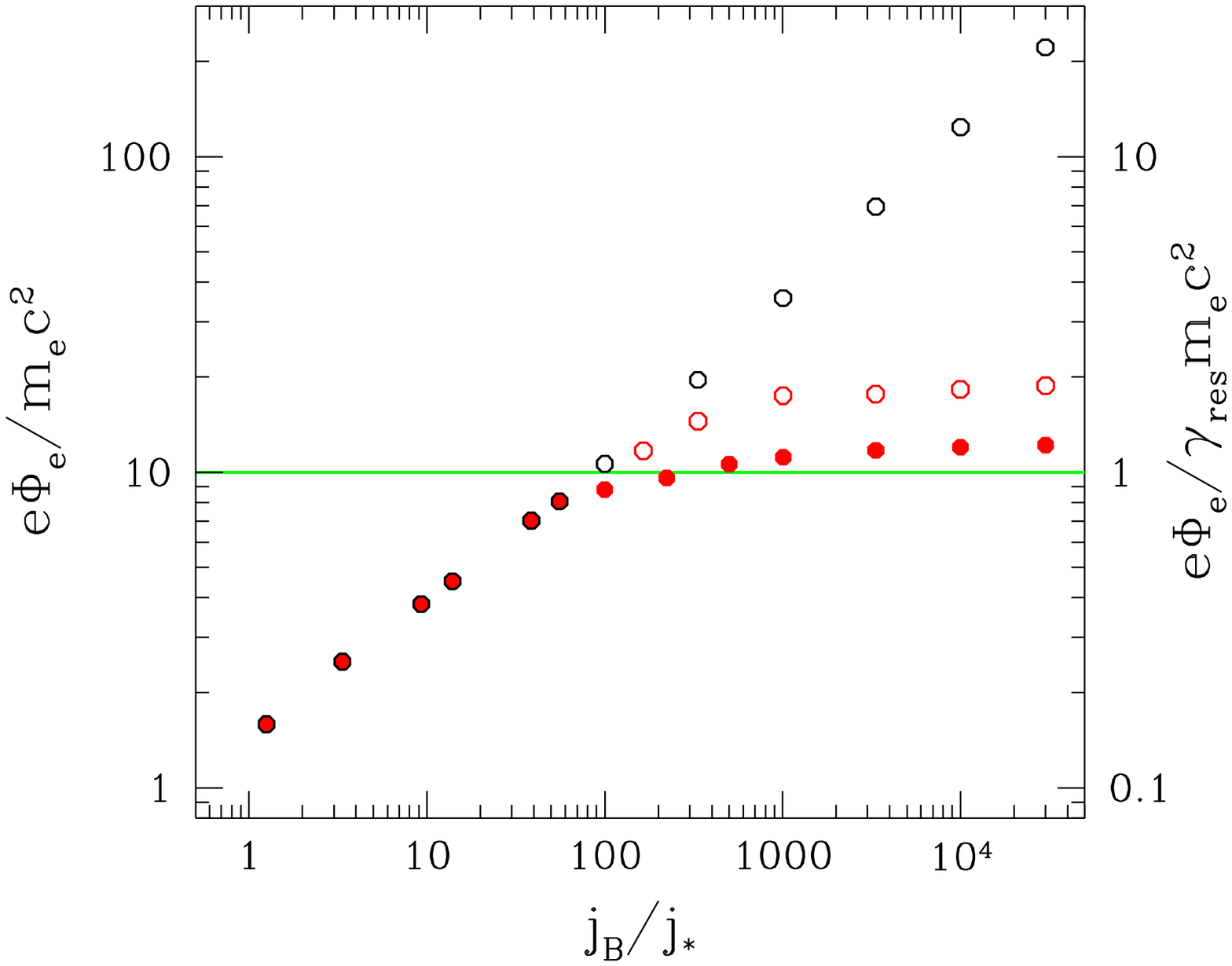}
\label{fig6}
\caption
{Voltage vs. current for circuits with (red circles) and without 
(black circles) $e^\pm$ creation. Model~A of pair production is represented
by filled red circles and Model~B by open red circles;
both models have the resonant Lorentz factor $\gres=10$.
Each point in the figure represents the result of one experiment
and shows the time-averaged voltage $\Phi_e$.
The current is expressed in units of $j_*$ (see the text).
}
\end{center}
\end{figure}

\subsubsection{Results}

The results of the numerical experiments are shown in Figures~6-14. 
Figure~6 displays the relation between current $j$ and voltage $\Phi_e$
in a circuit with $m_i=10m_e$ and $\gres=10$. 
The current is normalized to
a minimum current $j_*$ for which the characteristic Debye length 
$\lambda_{{\rm D}e}=c/\omega_P$ equals the 
size of the system $L$ (see eq.~[\ref{eq:j*}]). At small $\jB$, 
the double layer configuration is established in the circuit;
its voltage is not sufficient to ignite $e^\pm$ production.
With increasing $\jB$, the voltage of the double layer
increases as $(\jB/j_*)^{1/2}$ and reaches $\gres m_ec^2$ at 
$j_1\simeq j_*\gres^2$; then it stops growing. 
At $\jB>j_1$ the voltage saturates at the characteristic 
$e\Phi_e\sim\gres m_ec^2$ in both Models A and B. 
This asymptotic regime $\jB\gg j_1$ is of interest to us because 
magnetars are in this regime. All subsequent figures show
the asymptotic behavior at $\jB\gg j_1$. 

\begin{figure}
\begin{center}
\plotone{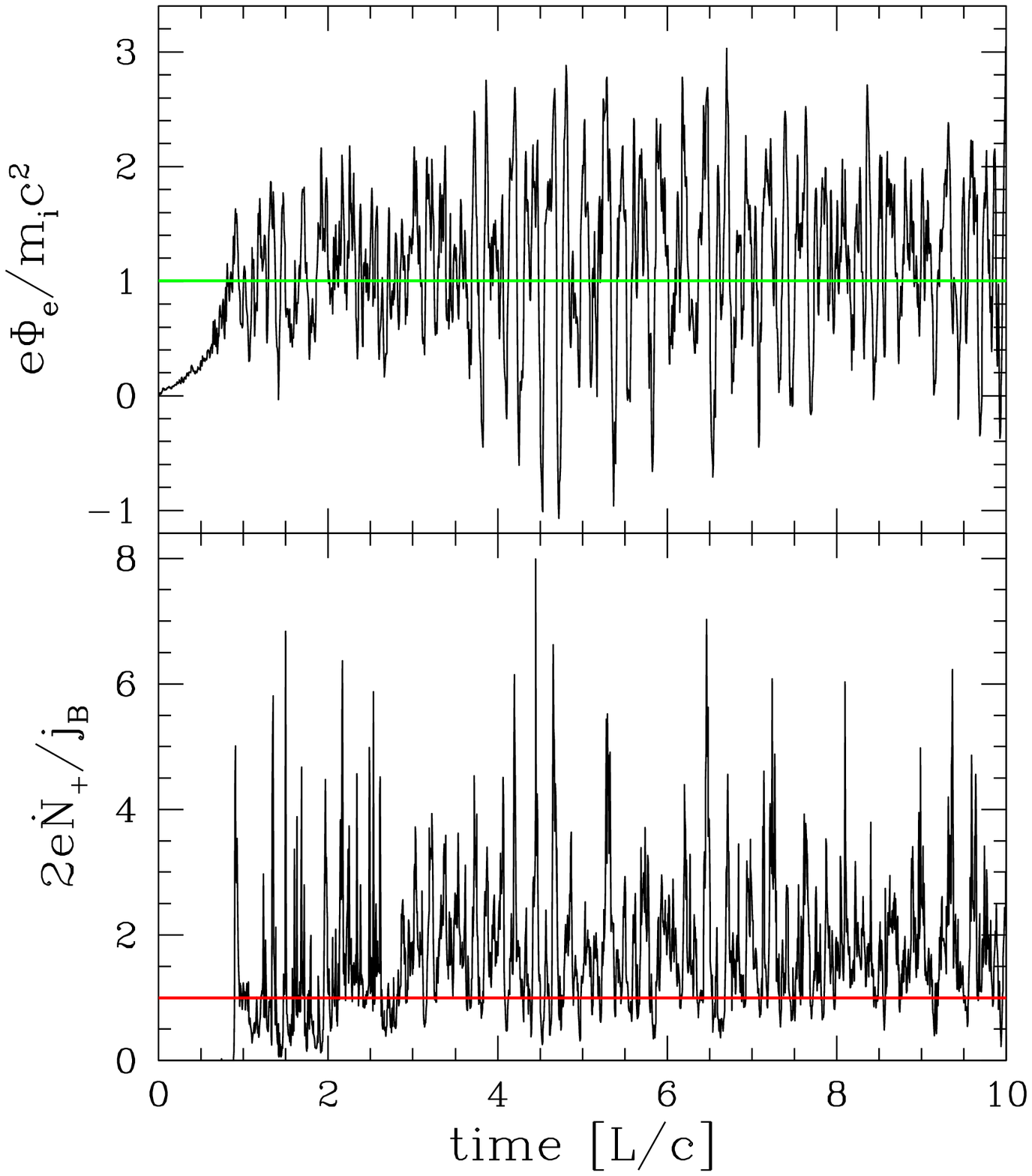}
\label{fig7}
\caption
{ Relaxation of the circuit during the first 10 light-crossing
times. The circuit parameters are: $m_i=10m_e$, $j=10^4j_*$,
$\gres=m_i/m_e$. Model A is used for pair production (infinite
window above $\gres$ with free path $\lres=0$).
A quasi-steady fluctuating state is established after a few $L/c$.}
\end{center}
\end{figure}
\begin{figure}
\begin{center}
\plotone{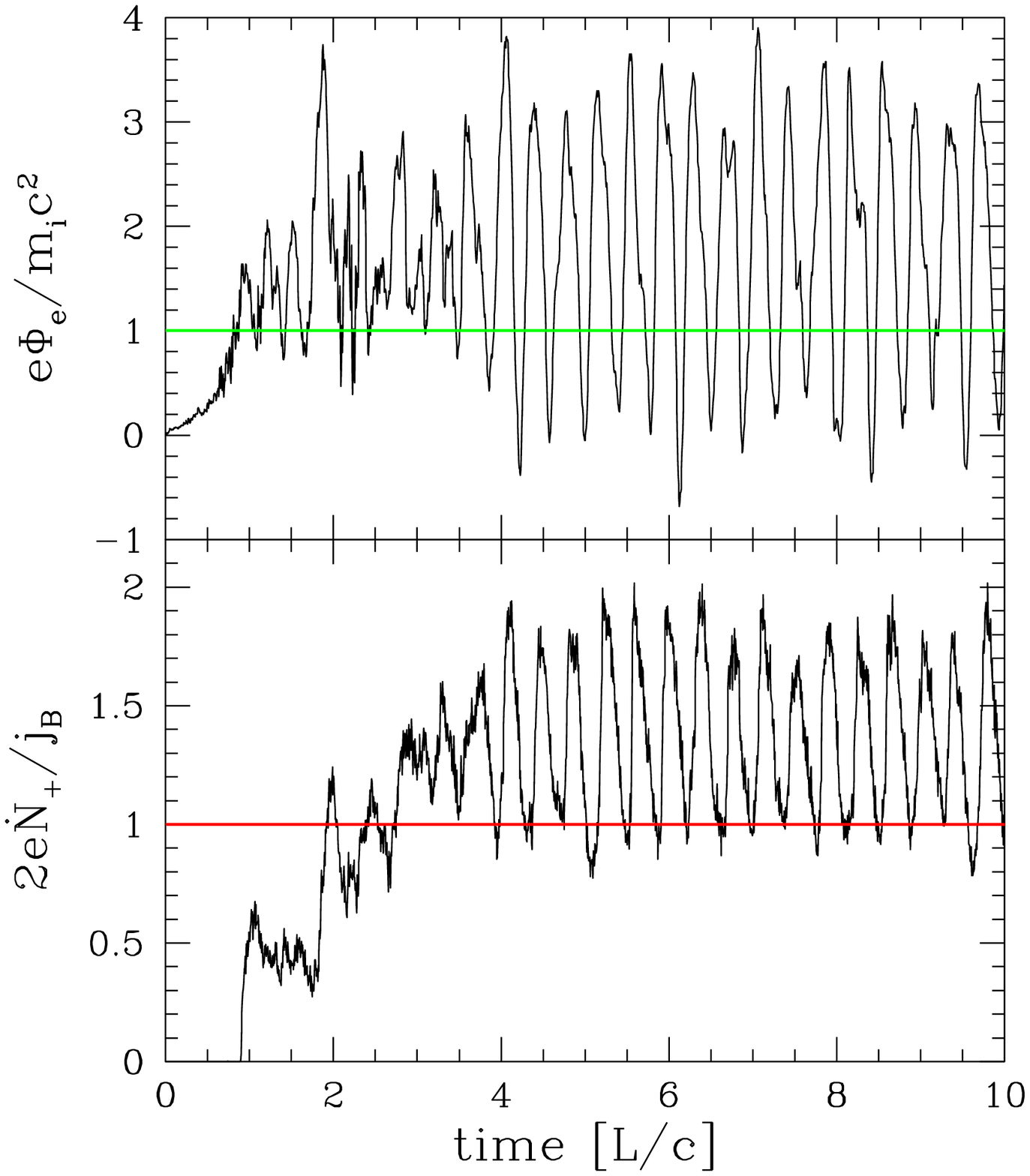}
\label{fig8}
\caption
{Same as Fig.~7, but Model B is used for pair production: narrow
window $\gres<\gamma_e<2\gres$ with free path $\lres=L/3$.
}
\end{center}
\end{figure}

Each point in Fig.~6 is obtained by running a time-dependent 
simulation that relaxes to a quasi-steady state. Figures~7 and 8 show 
the relaxation history in two simulations with $\jB=10^4j_*$.
The voltage $\Phi_e$ is zero in the initial state and grows on the dynamical
timescale $\tdyn=L/c$ while the plasma and electric field self-organize 
to maintain the required current. The voltage grows until electrons 
in the circuit get accelerated to Lorentz factors $\gamma_e\sim\gres$
and pair production begins. One can see that after a few dynamical times
the voltage $\Phi_e$ stops growing and the circuit enters an oscillatory 
regime. During each oscillation, an increased voltage leads to a 
higher $e^\pm$ production rate $\dot{N}_+$, 
then too many $e^\pm$ are produced which screen the voltage (discharge),
$\dot{N}_+$ drops, and $\Phi_e$ begins to grow again. 
The oscillations persist until the end of the simulation ($40L/c$), with a 
constant amplitude.

We see that a quasi-steady state is reached after a few $\tdyn$.
This state is time-dependent on short timescales, but it has a 
well-defined steady voltage when averaged over a few $\tdyn$. 
Models~A and B have similar histories of $\Phi_e$ and $\dot{N}_+$
(cf. Figs.~7 and 8) except that Model~A is noisier on short timescales
and shows less coherent oscillations. (The quasi-periodic oscillations are,
however, pronounced in the Fourier spectra of $\Phi_e[t]$ and $\dot{N}_+[t]$.)
It is not surprising that $\Phi_e$ may vary on short timescales $t\ll \tdyn$.
For example, if the current suddenly stopped, a characteristic
voltage $e\Phi_e=m_ec^2$ that is able to accelerate electrons would be 
created after time $t=m_ec^2/4\pi eL\jB=(\jB/j_*)^{-1}(L/c)\ll L/c$.

In all circuits of interest ($\jB\gg j_1$) we found 
the time-average voltage,
\be
 e\bar{\Phi}_e\sim \gres m_ec^2. 
\ee
This relation applies even to circuits with $\gres m_ec^2\gg m_ic^2$,
where lifting of the ions is energetically preferable to pair creation.
We conclude that {\it the voltage along the magnetic tube is self-regulated 
to a value just enough to maintain pair production and feed the current 
with $e^\pm$ pairs}. The robustness of this result is illustrated in 
Figure~9, which shows circuits with $m_i/m_e=10,30$ and various $\gres$. 
In all cases, $e\Phi_e\sim\gres m_ec^2$ is established, and the pair 
creation rate $2\dot{N}_+\sim \jB/e$ is maintained.
Thus, the near-critical state is self-organized as expected. 
Avalanches of $e^\pm$ creation are observed to happen anywhere
in the circuit between the dense boundary layers.

\begin{figure}
\begin{center}
\plotone{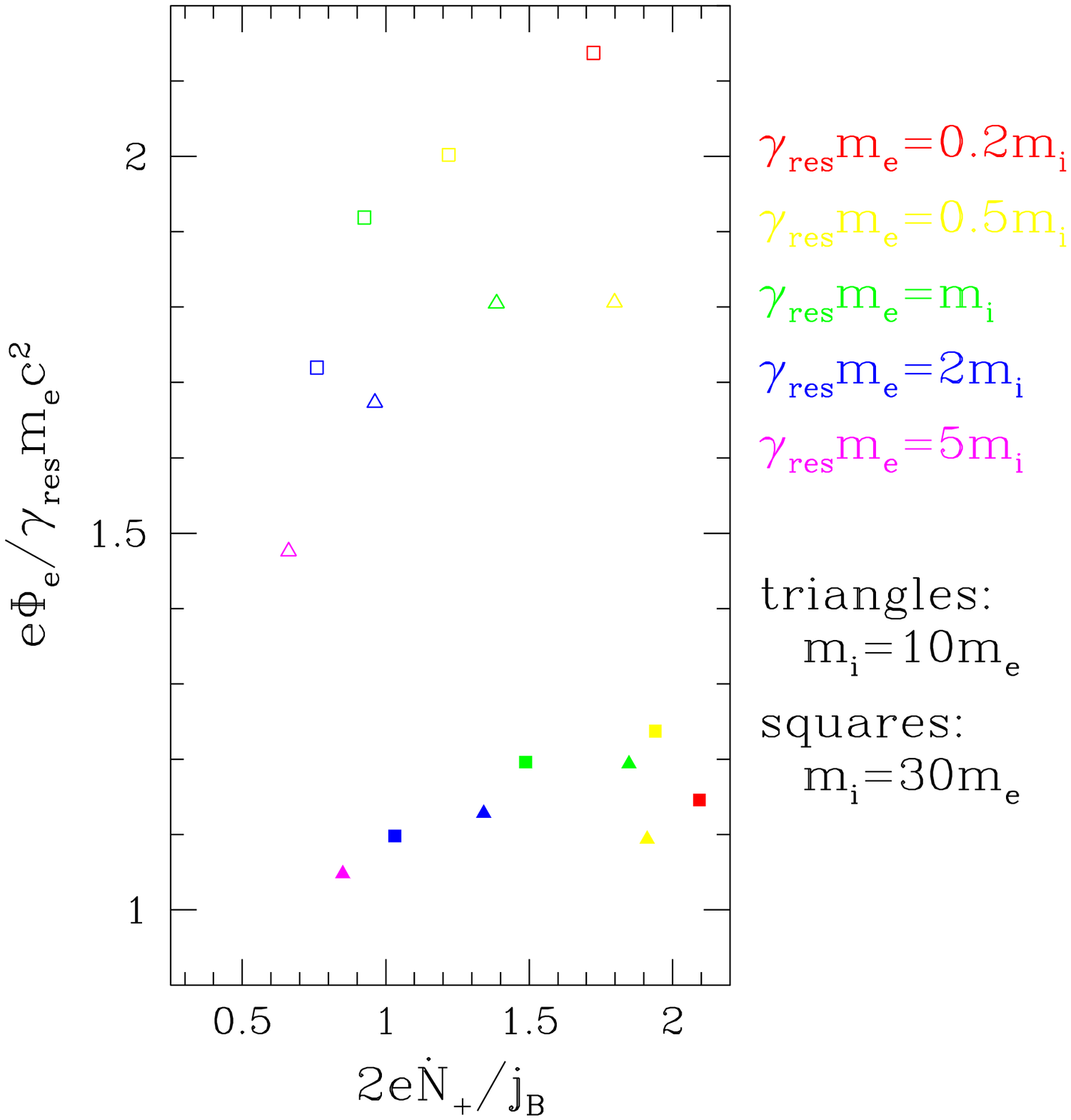}
\label{fig9}
\caption
{Circuits with $m_i/m_e=10,30$ and various values of $\gres$ ranging 
from $m_i/5m_e$ to $5m_i/m_e$. Here $\bar{\Phi}_e$ is the time-averaged
voltage and $2\dot{N}_+$ is the time-averaged rate of $e^\pm$ creation
in the circuit. Model~A of pair production is shown by open symbols, and 
Model~B by filled symbols. 
}
\end{center}
\end{figure}

\begin{figure}
\begin{center}
\plotone{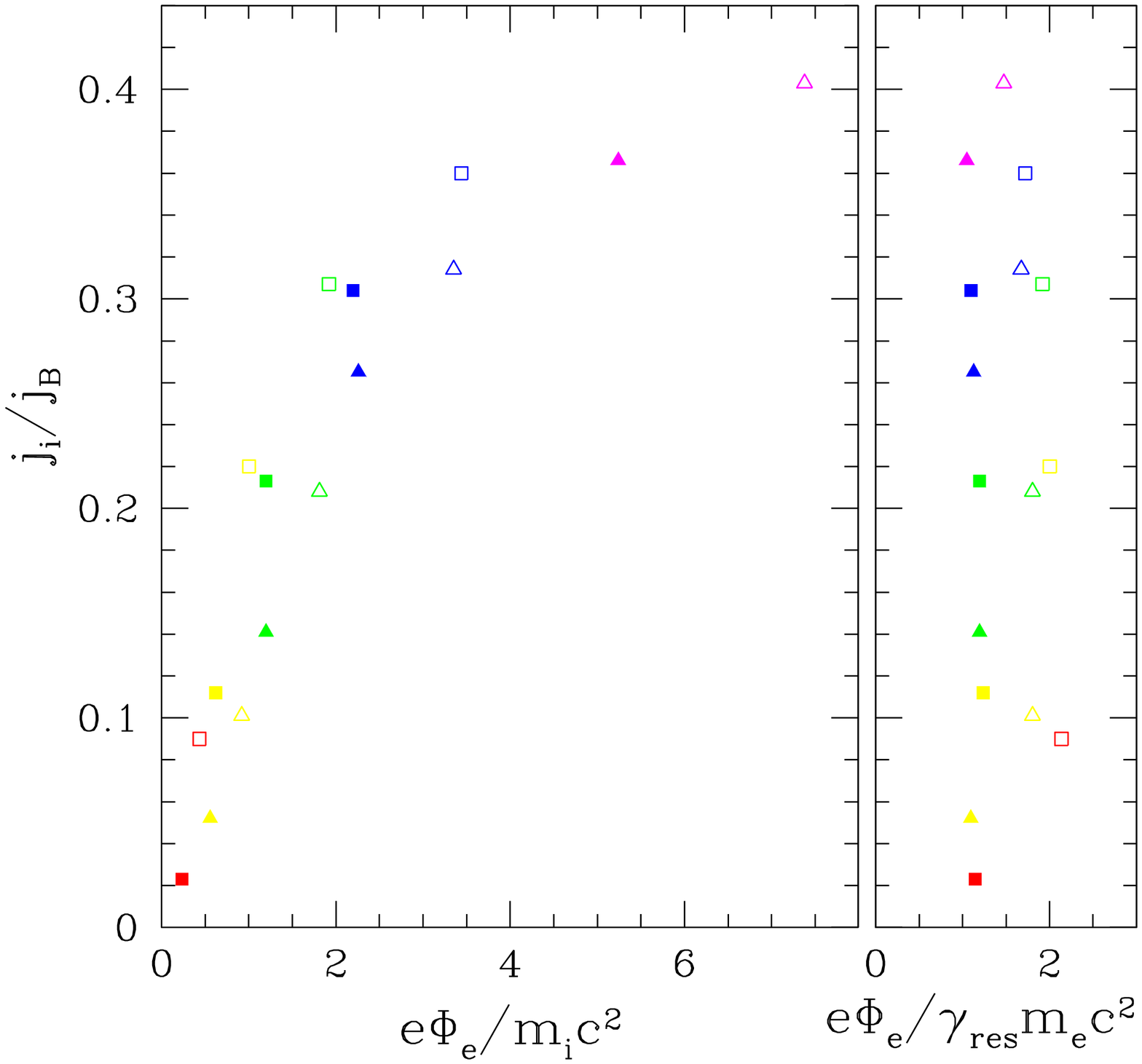}
\label{fig10}
\caption
{Fraction of the electric current carried by ions vs. voltage, 
for the same set of models as in Fig.~9. Colors have the same
meanings. In the left panel voltage is taken in units of $m_ic^2$,
and in the right panel in units of $\gres m_ec^2$.
}
\end{center}
\end{figure}

The current is carried largely by $e^\pm$ everywhere in the tube.
The ion fraction in the current depends on the ratio 
$e\Phi_e/m_ic^2\simeq\gres m_e/m_i$ (Fig.~10). If this ratio is small, 
pairs are easily produced with a small electric field, too small to lift 
the ions, and the ion current is suppressed. In the opposite case, ions 
carry about 1/2 of the current.
For typical parameters expected in the magnetosphere, $\gres\simlt m_i/m_e$,
ions carry $\simlt 10$\% of the current.

The simulations show that the released energy sinks through the anode and 
cathode boundaries at approximately equal rates, and that their sum equals
$j_B\bar{\Phi}_e$. For magnetars, this implies that both footpoints of 
a twisted magnetic tube will radiate energy bolometrically at comparable 
rates.

\subsubsection{Sample Model}

We now describe in more detail the plasma behavior in one sample
experiment. It assumes Model~A for $e^\pm$ creation and 
$\gres=m_i/m_e=10$.\footnote{A similar solution is obtained for 
$\gres=m_i/m_e=30$. We expect the same circuit solution to apply to 
$\gres=m_i/m_e=1836$ or any other $\gres=m_i/m_e\gg 1$.} 
The relaxation history of this experiment is shown in Figure~7, and
a snapshot of the circuit is shown in Figure~11. 

The charge density $\rho=e(n_i+n_p-n_e)$ and electric field are found to
fluctuate significantly in time, while the densities $n_i$, $n_p$, and $n_e$ 
are approximately steady. Deviations from charge neutrality are small: 
$\rho/e(n_i+n_p+n_i)\ll 1$ and $E^2/8\pi U_{pl}\ll 1$ where $U_{pl}$ is 
the plasma kinetic energy density.  The plasma turbulence may then be
described as a superposition of Langmuir waves. We do not, however, 
attempt a development of analytical theory in this paper.

The anode current $j_e+j_i+j_p$ is dominated by electrons.  The anode 
value of $j_e$ even exceeds $\jB$ by 17\%. Together with ions
($j_i\simeq 0.13\jB$), electrons 
balance the negative current of positrons $j_p\simeq -0.3\jB$
so that the net current $j_e+j_i+j_p$ equals $\jB$. 
Positrons contribute negative $j_p$ at the anode because 
they move in the ``wrong'' direction (they are created by accelerated 
electrons, and some of them leave the box against the mean electric field).
The cathode current is dominated by positrons (64\%); electrons contribute
23\%, and ions 13\%.

\begin{figure}
\begin{center}
\plotone{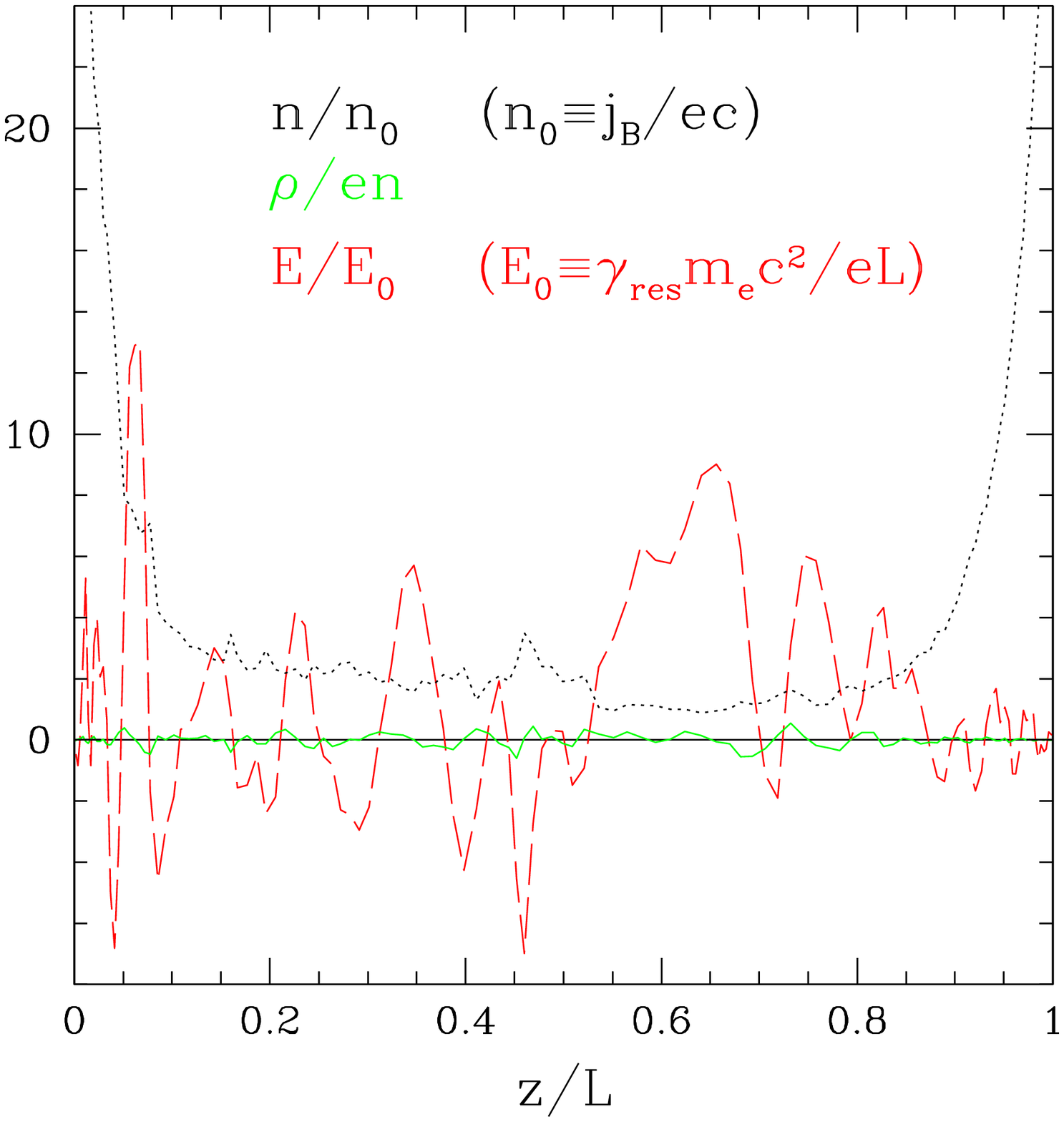}
\label{fig11}
\caption
{Snapshot of the circuit with $\gres=m_i/m_e=10$ and $\jB=10^4j_*$ 
($\lambda_{\rm D}/L=10^{-2}$). Model~A is assumed for $e^\pm$ creation.
The three curves show particle density $n=n_i+n_e+n_p$ (dotted), charge 
density $\rho=e(n_i-n_e+n_p)$ (solid), and electric field $E$ (long-dash).
}
\end{center}
\end{figure}

\begin{figure}
\begin{center}
\plotone{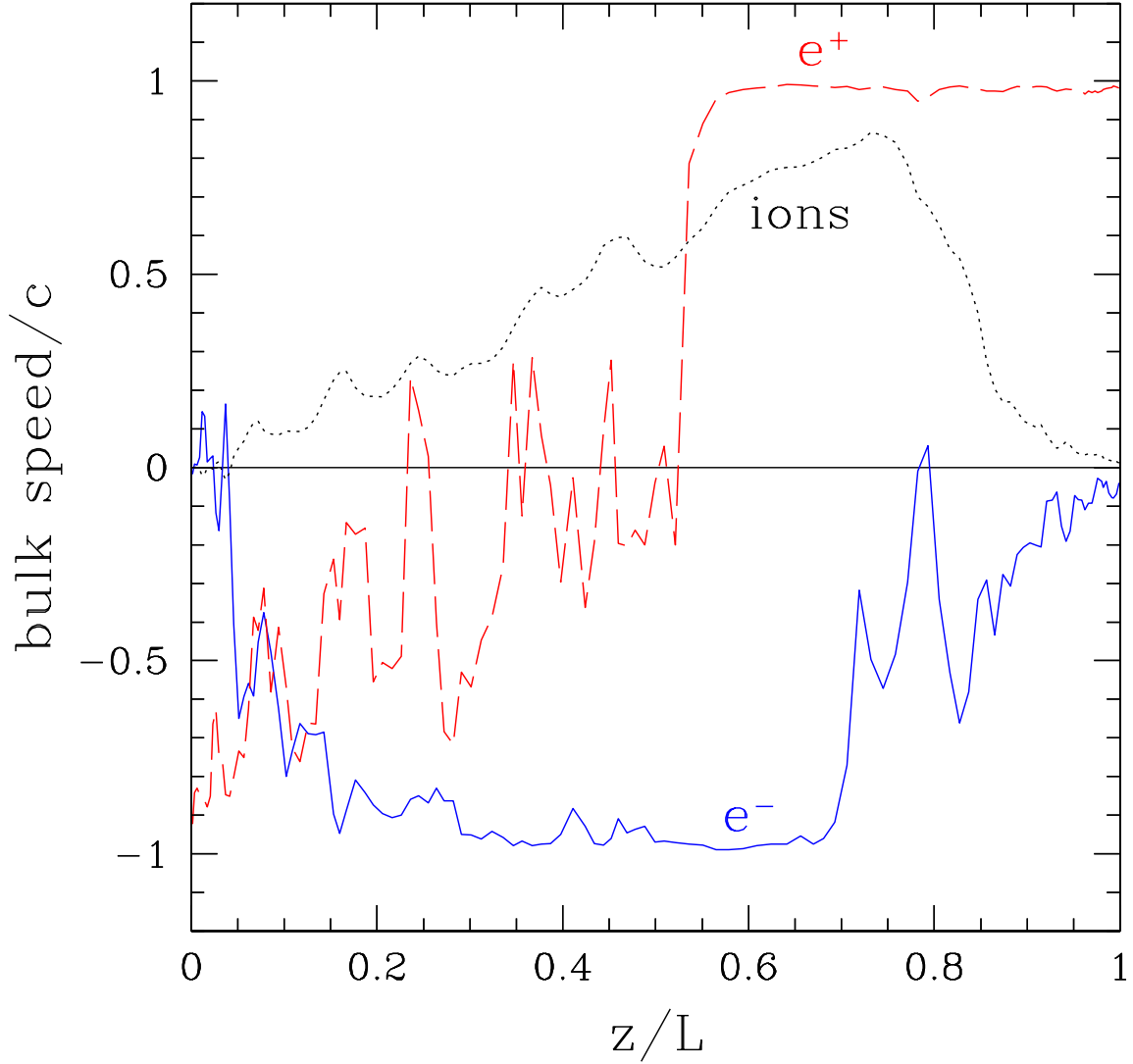}
\label{fig12}
\caption
{Mean velocities of ions, electrons, and positrons as a function of 
position in the circuit. The mean velocities $v_X$ are related to 
densities $n_X$ of the three species $X=i,e,p$ by $v_X=j_X/\bar{v}_x$
where  $j_X$ is the current carried by species $X$. The snapshot is 
taken of the same model and the same time as Fig.~11.
}
\end{center}
\end{figure}

\begin{figure}
\begin{center}
\plotone{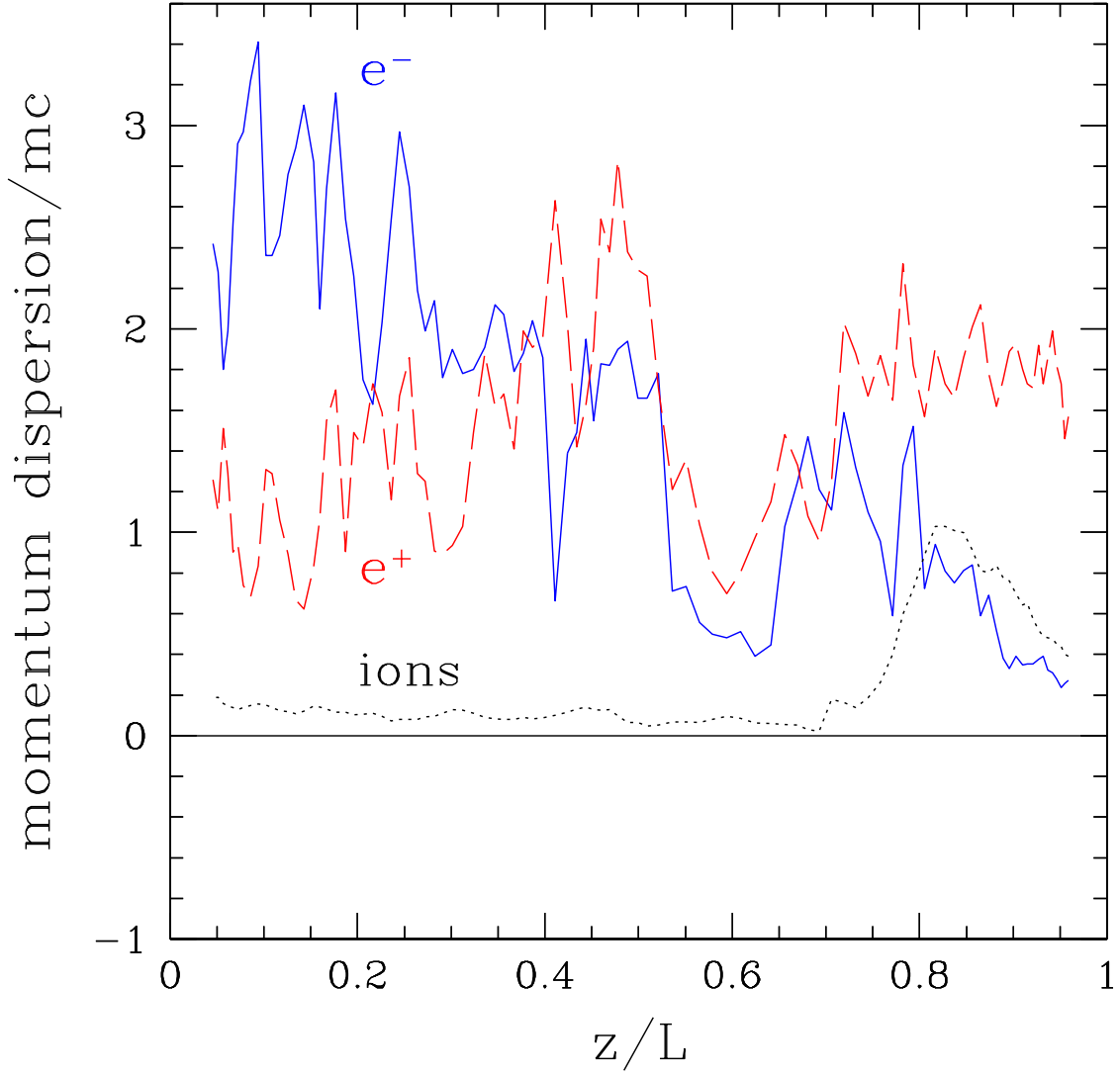}
\label{fig13}
\caption
{Dispersion of momentum (around the mean bulk momentum) for ions, 
electrons, and positrons as a function of position in the circuit. 
The snapshot is taken of the same model at the same time as in Fig.~11.
Only the corona is shown ($h<z<L-h$) and the surface layers of height
$h=0.04L$ are excluded from the figure.
}
\end{center}
\end{figure}

\begin{figure}
\begin{center}
\plotone{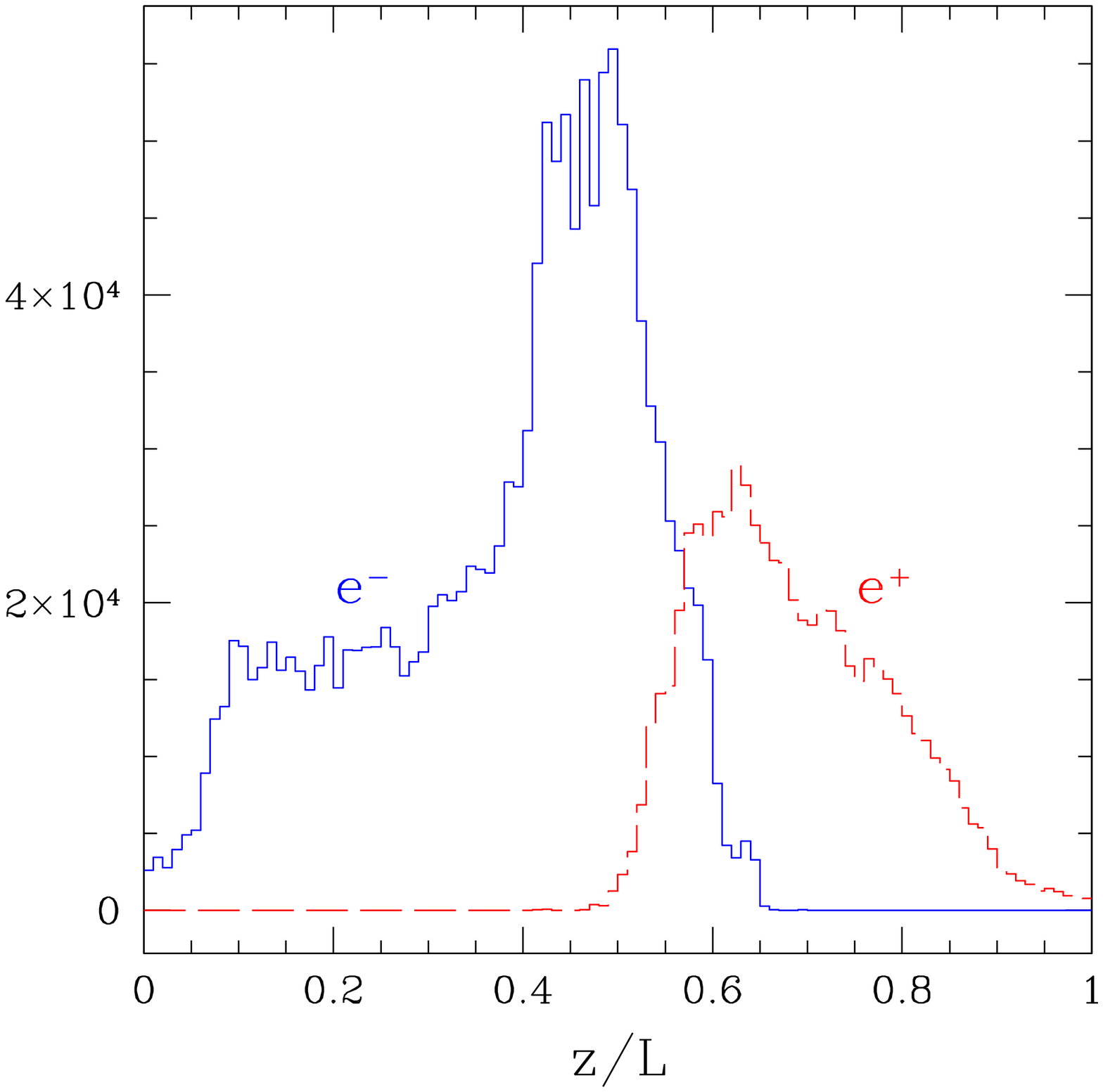}
\label{fig14}
\caption
{Distribution of pair-creation events at $10<ct/L<20$ over $z$.
The circuit model is the same as in Figs.~7 and 11-13. The figure shows 
separately pair creation by accelerated $e^-$ and $e^+$ 
(solid and long-dashed curves, respectively). 
}
\end{center}
\end{figure}

Bulk speeds of the three plasma components, $v_e$, $v_i$, $v_p$,
and the dispersions of their dimensionless momenta $p/mc=\gamma\beta$ 
are shown in Figures~12 and 13. The dispersions are significant, 
i.e. the corresponding distribution functions are broad.
The time-averaged distribution of pair creation over $z$ is shown in 
Figure~14. It has two 
peaks that correspond to pair production by accelerated $e^-$ and $e^+$. 
The net rate of pair creation by $e^-$ is about 2 times higher than
that by $e^+$ in this model.

A complicated kinetic state self-organizes in the phase space of electrons, 
ions, and positrons. The state keeps oscillating. The oscillations might 
be related to the fact that positrons move in a wrong direction at 
creation (negative current) and have to reverse direction
to maintain the net $j=\jB$
without strong deviations from neutrality. This requires 
an extra voltage which leads to ``overshooting'' in the pair creation rate 
followed by a temporary screening of the electric field.


\section{TRANSITION LAYER}\label{sc:layer}

A transition layer must be established between the corona and the 
relatively cold optically thick atmosphere of the star. It 
may play a crucial role for two reasons. 
(1) The layer is able to radiate away the energy received from
the corona because it is dense enough for efficient bremsstrahlung 
emission. The corona itself may not efficiently radiate the dissipated power 
because its bremsstrahlung output is negligible, and  
Compton cooling is inefficient as upscattered photons cannot escape ---
they convert to pairs in the ultra-strong magnetic field. 
(2) The layer may be hot enough to maintain thermal pair 
creation at an interesting rate and feed the charge flow in the 
coronal circuit (Thompson \& Beloborodov 2005).
A detailed analysis of the transition layer is left for future work. 
Below we describe some key mechanisms by which energy is deposited and
redistributed within the layer, and by which $e^\pm$ pairs may be 
created.

\subsection{Heating by Beam Instability}
\label{sc:bi}

The particles accelerated in the corona enter the hydrostatic atmosphere
with high Lorentz factors $\gb\sim e\Phi_e/m_ec^2$. 
This relativistic beam drives an electrostatic instability with the
growth length (e.g. Godfrey, Shanahan, \& Thode 1975),
\be
   \lb\sim\gb\left({n_e\over\nb}\right)^{1/3} 
{c\over\omega_{Pe}}
  =\frac{\gb}{(4\pi \nb r_e)^{1/2}}
  \left({\nb\over n_e}\right)^{1/6}.
\label{eq:lbeam}
\ee
Here $\omega_{Pe} = (4\pi n_e e^2/m_e)^{1/2}$ is the electron plasma
frequency in the atmosphere with density $n_e$,
and $r_e=e^2/m_ec^2\simeq 2.8\times 10^{-13}$~cm. 
The beam exponentially amplifies a seed Langmuir wave
whose phase speed is in resonance with the beam. The exchange of energy
with the atmospheric plasma occurs on a scale comparable to
$\lb$. As the beam propagates a distance $\sim\lb$
into the layer, the amplified Langmuir waves feed back on the beam and flatten
its distribution function to a plateau. If the initial distribution function 
of the beam is idealized by $\delta$-function, 
$f_b(\gamma_e)={\rm const}\,\delta(\gamma_e-\gb)$, its 
relaxation to the plateau $1<\gamma_e<\gb$ implies the 
deposition of ${1\over 2}$ of the beam energy into plasma turbulence. 
Since the relaxation occurs in a strong magnetic field, 
the particle momentum distribution remains one-dimensional. 
This is different from beam relaxation in a normal plasma --- in that 
case a large dispersion of perpendicular momenta develops before relaxing 
to the plateau in energy.

The coronal beam has a density 
$\nb\sim n_c=\jB/ec\sim 10^{16}-10^{17}$~cm$^{-3}$ 
and Lorentz factor $\gb\simeq\gres\sim 10^3$ (see \S~5; this would correspond
to a total energy deposition rate of $10^{36}\,\gamma_{b,3}n_{16}$~erg~s$^{-1}$ 
if integrated over the entire surface of the star).
Substituting these values to equation~(\ref{eq:lbeam}) one finds
$\lb\sim 1$~cm.  
The beam relaxation occurs at a much larger depth than the skin depth of 
the atmospheric plasma, $\lb\gg c/\omega_P$. Therefore, the beam 
instability may heat the atmosphere only well below the screening layer, 
i.e. well below the region where the induced electric field develops
and particles are lifted to the corona. 

In our numerical experiments $\lb\sim 10(c/\omega_P)\sim 0.1L$
was larger than the scale height of the atmosphere, $h$, so heating
by beam instability could not possibly operate in the computatonal box.
The possible atmospheric heating at larger depths was regulated in the 
experiment by hand, by setting the boundary temperatures $T_e$ and $T_i$.
Further investigation of beam heating is deferred to a future work.

Comparing the expected $\lb$ and $h$, one has
\be
   {\lb\over h} \sim    \left({\nb\over n_e}\right)^{1/6}
   \,\left[{k_{\rm B}(T_e+T_i)\over {\rm keV}}\right]^{-1}.
\ee
Note that $\lb/h$ is reduced if the plasma temperature grows.
It is possible that the atmosphere develops a hot layer with 
$\lb\ll h$, which absorbs a significant part of the beam energy. 
The excited Langmuir turbulence heats mostly electrons and $T_e$ may grow 
to $k_{\rm B}T_e \ga 100$~keV. Then hard X-ray emission and thermal pair 
creation may occur as discussed below.

\subsection{Thermal Conduction and Radiative Losses} 
\label{cond}

Dissipation without losses of energy would imply an unlimited 
growth of temperature. If the twisted magnetosphere
were thermally insulated, it would quickly heat up and remain filled with a 
hot plasma that would steadily conduct 
the current with no further dissipation. Losses of the dissipated energy 
are therefore crucial in the regulation of the state of the circuit.
In our numerical experiment we simply allowed the released energy to sink 
through the anode and cathode boundaries, assuming that it is radiated 
away outside our boundaries. Here we discuss a possible mechanism
of the energy loss.

We expect the energy of the accelerated coronal particles
to be deposited in an optically 
thin atmospheric layer by collisionless processes (such as beam instability
discussed in \S~\ref{sc:bi}).  This layer is still unable
to cool radiatively, and the deposited heat will be 
conducted to deeper and denser layers that can radiate the heat away 
by bremsstrahlung (Thompson \& Beloborodov 2005). 
The radiation flux emitted at that depth is $F_\gamma\sim h\epsilon_{\rm ff}$ 
where $h$ is a characteristic scale-height and $\epsilon_{\rm ff}$ is 
the free-free emissivity. Plasma with temperature $T_e$ and density $n_e$ 
has the bremsstrahlung emissivity
\be
\label{eq:epff}
   \epsilon_{\rm ff} \approx \alpha\,\theta_e^{1/2}
       n_e^2\sT m_ec^3,  \qquad \theta_e\equiv\frac{k_{\rm B}T_e}{m_ec^2},
\ee
where $\alpha=e^2/\hbar c=1/137$ is the fine structure constant.
The characteristic scale-height on which density and temperature 
change in the layer is the hydrostatic scale-height, $h=k_{\rm B}T_e/m_ig$
(we assume here $T_e<T_i$ since the collisionless dissipation heats
mostly electrons). Then the emitted radiation flux is
\be
\label{eq:Fg}
  F_\gamma=h\epsilon_{\rm ff}
     \approx\alpha\,\frac{m_e}{m_i} \,
                   \theta_e^{3/2} \frac{c^2}{g} n_e^2\sT m_ec^3.
\ee
It carries away a significant fraction of the energy flux received
from the corona when $F_\gamma\sim F=n_c\gb m_ec^3$, 
where $\gb=e\Phi_e/m_ec^2\simeq\gres$.
This gives one relation between temperature $\theta_e$
and Thomson optical depth $\tT=h\sT n_e$ at which the heat flux is converted 
to radiation, 
\be
\label{eq:rel1}
    \theta_e^{-1/2}\tT^2\simeq\frac{\tau_c\gb}
          {\alpha}\,\frac{m_e}{m_i}\,\frac{c^2}{g\RNS},
\ee
where
\be
\label{eq:tau0}
  \tau_c\equiv\RNS\sT n_c\sim 0.01-0.1.
\ee

Equation~(\ref{eq:rel1}) is not sufficient to determine $\theta_e$ and 
$\tT$ of the emitting layer because the mechanism of heat transport to 
this layer has not been specified yet.
A second relation is needed, which is the equation of thermal conductivity
(Thompson \& Beloborodov 2005).
It relates the heat flux $F_c\sim F\sim F_\gamma$ to the temperature gradient:
$F_c=\kappa dT_e/dz$ 
where $\kappa=n_ev_e\ell k_{\rm B}$, $v_e=(k_{\rm B}T_e/m_e)^{1/2}$,
and $\ell$ is the mean free path of electrons. In the absence of plasma
turbulence, the mean free path is determined by Coulomb collisions,
\be
\label{eq:lCoul}
  \lCoul=\frac{\theta_e^2}{n_e\sT}.
\ee
If strong plasma turbulence develops with energy density above the 
thermal level, $\Uturb = (\delta E_\parallel)^2/8\pi 
\gg k_{\rm B}T_e/\lDe^3$, the fluctuating electric field 
suppresses the mean free path. It is then given by
\be
\label{eq:ell}
  \ell\simeq\lDe\,\left(\frac{\Uturb}{n_ek_{\rm B}T_e}\right)^{-1}.
\ee
(We here assumed that the turbulence develops at frequencies 
$\omega\sim\omega_{Pe}$.) We will keep $\ell$ as a parameter in our 
estimates. It is bounded from above and below by $\lDe<\ell<\lCoul$.

Estimating $dT_e/dz\sim T_e/h$, one finds
\be
\label{eq:Fc}
  F_c=n_ev_e\ell \frac{k_{\rm B}T}{h}=n_e\,v_e\,\ell\, gm_i,
\ee
\be
\label{eq:rel2}
   \theta_e^{-3}\tT^2\simeq\frac{1}{\alpha}\,
                             \left(\frac{\ell}{\ell_{\rm Coul}}\right).
\ee
Combining equations~(\ref{eq:rel1}) and (\ref{eq:rel2}), one finds 
the temperature of the radiating layer,
\be
\label{eq:Tem}
     \theta_e \approx \left(\tau_c\gb\,\frac{m_e}{m_i}\,\frac{c^2}{g\RNS}\,
      \frac{\lCoul}{\ell}\right)^{2/5},  \qquad
  k_{\rm B}T_e \approx 200\,\left(\frac{\tau_c}{0.1}\right)^{2/5}
                \left(\frac{\gb}{10^3}\right)^{2/5}
                \left(\frac{\ell}{\lCoul}\right)^{-2/5} {\rm ~keV}.
\ee
If $\ell<\lCoul$ (suppressing heat conduction) the emission temperature 
will grow above 200~keV. The growth, however, will be quickly stopped at 
$\theta_e\sim 1$ because the hot layer starts to generate $e^\pm$ pairs
which can short out the circuit (see \S~\ref{pairatm}).

\subsection{Heated Atmospheric Layer as a Possible Source of $e^\pm$}
\label{pairatm}

\subsubsection{Channels of Pair Creation}\label{channel}

The dominant channels of thermal pair creation in a hot atmosphere
involve an electron-ion collision. Three such channels are summarized below. 

\noindent
1. An $e^\pm$ pair can be created directly in a collision between an
ion and an energetic electron (or positron),
$e^- + Z \rightarrow e^- + Z + e^- + e^+$.  
This reaction is possible when the kinetic energy of the incident electron
exceeds a threshold $E_{\rm thr}=2m_ec^2$
(both $e^-$ and $e^+$ can be created in the lowest Landau state).  
When the temperature $k_{\rm B}T_e\ll 2m_ec^2$, the reaction occurs 
in the high-energy tail of the electron distribution. 

\noindent
2.  A bremsstrahlung photon can be emitted in an electron-ion
collision, almost always in the O-mode.
The photon converts to an $e^\pm$ pair if satisfies the threshold
condition $E_{\gamma}>2m_ec^2/\sin\them$ where $\them$
is the emission angle (cf. \S~5.1).

This second channel of pair creation is similar to the first one, where a 
virtual photon converts directly to a pair during the electron-ion collision. 
The emission of a real bremsstrahlung photon makes the process two-step and
increases its threshold from $2m_ec^2$ to $2m_ec^2/\sin\them$. 
The exact cross sections of these processes in the ultrastrong magnetic 
field have not been calculated, to our knowledge. They may be roughly 
estimated using the nonmagnetic formulae (e.g. Berestetskii, Lifshitz, 
\& Pitaevskii 1982): $\sigma_{\rm bremss} \sim 0.3\,Z^2\alpha\,\sigma_T$ 
and $\sigma_{\rm direct} \sim \alpha \sigma_{\rm bremss}$.

Note that relativistic bremsstrahlung emission is preferentially beamed along 
${\bf B}$, in the direction of the incident electron, so $\them$ is small.
Then the threshold is substantially higher than $2m_ec^2$ and the 
bremsstrahlung photon will be unable to convert to $e^\pm$ in situ.
If the photon propagates away from the emission site where $\bB$ has a 
different direction, the threshold may be reduced because the pitch angle 
of the photon increases. Gravitational bending of the photon trajectory 
also helps to increase the pitch angle.
Thus, pair creation by bremsstrahlung photons can be non-local and
take place above the hydrostatic atmosphere.

\noindent
3. An electron can be excited to the first Landau state by a Coulomb 
collision with an ion.  The excited electron rapidly de-excites and emits 
an energetic photon that may convert to a pair (or split, see \S~5.1).
This channel requires the initial electron to have a minimum 
energy $\Ethr=E_B\approx 6.7\,B_{15}^{1/2}\,m_ec^2$ (eq.~\ref{eq:E_B}).
The side-scattering into the first Landau state involves a perpendicular 
momentum transfer $p_\perp = (2B/B_{\rm QED})^{1/2}m_ec$, 
and the corresponding Coulomb cross section may be estimated as
\be\label{eq:Coul}
\sigma_{\rm Coul}(0\rightarrow1) \approx {4\pi Z^2e^4\over (p_\perp c)^2}
   = {3Z^2\over 4}\left({B\over B_{\rm QED}}\right)^{-1}\,\sigma_T.
\ee

\subsubsection{Feeding the Corona with Atmospheric Pairs}

The rate of pair creation in the atmosphere may be roughly estimated as 
the rate of electron-ion collisions above the threshold $\Ethr$,
\be
  \dot{n}_+\sim n_i\,\fthr\,n_e\,\sigma\,c,
\ee
where $\fthr\sim\exp(-\Ethr/\kB T_e)$ is the fraction of thermal 1-D electrons 
that have energy $E_e>\Ethr$; $\sigma$ and $\Ethr$ are the cross section 
and the threshold energy of the relevant channel.
Atmospheric pair production is able to feed the coronal current $\jB$ if
$h\dot{n}_+>\jB/e$ or, equivalently,
\be
  \fthr\,\tT n_i\left(\frac{\sigma}{\sT}\right)>n_c,
\ee
where $\tau_T = \sT n_{e} h$. This condition may be rewritten using 
$\tau_c=\RNS\sT n_c$ and $h=\theta_e (m_e/m_i)(c^2/g)$,
\be
\label{eq:condit}
     \theta_e^{-1}\tT^2
    >\frac{m_e}{m_p}\,
      \left(\frac{c^2}{\RNS g}\right)\,
     \,\frac{\sT}{\sigma}\,\frac{\tau_c}{\fthr}
    \approx 2\times 10^{-3}\,\frac{\tau_c}{\fthr}\,\frac{\sT}{\sigma}.
\ee
If this condition is met at some depth $\tT$, the atmospheric pairs 
can dominate the charge supply to the coronal circuit. 
At $\tT\gg 1$ the parameter $\theta_e\tT^2\simlt 1$ is limited by 
Compton cooling of the plasma, which implies a low $T_e$ and negligible
$\fthr$. The condition~(\ref{eq:condit}) may be satisfied only at 
$\tT\simlt 1$ and requires at least mildly relativistic $T_e$.
The created pairs have kinetic energies $\sim k_{\rm B}T_e$, so they may 
elevate, overflow the gravitational barrier, and maintain the current.

\subsubsection{Numerical Models}

In order to see how atmospheric pair production could affect the circuit,
we have done a few numerical simulations using the following toy model.
We assume that the atmosphere responds to the heat flux from the corona by
producing $e^\pm$ pairs. Therefore, in addition to the ion-electron
atmosphere, we now maintain a thermal $e^\pm$ atmosphere. Its density
at each boundary is regulated in the experiment so that it is
proportional to the flux of lost energy through the boundary.
                                                                                
If the atmospheric pairs 
were cold, the circuit would be in the same regime
as in their absence. The fact that positrons are now available 
at the boundaries
to carry the current instead of ions does not lead to any significant 
reduction of the voltage: it still evolves to 
the state with $e\Phi_e\simeq\gres m_ec^2$, in which energetic $e^\pm$ are 
supplied via the resonant-scattering channel. We have checked this by 
running the models with cold $e^\pm$ atmospheres. 
The key property of atmospheric pairs is, however, that they are created
hot enough to feed the current thermally (cf. \S~4.1).
Their expected kinetic energy is comparable to the height of the
gravitational barrier (\S~6.3.2) and they
can overflow the barrier. Therefore, the required current $\jB$ can
be conducted at a small voltage $e\Phi_e<\gres m_ec^2$.
                                                                     
The thermally-driven regime was indeed established
in our experiment with a realistic $k_{\rm B}T_\pm=0.5m_ec^2$.
The established voltage then depends on the efficiency of pair supply
in response to the energy flux from the corona.  We define 
the efficiency $\xi$ as the fraction of the energy flux through the 
boundary that is returned to the corona in the form of $e^\pm$ 
kinetic and rest-energy. We find that the voltage $\Phi_e$
across the circuit adjusts so that the flux of released energy, $\jB\Phi_e$, 
can generate the minimum flux of atmospheric pairs to feed the coronal 
current, $\jB/e$. Then voltage is related to $\xi$ roughly as 
$e\Phi_e\sim \xi^{-1}m_ec^2$.
For example, a maximum efficiency $\xi\sim 1$ would reduce the voltage 
to $e\Phi_e\sim m_ec^2$.
We have done the corresponding experiment that confirmed this estimate.
The voltage was found to fluctuate significantly, but it did not
get high enough to initiate the resonant channel of
$e^\pm$ production ($\gres=m_i/m_e=10$ in the simulation).
The ion current was small, $j_i/\jB\simeq 10^{-2}$.

When $\xi$ is reduced, a higher voltage is required in the circuit
according to the relation $e\Phi_e\sim \xi^{-1} m_ec^2$.
This estimate is applicable as long as $\xi\simgt \gres^{-1}$, so
that $e\Phi_e<\gres m_ec^2$. A further reduction of $\xi$ does not
lead to an increase of voltage: then the circuit is mainly fed by pairs
created via the resonant channel, and 
$e\Phi_e\sim \gres m_ec^2$ as described in \S~5. The thermal pair
creation in the atmosphere then has a negligible effect on the circuit.
                                                                                
We have also studied one modification of this experiment by allowing
only the anode to supply pairs with efficiency $\xi=1/2$. In this case, 
the $e^\pm$ atmosphere created by the anode filled the whole tube,
and conducted the current. Ions lifted from the anode surface
contributed a small fraction to the current $j_i/\jB\simeq 3\times 10^{-2}$.
The established voltage was $\sim 5$ times higher than in the model with
pair supply at both ends. However, the circuit was able to operate without
igniting the resonant pair creation (which we checked by setting
$\gres=\infty$). This shows that the thermally-driven regime can operate
even when pair creation takes place at one footpoint of the magnetic tube.
                                                                                
In summary, the coronal current may be fed by atmospheric $e^\pm$ 
instead of the pair discharge via the resonant channel if 
$\xi\simgt\gres^{-1}$. The numerical coefficient in this condition 
increases if one of the surfaces (anode or cathode)
dominates the pair supply.

\subsection{Anomalous Resistivity and Hydrostatic Equilibrium}\label{anom}

In addition to the heating by the relativistic beam that comes from the 
corona, the atmosphere may experience anomalous heating in 
response to the strong electric current. This type of heating is known to 
occur in astrophysical and laboratory plasmas when the drift speed of the 
current-carrying electrons, $v_d$, exceeds the sound speed  
$c_s = (k_{\rm B}T_e/m_i)^{1/2}$ (e.g. Bernstein \& Kulsrud 1961).
If the electron current can be described as 
a shift of a Maxwellian distribution with respect to the ions
in velocity space, $\Delta v_e=v_d$, a positive gradient 
$df_e/dv_e>0$ appears in the electron distribution function viewed from 
the rest-frame of the ion component. The positive gradient exists at 
$v_e<v_d$ and will amplify slow plasma modes with phase speed 
along the magnetic field  $\omega/k_\parallel<v_d$.
The dispersion relation of ion-sound waves gives 
$\omega/k_\parallel=c_s\,(1+k^2\lambda_{{\rm D}e}^2)^{-1/2}$
and they are easily amplified when $v_d>c_s$.
The excited ion-sound turbulence creates the anomalous resistivity 
and heating.  

With an increasing level of turbulence, $\Uturb/n_ek_{\rm B}T_e$, 
the mean free path 
of the electrons $\ell$ decreases according to equation~(\ref{eq:ell}). 
This implies a higher resistivity and a lower thermal conductivity. 
Thus, a reduced $\ell$ leads to a strong anomalous heating and 
the suppression of heat conduction out of the layer.  
It is then possible that the
layer is overheated, loses hydrostatic balance, and 
expands in a runway manner. We now estimate the critical $\ell$ for this
to happen. We neglect the radiative losses in this estimate.

The ohmic heating rate per unit area of the layer is
\be
   F_{\rm Ohm}\simeq h\,\frac{j^2}{\sigma},
\ee
where $h = k_{\rm B}T_e/m_ig$, $\sigma = n_e e^2 \ell/m_e v_e$
is electrical conductivity, and $v_e=(\kB T_e/m_e)^{1/2}$ is electron 
thermal speed. The conductive heat loss $F_c$ is given by 
equation~(\ref{eq:Fc}), and we find 
\be\label{fluxrat}
  \frac{F_{\rm Ohm}}{F_c}=
    \left(\frac{m_e}{m_i}\right)^2\,
    \left(\frac{c^2}{g\RNS}\right)^2\,
    \frac{\tau_c^2}{\theta_e^3}\,\left(\frac{\ell}{\lCoul}\right)^{-2}.
\ee
Hence the layer must be overheated if
\be\label{lratio}
  \frac{\ell}{\lCoul} < 2\times 10^{-3}\,
   \theta_e^{-3/2}\,\tau_c.
\ee
Here $\tau_c=\sT\RNS\,j/ce\sim 10^{-1}-10^{-2}$ (\S~6.2), and
the temperature is limited by pair creation to $\theta_e\simlt 1$
(see \S~\ref{pairatm}). One can see that the ohmic 
overheating requires a high level of plasma turbulence, when the 
mean-free path is strongly reduced compared to the Coulomb value.
This level is achievable in principle because it is still 
below the upper limit $\Uturb/n_ek_{\rm B}T_e\sim 1$ that corresponds to 
the minimum $\ell\sim\lDe$. Indeed, 
\be
 \frac{\lDe}{\lCoul}\simeq \frac{4}{3}\,\pi^{1/2}
      \left(\frac{n_er_e^3}{\theta_e}\right)^{1/2}
   =\left(\frac{2}{3}\right)^{1/2}\,\frac{\tT^{1/2}}{\theta_e}\,
    \left(\frac{gr_e}{c^2}\,\frac{m_i}{m_e}\right)^{1/2}
   \simeq 2\times 10^{-7}\,\frac{\tT^{1/2}}{\theta_e}.
\ee

Note that anomalous heating becomes interesting only when the thermal
heat flux is strongly suppressed.  In other words,
$F_{\rm Ohm}$ may compete with $F_c$ and overheat the layer 
only when $F_c$ is small. One can derive from the above formulae,
\be
  \frac{F_{\rm Ohm}}{F_c}\simeq
   \left(\frac{F_c}{4\times 10^{22}\,\tau_c\theta_e\,
         {\rm~erg~s}^{-1}{\rm~cm}^{-2}}\right)^{-2}.
\ee
Substituting $\tau_c\sim 10^{-1}-10^{-2}$ and $\theta_e\simlt 1$ one finds 
that the maximum energy flux from the ohmically-heated
layer is $F_c\sim 10^{21}$~erg~s$^{-1}$~cm$^{-2}$. Even if the layer covers
the whole star with area $4\pi\RNS^2\simeq 10^{13}$~cm$^2$, its total power 
would not exceed $10^{34}$~erg~s$^{-1}$. This is two orders of magnitude 
smaller than the observed luminosity of the corona, and hence the anomalous
heating of the atmosphere is unlikely to generate the observed activity.

The marginally stable state $F_{\rm Ohm}=F_c$ has an electrostatic potential 
drop $e\delta\Phi_e\sim k_{\rm B}T_e$ across the layer, which equals the
energy gained by an electron crossing the layer. Since $T_e$ is limited 
by pair creation, the maximum potential drop that can be sustained by the 
layer is $\sim m_ec^2$. This is at least 2 orders of magnitude lower 
than the voltage $\gres m_ec^2$ that is developed in the corona with 
$e^\pm$ discharges. If the turbulence in the transition layer reaches the 
level required for runaway heating, the plasma may be supplied to the corona 
at a much lower voltage compared with the pair discharge described in \S~5.

Pair creation in the atmosphere (see \S~\ref{pairatm}) would tend to 
reduce its mean molecular weight and increase $h$.
When the expressions for $F_{\rm Ohm}$ and $F_c$ are modified
to include the pair density, $n_e\rightarrow n_e+n_\pm$, one finds that
the critical value of $\ell/\lCoul$ for runaway heating (eq.~\ref{lratio})
increases as $1+n_\pm/n_e$. The critical value of pair density is
$n_*\sim n_c(h/\ell)$, at which the diffusive flux of hot $e^\pm$ out 
of the atmosphere dominates the corona and shortens out the electric 
circuit. 
One finds the relation
\be
\label{eq:Ohmpair}
  \frac{F_{\rm Ohm}}{F_c}=\left(\frac{n_*}{n_e+n_\pm}\right)^2
                          \,\frac{1}{\theta_e}.
\ee
If the atmosphere is the main source of charges for the coronal current
then $n_\pm\approx n_*$ is expected. Ohmic heating is unimportant in 
such an ``electrically regulated'' atmosphere if $n_\pm\approx n_*<n_e$.
It can become important if the atmosphere happens to enter the 
pair-dominated regime, $n_\pm >n_e$. Then the overheating with 
$F_{\rm Ohm}/F_c\sim \theta_e^{-1}>1$ is possible, leading to 
over-production of pairs, $n_\pm>n_*$, which will suppress 
$F_{\rm Ohm}/F_c$ and quickly shut the instability. 
Perhaps a cyclic overheating with pair enrichment would occur.

The effects discussed in this section indicate the possibility 
of an alternative way of forming the corona, through heating and pair 
creation in the atmosphere. This would require a high level of turbulence 
in the atmospheric plasma, and it is unclear whether this level is reached. 
Observations of the high-energy tail of the X-ray spectrum may help 
constrain the composition and temperature of the transition layer.


\section{CRUST EXCAVATION AND FORMATION OF A LIGHT-ELEMENT SURFACE LAYER}
\label{spall}

\subsection{Mass Transfer in the Coronal Circuit}

The existence of an ion current in the magnetosphere implies the transfer
of mass from the anode footpoint of a magnetic line to its cathode footpoint.   
The rate of transfer may be easily estimated for the idealized configuration 
of a twisted dipole (\S~2),
\be
  \dot{M}\sim m_p\frac{j_i}{e}\,4\pi\RNS^2 
         \sim 
   \,\left(\frac{j_i}{j}\right)\,
 \frac{m_p\,c}{e}\, \varepsilon_j\, B\RNS \sim 10^{17}\,
   \,\left(\frac{j_i}{j}\right)\,
\varepsilon_jB_{15}\, {\rm ~g~s}^{-1}.
\ee
Here the dimensionless coefficient 
$\varepsilon_j\equiv\Delta\phi \sin^2\theta$ 
measures the strength of the twist in the dipole magnetosphere 
(see eq.~[\ref{eq:j}]). During the active life of a magnetar, $t\sim 10^4$~yr, 
a significant mass is transferred through the magnetosphere,
$\Delta M\sim 3\times 10^{28}\,(j_i/j)\,\varepsilon_j\,B_{15}$~ g.
It is much larger than the mass of the stellar atmosphere, and the 
atmosphere is quickly removed from the anode part of the stellar surface.

There is, however, a mechanism which can re-generate the atmosphere 
and regulate its column density. A thin atmosphere is unable to stop the 
energetic particles flowing from the corona and these particles will hit 
the crust and knock out new ions. In a steady state, the atmosphere column 
density can adjust so that the rate of ion supply equals the rate of their
transfer through the corona. 
Then the rate of crust excavation at the anode footpoints is 
proportional to the ion current through the corona.

Excavation depends on the chemical composition of the uppermost crust of 
the neutron star. If it is made of light elements, the atmosphere is easily 
re-generated by simple knock out of ions from the surface. If it is made of 
carbon, oxygen, or any heavier elements, the ions are condensed into long 
molecular chains in a $\sim 10^{15}$~G magnetic field 
(Neuhauser, Koonin, \& Langanke 1987; Lieb, Solovej, \& Yngvason 1992;
see Thompson et al. 2000 for estimates of the chain binding energy in 
the case of $\sim 10^{15}$~G magnetic fields).
The knock out of light elements by the beam is still possible in this case
because of spallation reactions. We discuss the excavation process in more 
detail below.

\subsection{Excavation of Hydrogen and Helium Layers}

Light elements --- hydrogen and helium --- may be present only above a 
certain depth that may be estimated as follows.
The column mass density $\Sigma$ above a given depth in the crust is 
related to pressure $P$ at this depth through the equation of hydrostatic 
equilibrium,
\be
   \Sigma g = P = {eB_{\rm NS}\over 2\pi \hbar^2} {p_F^2   \over 2\pi}.
\ee
Here it is taken into account that the pressure is dominated by degenerate
electrons that are confined to their lowest Landau state 
($E_F<E_B$ in the density range of interest); 
$p_F\sim m_ec\, \rho_7 B_{15}^{-1}$ and
$E_F=[p_F^2c^2+m_e^2c^4]^{1/2}$ is the Fermi energy. A conservative
upper bound on the depth of a hydrogen layer is obtained by balancing 
$E_F$ with the threshold for electron captures on protons,
$E_F\sim 1.3$~MeV. This corresponds to $\rho\sim 10^7$~g~cm$^{-3}$.
At larger depths, free protons are rapidly consumed by 
reaction $e^-+p\rightarrow n+\nu$. 

A more realistic upper bound takes into account that hydrogen is burned
into helium, and the
maximum depth of a helium layer is limited by the triple-alpha
reaction $3\alpha \rightarrow ^{12}$C, which is very 
temperature-dependent. The interior of a magnetar probably sustains a 
temperature of $\sim 3\times 10^8$~K at an age of $\sim 10^3-10^4$~yr 
(e.g. Arras, Cumming, \& Thompson 2004). In the presence of a light-element 
atmosphere, the crust temperature remains close to the central temperature 
at depths $\rho>10^6$~g~cm$^{-3}$ (e.g. Potekhin et al. 2003). 
At such densities and temperatures helium (and hydrogen)
cannot be present --- its lifetime would be only 
$\sim 10^{-3}~(Y_e\rho_6)^{-2}$~yr where $Y_e\approx 0.5$ is the 
electron fraction. Helium may survive only in an upper layer 
of density $\rho<10^5$~g~cm$^{-3}$, where temperature is below 
$\sim 1\times 10^8$~K.

The maximal layer of light elements will be drained rapidly from the anode 
surface and deposited at the cathode surface in the presence of a magnetic 
twist. The excavation time down to a depth with a Fermi momentum $p_F$ is
\be
   t_{\rm exc} = \frac{\Sigma}{A m_pj_i/e}
               = 0.3\;{Z/A\over\varepsilon_j} \left(\frac{j_i}{j}\right)^{-1}
         \left({p_F\over m_ec}\right)^2 {R_{\rm NS,6}\over g_{14}}\;\;\;\;
{\rm yr}.
\ee
Substituting $j_i/j\sim 0.1$ and $\varepsilon_j \sim 0.1$ 
shows that the excavation time of the maximal hydrogen layer 
($E_F\simeq 1.3$~MeV) is much shorter than the active $\sim 10^4$-yr 
lifetime of a magnetar. Taking a more realistic maximum 
density $\rho \sim 10^5$~g~cm$^{-3}$ and
$p_F/m_ec \simeq 6\times 10^{-3}\rho_5B_{15}^{-1}$, 
one finds the excavation time of hours or days. It is 
faster than hydrogen consumption through
diffusion to a deeper layer of carbon (Chang, Arras, \& Bildsten 2004). 

We conclude that the anode surface of the star should be 
made of heavier elements --- carbon or iron. 
A gaseous atmosphere of light elements may be maintained above such a 
condensed, heavy-element surface due to the beam of energetic particles 
from the corona.

\subsection{Beam Stopping and Spallation of Nuclei in a Solid Crust}

When the coronal beam enters the crust, it is stopped by
two-body collisions, and collective processes become negligible. In particular,
the growth length of beam instability $\lb$ (eq.~[\ref{eq:lbeam}]) is 
much larger than the mean free path to two-body collisions (their ratio scales 
as $n_e^{5/6}$ and is large in condensed matter because $n_e$ is large). 
The stopping length of the beam is then mainly determined by Coulomb 
collisions with the nuclei.

Consider a downward-moving electron with an energy $E_e=\gamma m_ec^2\gg E_B$ 
(eq. [\ref{eq:E_B}]).
Collision with an ion can excite the electron to any Landau level 
$n>0$ with energy $E_n=(2nB/\BQED+1)^{1/2}<E_e$. Most 
frequently, it will be the first level $E_1=E_B$ 
because this process has the largest cross section (eq.~\ref{eq:Coul}).
The probability of excitation to levels $n^\prime\geq n$ is reduced
by a factor $\sim\frac{1}{n}$ (it requires a smaller impact
parameter; see also Kotov \& Kelner 1985).
One may assume a two-level system $n=0,1$ in simplified estimates.

The excited electron moves along the magnetic line with a new Lorentz 
factor $\gamma_\parallel$, which is found from the energy conservation,
$\gamma_\parallel E_B=E_e$. It immediately de-excites and emits a photon
at pitch angle $\them$ with energy $E_\gamma(\them)$ (\S~5.1).
The normal modes of radiation in the uppermost crust are determined 
by both vaccuum and matter polarization, which gives elliptical polarization 
states. A photon in either elliptical mode immediately converts to $e^\pm$ 
off the magnetic field if it is above the threshold, 
$E_\gamma>2m_ec^2/\sin\them$.
This condition is satisfied for a fraction $\fpair$ of de-excitation 
photons, which is estimated using the results of Herold et al. (1982). 
When $B \gg B_{\rm QED}$, the angular distribution of the emitted
photons is proportional to  $E_\gamma^2\,\exp[-{1\over 2}
(B/B_{\rm QED})^{-1}\,(E_\gamma/m_ec^2)^2]$ in both polarization modes.  
Combining this expression with $E_\gamma(\theta_{\rm em})$ 
(eq. [\ref{thr_O}]), we find that $\fpair$ can be well approximated by 
\be
      \fpair\approx 1- 
{2e\over (e-1)}\,\left({B\over B_{\rm QED}}\right)^{-1}
    \approx 1-0.14\,B_{15}^{-1},  \qquad B\gg \BQED.
\ee
The net result of collisional excitation, de-excitation, and 
photon conversion to $e^\pm$ is the 
sharing of initial electron energy $E_e$ between three particles, and 
each of them will again collide with an ion and produce more $e^\pm$.
Thus, a pair cascade develops as a beam with initial $\gamma=\gamma_0$ 
propagates into the crust, giving more $e^\pm$ with lower Lorentz factors 
$\gamma<\gamma_0$.

This branch of the cascade operates for a large 
fraction of the de-excitation photons. Photons that cannot immediately 
convert off the field (a fraction $1-\fpair$) will wait until they 
interact with an ion in the crust
(photon splitting is slow compared with photon-ion interaction).
Normally, such a photon will convert to $e^\pm$ in the Coulomb field 
of the nucleus. Less frequently, it can knock out a neutron, proton, or 
(rarely) alpha particle. This second branch of the cascade is most 
important for us, since it produces light elements.

The probability of proton knockout instead of normal conversion to 
$e^\pm$ may be roughly estimated using the known nonmagnetic cross 
sections for photon-nucleous collision. Knockout requires a minimum 
photon energy of $\sim 15-20$ MeV, and its cross section typically peaks 
at $\sim 20-30$ MeV (e.g. Dietrich \& Berman 1988). The cross section
for proton knockout from oxygen is
$\sigma(\gamma,^{16}{\rm O}\rightarrow p,^{15}{\rm N})\approx 12\;\;{\rm mb}$
at $E_\gamma \simeq 25$~MeV. It scales approximately in proportion to $Z$, 
and so we take
\be
   \sigma\left(\gamma,Z\rightarrow p,Z-1\right) = 1.5\, Z\;\;{\rm mb}.
\ee
The cross section for pair creation is
\be
    \sigma^Z_{e^+e^-} = {7\over 6\pi}\,Z^2 \alpha
 \left[\ln\left({2E_\gamma\over m_ec^2}\right)-{109\over 42}\right]\,\sT
  \approx 3.2\,Z^2 {\rm ~mb}.
\ee

We can now estimate the number of protons knocked out per incident 
energetic electron. We will assume in this estimate an idealized surface 
layer that is composed of ions of a single charge $Z$. We roughly
estimate the number of generations in the cascade down to an energy of 
$\sim 25{\rm ~MeV}\simeq 50m_ec^2$ as $N_g\sim\log_2(\gamma_0/50)\sim 5$. 
The number of $\sim 25$ MeV photons created per incoming electron is about 
$\gamma_0/50$. A fraction $1-\fpair$ of these photons hit a nucleus, 
yielding a proton with a probability 
$\sigma\left(\gamma,Z\rightarrow p,Z-1\right)/\sigma^Z_{e^+e^-}$. The 
resulting yield of protons per incident primary electron is 
\be\label{pyield}
     N_p \approx (1-\fpair)\,
   B_{15}^{-1} \left({\gamma_0\over 50}\right)\,
  {\sigma\left(\gamma,Z\rightarrow p,Z-1\right)\over \sigma^Z_{e^+e^-}}
   \sim 0.1\,
   \left({\gamma_0\over 10^3}\right)\,\left({Z\over 26}\right)^{-1}.
\ee

The $e^\pm$ cascade develops at a significant depth that buries the 
created positrons --- they annihilate in the crust and cannot escape 
back through the surface. The first Coulomb collision occurs at a 
Thomson depth $\tT\sim Z\sigma_T/\sigma_{\rm Coul}\approx 20\,B_{15}/Z$.
The relativistic cascade so initiated is strongly beamed 
into the star, and noticeable backscattering may occur only after the last 
stage of the cascade that gives $e^\pm$ with 
$\gamma\sim E_B/2m_ec^2\sim 3B_{15}^{1/2}$, after 
$\log_2(\gamma_0m_ec^2/E_B)\sim 8$ generations.
Even the first branch of the cascade, which involves immediate pair 
creation off the magnetic field, ends at a significant
$\tT\simgt 100Z^{-1}B_{15}$. 
This layer is optically thick to positron annihilation.
Therefore, the anode surface can supply positive charges only by emitting 
light ions, mostly protons that have been produced by spallation.

Spallation reactions occur down to a Thomson depth 
$\tT\sim N_g\,Z\sigma_T/\sigma^Z_{e^+e^-}\sim 10^3/Z$.
The knocked-out protons diffuse through the spallation layer 
on a short timescale, much shorter than the excavation time of this layer, 
and are quickly sublimated to the atmosphere.\footnote{
Protons have a binding energy $E_I$ of a few keV and are 
easily sublimated at the temperature 
$k_{\rm B}T\sim 0.5$~keV, in contrast to the heavy elements 
that are locked in molecular chains with $E_I\sim 10^2$~keV.} 
Thus, excavation of the crust is driven by spallation followed immediately
by sublimation. 

Our preferred circuit model has voltage $e\Phi_e\sim 1$~GeV and
ion current $j_i\sim 0.1j$ (\S~5). If the gaseous atmosphere does not 
decelerate the particles flowing from the corona, they will impact the 
solid crust with $\gamma_0\sim 10^3$. The estimated proton yield 
$N_p\sim 0.1(Z/26)^{-1}$ is then sufficient to feed $j_i\sim 0.1j$ 
and can supply more ions than transferred through the corona. Then
the column density of the atmosphere will 
grow until it is able to damp the relativistic beam from the corona and 
reduce the crustal excavation rate. 

The profile of the atmosphere is determined by its 
efficiency of stopping the relativistic $e^\pm$. If collisionless processes 
do not decelerate the relativistic beam but just form a plateau 
in its distribution function (\S~6.1), then a 
thick atmosphere must be accumulated to decelerate
the beam through Coulomb collisions. Collisions initiate a cascade 
similar to that in the crust: the beam particles collide with 
ions and emit photons that convert to pairs.\footnote{
The relatively low density of the atmosphere implies two differences with
the cascade in the crust: (1) The polarization states of photons are now
linear --- vaccuum O- and E-modes -- and the E-mode does not 
convert to $e^\pm$ off the field. (2) Splitting of E-mode occurs faster 
than photon-ion interaction. This leaves only one branch for the 
cascade, with $e^\pm$ production off the magnetic field.}
The crustal spallation rate will be reduced if 
the beam is decelerated to $\gamma\simlt 50$
before reaching the solid surface.
This requires
$\log_2(\gres/50)\sim 5$ generations of pair-creation, and so 
the optical depth to Coulomb collisions should be $\sim 5$. Thus, the 
column density of the atmosphere will be regulated to the value
\be
    N_{  \rm atm}\sim \frac{\,
     \log_2(\gres/50)}{\sigma_{\rm Coul}}
      \sim 100\,B_{15}\,\sT^{-1}.
\ee

If $N_p/(1+N_p)<j_i/j$, the crust will not supply the ion flux
implied by our circuit solution with $e\Phi_e\sim \gres m_ec^2$.
We then expect the voltage to grow, giving a higher spallation rate
of the crust. We leave this regime for a future study.

The thickness of the plasma layer that is accumulated near the cathode 
surface is limited by burning processes and diffusion to greater depths 
(Chang et al. 2004).


\section{DISCUSSION}

The basic finding of this paper is that an $e^\pm$ corona must be 
maintained around a magnetar. It exists in a state of self-organized
criticality, near the threshold for $e^\pm$ breakdown. The stochastic
$e^\pm$ discharge continually replenishes the coronal plasma, which 
is lost on the dynamic time $t_{\rm dyn}\sim 10^{-4}$~s. 

The solution for the plasma dynamics in the corona is strongly 
non-linear and time-dependent. It is essentially global in the sense that 
the plasma behavior near one footpoint of a magnetospheric field line is 
coupled to the behavior near the other footpoint. Remarkably, this 
complicated global behavior may be described as essentially one-dimensional 
electric circuit that is subject to simple boundary conditions $E\approx 0$ 
on the surface of the star and can be studied using a direct numerical 
experiment.

The established voltage across the entire magnetospheric circuit is 
marginally sufficient to accelerate an electron (or positron) to the 
energy where collisions with ambient X-ray photons spawn new pairs.
This physical condition determines the rate of energy dissipation in 
the corona. The voltage is maintained by small deviations from charge 
neutrality in the coronal plasma as required by Gauss' law. 
The generation of this voltage is precisely the 
self-induction effect of the gradually decaying magnetic twist, and 
the energy release in the corona is fed by the magnetic energy
of the twisted field.

\subsection{Comparison With Canonical Radio Pulsars}

Canonical radio pulsars are usually assumed to have a magnetosphere that
is almost everywhere potential, $\nabla\times\bB=0$, except for narrow
bundles of open field lines that connect the polar caps of the star to its
light cylinder. The plasma content of the magnetosphere then depends on 
its rotation: the charge density 
$\rho_{\rm co}=-{\mathbf \Omega\cdot{\bf B}}/2\pi c$
must be maintained to screen the electric field induced by rotation
(Goldreich \& Julian 1969). In this picture, 
the persistent current of a canonical pulsar 
only flows along the open field lines out to the light cylinder.
The sign of this current $I_{\rm open}$ is determined by
the corotation charge density.  
The closure of the electric circuit back to the star remains poorly 
understood.

Previous studies of plasma
near neutron stars focused on open magnetic lines. It was shown that 
pair discharge can be sparked near the polar cap 
(Ruderman \& Sutherland 1975; Arons \& Scharlemann 1979; 
Muslimov \& Tsygan 1992), and the created pairs carry the current 
$I_{\rm open}$ toward the light cylinder. Recent calculations of the 
discharge (Harding \& Muslimov 2002) employ an 
approximate analytic prescription for the decay of the electric field with 
distance from the star. (Self-consistent calculations of the electric field 
and pair creation are now available for the outer gap, see Takata et al. 2006.)  
The global dynamics 
of the plasma on open magnetic lines is poorly understood because of the
complicated behavior at the light cylinder, so only one boundary condition
is known --- at the polar cap.

The solution presented in this paper applies to 
currents along {\it closed} magnetic 
lines of the inner magnetosphere. In this case, it is easier to find the 
global solution for the plasma dynamics because both boundary conditions 
are defined at the two footpoints of the magnetic lines. 
This problem has important applications to the magnetar physics 
for two reasons: 
(1) The observed radiative output is generated in the closed 
magnetosphere --- the dissipation rate 
on the open field lines is orders of magnitude smaller and cannot 
explain the observed emission.
(2) A current is naturally created on closed magnetic lines
because they are twisted by the starquakes of magnetars, 
and our solution shows that such twists have a long lifetime.

One may compare the twist of open magnetic lines at the polar cap,
as sustained by the rotation of the star, 
$B_\phi/B_P\sim (\RNS/R_{\rm lc})^3$, with the twist that 
can be created in the closed magnetosphere by the starquakes.
Consider a current $I$ flowing along a magnetic flux tube of radius $a$ 
and let $B_\phi$ be the twisted component of the field. Then
\be
  I = {c\over 2}a B_\phi = {c\over 2} a^2 B_P {\dd \phi\over \dd l},
\ee
where $l$ is the length measured along the tube.
Since the current $I$ and the poloidal magnetic flux $\pi a^2 B_P$ 
are constants along the tube, the twist 
angle $\phi$ satisfies $\dd\phi/\dd l=const$. Hence,
\be
{\dd\phi\over \dd l} \sim {\Delta\phi\over R_{\rm max}},
\ee
where $R_{\rm max}$ is the maximum radius to which the tube extends and
$\Delta\phi$ is the net twist angle integrated along the tube.
The net twist angle is limited by the global stability of magnetosphere
to $\Delta\phi\simlt 1$. 
On the star's surface one then finds $B_\phi/B_P=a\dd\phi/\dd l\simlt 
a/R_{\rm max}$. This implies that the closed field lines with 
$R_{\max}\sim 2\RNS$ and $a\sim \RNS$
can bear much stronger twists than the open lines.

Note also one more differrence 
of the magnetar currents: they impact the stellar surface with
a modest Lorentz factor $\gb\sim 10^3$. Such a beam can deposit 
its energy in an optically thin atmospheric layer through a plasma 
instability (\S~\ref{sc:bi}).
By contrast, $e^\pm$ from the discharge in an ordinary radio pulsar
impact the star with high $\gb=10^6-10^7$, which suppresses
the beam instability in the atmosphere: its length scale $\lb$ 
is proportional to $\gb$ (eq.~\ref{eq:lbeam}) and becomes much
larger than the atmosphere scale-height. Such a beam must
be stopped by Coulomb collisions and relativistic bremsstrahlung 
in the solid crust. Therefore, the polar cap of an ordinary pulsar is
expected to release the beam energy in the form of blackbody radiation,
in agreement with observations (e.g. De Luca et al. 2005).
By contrast, magnetars may have a strongly overheated hydrostatic 
atmosphere, with $\kB T > 100$~keV (\S~6.2).

\subsection{Comparison with Solar Corona}

The corona of a magnetar may be viewed as a collection 
of closed current-carrying flux tubes with footpoints anchored in a 
relatively cold stellar surface.  As in the Solar corona, a 
current is created by the twisting motions of the footpoints.
Another similarity with the Solar corona 
is the important interaction of the coronal plasma with a dense
atmospheric layer on the stellar surface where the plasma becomes 
collisional and can efficiently radiate the energy flux received from 
the corona.

However, the magnetar corona is
distinguished from the Solar corona in a few  important respects. 
The differences in the behavior of the magnetic field are as follows.

\noindent
1. The footpoints of magnetic lines of the Solar corona are constantly 
pushed and twisted because the outer envelope of the Sun is convective 
and in constant motion. 
In addition, magnetic flux tubes are observed to emerge from the
Solar convection zone with a twist already implanted in them.
By contrast, the neutron star's rigid crust is 
subject only to rare and sporadic yielding events (starquakes).
Because the crust is stably stratified, the transfer of a toroidal
magnetic field to the exterior occurs 
through a differential rotation of the crustal material along the
gravitational equipotential surfaces.

\noindent
2. The magnetic twists have long lifetimes between starquakes, and their 
evolution differs from the evolution of twists in the Solar corona.
Once a current $\bj=(c/4\pi)\nabla\times\bB$ is created in the magnetosphere 
of a neutron star, it lives a long time because the net current 
$I\sim \RNS^2j$ is enormous and maintained by its self-induction.
The gradual decay of the current induces an e.m.f. along the magnetic field 
lines, which forces the plasma and electric field to self-organize into a 
quasi-steady electric circuit. The ease with which charges can be supplied 
by $e^\pm$ production when the induced voltage exceeds $e\Phi_e\sim 1$~GeV 
implies a ``bottleneck'' for the decay of the twist (\S\S~2 and 5). 

\noindent
3. The evolution of the magnetic field between the starquakes involves
a gradual release of the twist energy through Joule dissipation and 
spreading of the twist toward the magnetic polar axis. The latter
process causes a gradual flaring of the poloidal field lines.
Explosive magnetic reconnection similar to Solar flares is likely to 
be triggered by starquakes, but not during the following slow-dissipation 
stage.

\noindent
4. The external magnetic field of a magnetar is likely to have 
less small-scale structure compared with the Solar corona.
The small-scale braiding of magnetic lines followed by reconnection 
and dissipation is one of the possible heating mechanisms for 
the Solar corona (Parker 1979). 
Precisely what creates this structure is unclear: it could be driven 
by the hydrodynamic turbulence that is excited at the top of the Solar 
convection zone or by the Taylor relaxation of a helical magnetic field 
(e.g. Diamond \& Malkov 2003).
In magnetars, the small-scale structure of magnetic field can be expected 
to dissipate quickly after the starquakes if it is created in these 
events. The dissipation timescale is longest for the large-scale twists
expected from large-scale motions of the crust.
Indirect evidence that small-scale instabilities are less important 
comes from the observed distribution of energies of SGR bursts:
the cumulative energy in bursts $E^2dN/dE \sim E^{0.4}$ is weighted toward the
largest events (Cheng et al. 1996; Gogus et al. 2001).  By contrast,
several recent measurements of the distribution of Solar flare energies
suggest a power-law $dN/dE = E^k$ with $k< -2$, and are therefore
consistent with the basic picture of coronal heating by `nano-flares'
(Walsh \& Ireland 2003).

Let us also note a few differences in the plasma behavior.

\noindent
5. The rapid cooling of the transverse motion of charged particles
in the ultra-strong magnetic field of magnetars makes the particle distribution 
functions one-dimensional. This reduces the number of relevant plasma modes
in the corona. 

\noindent
6. The coronal plasma of magnetars is continually lost to the surface 
and re-generated. By contrast, the Solar corona is in hydrostatic equilibrium.
Its density $n$ is much higher than the minimum needed to maintain 
the current, and the mean drift speed $j/en$ of the coronal currents 
is much below the electron and ion thermal speeds.
So, the {\it mean} current does not experience a strong anomalous resistivity.
Strong plasma turbulence is expected in localized current sheets and 
can develop explosively, giving rise to the Solar flares. In the magnetar
corona, a strong turbulence is maintained routinely, everywhere where 
current flows. The very mechanism of the current supply involves
a strongly non-linear phenomenon --- $e^\pm$ discharge. 

\noindent
7. The magnetar corona is dominated by relativistic $e^\pm$ pairs. 
As indicated by our numerical experiments, ions lifted from the surface
contribute a modest fraction, perhaps $\sim 10$\%, to the charge flow.

\subsection{Magnetospheric Circuit as a Double-layer Problem}\label{double}

From the plasma physics point of view, there is an interesting connection
between the magnetospheric circuit and the double-layer problem that
was first studied by Langmuir (1929) and later by many authors (for 
reviews see Radau 1989; Block 1978). 
The magnetospheric current flows along a magnetic tube with footpoints on 
a cold conductor and is subject to the boundary condition
$E \simeq 0$ at the footpoints.
We argued that the tube curvature is not essential 
to understanding many features of the circuit, and studied the charge flow
in a straight tube
where gravity is replaced by a force $m{\mathbf g}$ that attracts
plasma to the anode and cathode boundaries.
The longitudinal electric field must develop between the 
two boundaries
because of gravity. This is a significant difference from
the normally invoked mechanisms for double-layer formation (such as 
current-driven instability).  In our case, gravity
traps the thermal particles at the magnetic footpoints
and impedes the flow of the current that is required by the 
self-induction of the twisted magnetic field. 

Thus, perfect conditions for double-layer formation
are created. And indeed, our numerical experiment showed the formation 
of a huge relativistic double layer between the footpoints, of size 
$L\sim\RNS$. 
This result is obtained if the current can be maintained only 
by particles lifted from the anode and cathode surfaces. It could have
dramatic consequences for the magnetar physics because of the huge 
dissipation rate in the double layer: the energy of magnetic twist 
would be quickly released in a powerful burst.

However, there is one more ingredient in the problem, which again makes the 
magnetar circuit different from the previously studied double layers:
new charges ($e^\pm$) can be created by accelerated particles in the tube. 
This process qualitatively changes the problem because voltage is now 
limited by $e^\pm$ discharge (\S~5). As a result, the circuit enters 
the new state of self-organized criticality where voltage repeatedly
builds up and is screened by the $e^\pm$ creation. This allows the 
current to flow and maintain the required $\nabla\times\bB$.

\subsection{Observational Implications}

Two important observational implications of the magnetospheric twist
were already pointed out by TLK, based on the force-free model of a 
twisted dipole:

\noindent
1. The poloidal field lines are inflated compared with a pure dipole, so 
that a larger magnetic flux connects the star to its light cylinder 
$R_{\rm lc}$. This implies a higher spin-down torque. Thus, the twisting 
leads to a higher spindown rate $\dot{P}$.

\noindent
2. The minimum particle density needed to maintain the currents, 
$n_c=j/ec$, is high enough to make the magnetosphere semi-opaque. 
The main source of opacity is resonant cyclotron scattering.
If the magnetic field lines are twisted, then
the optical depth to scattering happens to be $\sim n/n_c$
(independent of $r$, $B$, and nature of the flowing charges). 
Scattering by the magnetospheric charges must modify 
the X-ray spectrum and pulse profile.

TLK also pointed out that if the magnetospheric current is carried by 
ions and electrons lifted from the stellar surface by a minimum voltage 
$e\Phi_e\sim gm_p \RNS$, then one expects a dissipation rate comparable 
to the observed persistent luminosity of magnetars.
   
Further implications of our model for the coronal electric circuit
are as follows. 

\noindent
3. The density of the plasma corona is found to be close to its 
minimum $n_c=j/ec$ as conjectured by TLK. 
It is largely made of $e^\pm$ pairs rather than 
electron-ion plasma. This suggests that the optical depth of the corona 
is determined mostly by the cyclotron resonance of $e^\pm$. 
All species of particles ($e^-$, $e^+$, and even ions) 
flow with relativistic bulk speeds and large temperatures $\sim e\Phi_e$. 
The cyclotron resonance is very broad, which will have important implications 
for radiative transfer through the corona. 
Note, however, that our calculations focused on small radii $r\sim\RNS$ 
where most of the energy release occurs. At larger radii $r\sim(10-20)\RNS$, 
where electron cyclotron energy is in the keV range, $e^\pm$ feel a strong 
radiative drag force and can decelerate to a small velocity. 
We defer a detailed analysis of this part of the corona to another paper.

\noindent
4. The voltage $e\Phi_e\simeq \gres m_ec^2\sim 1$~GeV
along the twisted magnetic lines
of the corona is regulated by the threshold for $e^\pm$ discharge,
and its exact value depends on 
$B$ and the spectrum of target photons involved 
in $e^\pm$ creation, especially on its high-energy tail at $10-300$~keV
(\S~5).  
By coincidence, the mean electric field developed along the magnetic lines,
$E\sim \Phi_e/\RNS$, is marginally sufficient to lift ions from the anode 
footpoint. Therefore, in addition to $e^\pm$ pairs, some ions are present in
the corona. Their abundance depends on the exact $\Phi_e$ and is about 10\% 
for $e\Phi_e=1$~GeV (see Fig.~10). The coronal ions
may produce interesting features in the magnetar spectrum 
and likely to play a key role in the low-frequency emission (\S~8.4.2).

\noindent
5. The rate of energy dissipation in the twisted magnetosphere is given by 
$L_{\rm diss}=I\Phi_e$ where $I$ is the net current through the corona. 
The current may be estimated as 
\be
   I\sim \jB a^2\sim \frac{c}{4\pi}\,\Delta\phi\,\frac{B}{\RNS}\,a^2, 
\ee
where $a$ is the size of a twisted region on the stellar surface and 
$\Delta\phi=|\nabla\times\bB|(B/\RNS)^{-1}\simlt 1$ characterizes the
strength of the twist.
The calculated $e\Phi_e$ is comparable to 1~GeV and implies
\be
\label{eq:Ldiss}
   L_{\rm diss}=I\Phi_e\sim 10^{37}\,\Delta\phi\,
    \left(\frac{B}{10^{15}{\rm ~G}}\right)
    \left(\frac{a}{\RNS}\right)^2\left(\frac{e\Phi_e}{\rm GeV}\right)
           {\rm ~erg~s}^{-1}.
\ee
The observed luminosity $\Ldiss \sim 10^{36}$~erg~s$^{-1}$ 
is consistent with a partially twisted magnetosphere, $a\sim 0.3\RNS$, or 
a global moderate twist with $\Delta\phi\,(e\Phi_{e}/\rm GeV)\sim 0.1$.  
(Larger dissipation rates and larger twists are not
yet excluded by the data: e.g. the output at 1~MeV is presently unknown.)
More precise formulae may be obtained for concrete magnetic configurations.
In the simplest case of a twisted dipole (TLK), one can use 
equation~(\ref{eq:j}) and find $I$ by integrating $\bj\cdot d{\mathbf S}$ 
over one hemisphere of the star. This gives 
$I=\frac{1}{8}\,\Delta\phi c\Bpole\RNS$ and 
$\Ldiss=1.2\times 10^{37}\Delta\phi B_{{\rm pole},15}R_{{\rm NS},6}
\Phi_{e,\rm GeV}$.

\noindent
6. Once created, a magnetospheric twist has a relatively long but finite 
lifetime. The energy stored in it is ${\cal E}_{\rm twist}\sim(I^2/c^2)\RNS$.  
This energy dissipates in a time
\begin{eqnarray} 
\label{eq:tdecay}
     t_{\rm decay}=\frac{{\cal E}_{\rm twist}}{\Ldiss}\simeq 
          \frac{IR_{\rm NS}}{ c^2\Phi_e}=\frac{R_{\rm NS}}{ c^2}
                                    \frac{L_{\rm diss}}{\Phi_e^2} 
     \simeq 3\,
           \left(\frac{L_{\rm diss}}{10^{36}\rm ~erg~s^{-1}}\right)
                \left(\frac{e\Phi_e}{\rm GeV}\right)^{-2} {\rm ~yr}.
\end{eqnarray} 
The exact numerical coefficient in this equation again depends on the 
concrete magnetic configuration. For example, the energy stored in the 
global twist of a dipole is
${\cal E}_{\rm twist}=0.17(\Delta\phi)^2 E_{\rm dipole}$ (TLK), where 
$E_{\rm dipole}=\frac{1}{12}\,\Bpole^2\RNS^3$ is the energy of a normal
dipole. In this case, one finds 
$t_{\rm decay}=0.8\,(\RNS/c^2)\,\Ldiss/\Phi_e^2$.

The characteristic timescale of decay is determined by 
voltage $\Phi_e$ and
the corona luminosity $L_{\rm diss}$. 
Note that a stronger twist (brighter corona) lives longer.
Given a measured luminosity and its decay time, one may infer $\Phi_e$
and compare it with the theoretically expected value. Since $\Phi_e$ does 
not change as the twist decays in our model, 
one can solve equation~(\ref{eq:tdecay}) for ${\cal E}_{\rm twist}(t)$ and 
$L_{\rm diss}(t)=-\dd{\cal E}_{\rm twist}/\dd t$. 
This gives 
\be
\label{eq:Ldecay}
  L_{\rm diss}(t)=5\times 10^{35}\left(\frac{e\Phi_e}{\rm GeV}\right)^{-2} 
               \left(t_0-t\right) {\rm ~erg~s}^{-1},
\ee
where time $t$ is expressed in years. 
The final decay of the twist should occur at a well-defined time 
$t_0=t+t_{\rm decay}$.
The finite time of decay implies that the emission called ``persistent''
in this paper is not, strictly speaking, persistent. If the time between
large-scale starquakes is longer than $t_{\rm decay}$, the 
magnetar should be seen to enter a quiescent state and the observed
luminosity should then be dominated by the surface blackbody emission.

\noindent
7. The voltage $\Phi_e$ implies a certain effective resistivity
of the corona, ${\cal R}=\Phi_e/I$. This resistivity leads to 
spreading of the electric current {\it across} the magnetic lines. 
Equivalently, the magnetic helicity is redistributed within the magnetosphere.
This spreading is described by the induction equation 
$\partial\bB/\partial t=-c\nabla\times\bE$.  The magnitude of
$\nabla\times\bE$ depends on the gradient of $E_\parallel$ transverse
to the twisted magnetic field lines. Our model of the magnetospheric
voltage implies that $\Phi_e$ has a gradient on a lengthscale $\sim \RNS$,
so that  $\partial B_{\rm twist}/\partial t\sim c E_\parallel/\RNS$. 
The characteristic spreading time of the twist 
$B_{\rm twist}=\Delta\phi\,B\,(a/\RNS)$ is 
$t_{\rm spread}\sim B_{\rm twist}\,(\partial B_{\rm twist}/\partial t)^{-1}$
and can be roughly estimated as
\be
\label{eq:tspread}
  t_{\rm spread}\sim 
    \frac{\Delta\phi\,B  }
    {E_\parallel}\,\frac{\RNS}{c}\,\left({a\over \RNS}\right)
            \sim 300\,\Delta\phi\, \left(
    \frac{B  }
{10^{15}\rm ~G}\right)
                    \left(\frac{e\Phi_e}{\rm GeV}\right)^{-1}\,
\left({a\over \RNS}\right)
{\rm ~yr}.
\ee
If a twisting event happens near the magnetic equator, it initially 
does not affect the extended magnetic lines with $R_{\max}\sim R_{\rm lc}$. 
Its impact on spindown may appear with a delay as long as $t_{\rm spread}$. 
Although the exact numerical coefficient in equation~(\ref{eq:tspread}) is
unknown, the rough estimate suggests a timescale that is comparable
to the decay time (eq.~[\ref{eq:tdecay}]). Their ratio is 
$\propto (a/\RNS)^{-1}$ and depends sensitively on geometrical factors.
It is possible that starquakes occurring far from the poles will have 
little effect on the spindown.

\noindent
8. Even a small ion current implies a large transfer of mass through the 
magnetosphere over the $\sim 10^4$-yr active lifetime of a magnetar.
Essentially all hydrogen and helium that could be stored in the upper crust 
is rapidly transferred from the anode surface to the cathode
surface, thereby exposing heavier elements (e.g. carbon)
that are strongly bound in molecular chains. A continuing
flux of protons can, nonetheless, be maintained 
in the circuit. The bombarding relativistic electrons from the corona 
cause the gradual disruption of ions just below the solid surface.
They trigger a cascade of $e^\pm$ pairs and gamma-rays that knock out 
nucleons and occasionally $\alpha$-particles from the heavier nuclei. 
This process re-generates the
light-element atmosphere on the surface and regulates its column density
(\S~7). As a result, the Thomson depth of the atmosphere equilibrates to 
a value $\sim 100$. 
The chemical composition of the upper crust (carbon or heavier elements)
and its light-element atmosphere affect radiative transfer and the 
emerging spectrum of the star's radiation (e.g. Zavlin \& Pavlov 2002).

\noindent
9. Our study of the transition 
layer between the corona and the star (\S~\ref{sc:layer}) suggests that 
the energy dissipated in the corona is radiated below it in two forms
(see also Thompson \& Beloborodov 2005). First, the high-energy radiation 
with a hard spectrum possibly extending up to $\sim 1$~MeV; this 
radiation is caused by collisionless dissipation of the coronal beam 
in the transition layer.
Second, a blackbody radiation component that is caused by thermalization 
of the remaining energy of the beam when it enters the dense atmosphere
and then the crust; two-body collisions and development of an $e^\pm$ 
cascade are responsible for this thermalization. A crude estimate
gives comparable luminosities in the high-energy and blackbody components. 
Further investigation of the transition layer
may show that the collisionless dissipation is more efficient.
This would be in a better agreement with observations of SGR~1806-20 
whose high-energy component was found more luminous than the blackbody 
component (Mereghetti et al. 2005; Molkov et al. 2005).


\subsection{Emission from the Magnetar Corona}

\subsubsection{Hard X-ray Emission}

The radiative output from the corona is observed to peak above 100~keV 
(Kuiper, Hermsen \& Mendez 2004; Mereghetti et al. 2005; Molkov et al. 2005;
den Hartog et al. 2006). It is likely fed by the inner parts of the corona:
in the model investigated in this paper,
one expects that the bulk of the coronal current is concentrated fairly close
to the star, at radii $\la 2\RNS$. Currents in the outer 
region $r\gg \RNS$ (even if it is strongly twisted) flow through
a small polar cap on the star and therefore are small (\S~8.1).

Emission from the inner corona must be suppressed above $\sim 1$~MeV, 
regardless the mechanism of emission, because photons with energy 
$\simgt 1$~MeV cannot escape the ultra-strong magnetic field.
There are two possible linear polarizations of photons in the corona:
$E$-mode (electric vector perpendicular to $\bB$) and O-mode 
(electric vector has a component along $\bB$). An energetic
$E$-mode photon cannot 
escape because it splits into two photons. An O-mode photon propagating 
at angle $\theta$ with respect to $\bB$ cannot escape when its energy  
$E_\gamma>2m_ec^2/\sin\theta$ because it converts to a pair.
Therefore, the spectrum escaping the inner corona 
is limited to $\sim 1$~MeV. The available data show that 
the high-energy radiation extends above 100~keV. There is an indication
for a cutoff between 200~keV and 1~MeV from  
COMPTEL upper limits for AXP 4U 0142$+$61 (den Hartog et al. 2006).

The possible mechanisms of emission are strongly constrained.
The corona has a low density 
$n\sim n_c=\jB/ec\simlt B(4\pi\,e\RNS)^{-1}\sim 10^{17}B_{15}$~cm$^{-3}$ 
and particle collisions are rare, so two-body
radiative processes are negligible. The only available emission process 
in the corona is upscattering of surface X-rays. The scattering can occur
resonantly (excitation to the first Landau level followed by de-excitation)
or non-resonantly (Compton scattering). The scattering, however, cannot 
explain the observed 100-keV emission:

\noindent
1. {\it Resonant upscattering by electrons (or positrons)} with 
Lorentz factor $\gamma_e$ gives high-energy photons with energy 
$\sim\gamma_e\hbar eB/m_ec\gg 1$~MeV (\S~5.1), which cannot escape. 
They either split (if E-mode) or convert to pairs (if O-mode).
Splitting may give escaping photons below 1~MeV, but their number 
is not sufficient to explain the observed emission. 
The voltage along the magnetic lines adjusts so that the 
coronal particles experience $\sim 1$ resonant scattering before they are
lost to the surface, while the observed emission requires the production of
$10^3-10^4$ 100-keV photons per particle.

\noindent
2. {\it Resonant upscattering by ions} with Lorentz factors
 $\gamma_i$ gives photons of energy $\gamma_i eB/m_pc$, 
which is too low to explain the 100-keV emission (ions are mildly
relativistic in the circuit with $\sim$~GeV voltage, $\gamma_i\sim 1$).

\noindent
3. {\it Non-resonant Compton scattering} is negligible for 
$E$-mode photons.
The cross section of O-mode scattering by electron with Lorentz factor
$\gamma_e$ is $\sim \sT/\gamma_e^2$.\footnote{This assumes Thomson 
regime of scattering, $\gamma_e\hbar\omega\ll m_ec^2$. At larger energies,
the electron recoil becomes important (Klein-Nishina regime) and the cross 
section is reduced.} 
This may be seen by 
transforming the photon to the rest frame of the electron. 
Then the photon propagates at angle $\sin\theta^\prime\sim 1/\gamma_e$ 
with respect to the magnetic field and its electric vector is nearly 
perpendicular to $\bB$. Viewing 
it as a classical wave that shakes the 1D electron, one finds the cross 
section of scattering $\sigma=\sT\sin^2\theta^\prime=\sT/\gamma_e^2$.  
The corresponding optical depth of the corona is 
$\tau=n_c\RNS\sigma=\tau_c/\gamma_e^2$. The upscattered photon is boosted 
in energy by $\gamma_e^2$ and the net Compton amplification factor of 
O-mode radiation passing through the corona is $\tau\gamma_e^2=\tau_c$, 
i.e. does not depend on $\gamma_e$. Since $\tau_c=\sT\RNS n_c$ is small, 
$\sim 0.1-0.01$, one concludes that the energy losses of the corona 
through non-resonant Compton scattering are a small fraction of the 
surface O-mode luminosity.

A possible source of observed $20-100$~keV emission is the transition 
layer between the corona and the thermal photosphere
(see Thompson \& Beloborodov 2005 and \S~6). This layer
can radiate away (through bremsstrahlung) a significant 
fraction of the energy received from the corona. Its temperature is 
$k_{\rm B}T\sim 100(\ell/\lCoul)^{-2/5}$~keV (eq.~[\ref{eq:Tem}]), where 
$\ell/\lCoul<1$ 
parameterizes the suppression of thermal conductivity by plasma turbulence. 
This suppression is likely to be significant
and increase the temperature of the layer. The emitted hard X-ray 
spectrum may be approximated as single-temperature optically thin 
bremsstrahlung (Thompson \& Beloborodov 2005). Its photon index below 
the exponential cutoff is close to $-1$, in agreement with magnetar 
spectra observed with INTEGRAL and RXTE.
 
The observed pulsed fraction increases toward the high-energy end of the 
spectrum and approaches 100\% at 100~keV (Kuiper et al. 2004).
If the hard X-rays are 
produced by one or two twisted spots on the stellar surface, the large 
pulsed fraction implies that the spots almost disappear during some phase 
of rotation. Then they should not be too far from each other 
and, in particular, cannot be antipodal because a large fraction of the
stellar surface is visible to observer due to the gravitational bending 
of light: $S_{\rm vis}/4\pi\RNS^2=(2-4GM/c^2\RNS)^{-1}\approx 3/4$ 
(Beloborodov 2002). The decreasing pulsed fraction at 
smaller energies may be explained by contamination of the strongly 
pulsating component by another less pulsed and softer component of the 
magnetospheric emission (see \S~8.4.2).

Hard X-ray emission may also be produced at about 100~km
above the star surface (Thompson \& Beloborodov 2005). 
The magnetic field is about $10^{11}$~G in this region, and the
electrons can experience a significant radiative drag as a result of 
resonant cyclotron scattering. They cannot close the circuit unless an 
additional electric field develops that helps them flow back to the star
against the radiative pressure. Then some positrons in this region
can undergo a runaway acceleration and Compton scatter keV photons 
to a characteristic energy $E_\gamma\simeq (3\pi/8)\alpha^{-1} m_ec^2$,
where $\alpha=\frac{1}{137}$ is the fine structure constant. Such a photon 
is immediately converted to an $e^\pm$ pair in the magnetic field, 
and the created pairs emit a synchrotron spectrum consistent with the 
observed 20-100~keV emission.

\subsubsection{2-10 keV Emission}

A promising emission mechanism for the observed soft X-ray tail of the 
blackbody component is resonant upscattering of keV photons in the corona
(TLK). The emergent soft X-ray spectrum and pulse profile is likely to form 
at radii where $B\sim 10^{11}$~G and 
the $e^\pm$ can be decelerated to mildly 
relativistic speeds by the resonance with keV photons.

Ions flowing through the inner magnetosphere can also upscatter the 
keV photons. The ions are mildly relativistic
and their cyclotron resonance near the star is $6B_{15}$~keV --- just in 
the $2-10$~keV band. On the other hand, the ion current may be suppressed 
(see Fig.~10), which would imply their small optical depth in the corona, 
except the first $10^3-10^5$ seconds following an SGR flare when many ions 
are lifted to the magnetosphere (Ibrahim et al. 2001).

\subsubsection{Optical and Infrared Emission}

The observed infrared ($K$-band) and optical luminosities of magnetars are 
$\sim 10^{32}$~erg/s (Hulleman et al. 2004), which is far above the 
Rayleigh-Jeans tail of the surface blackbody radiation.  The 
inferred brightness temperature is $\sim 10^{13}$~K if the radiation 
is emitted from the surface of the star. Four principle 
emission mechanisms may then be considered.

\noindent 
1. Coherent plasma emission.
Eichler, Gedalin, \& Lyubarsky (2002) proposed that the optical emission 
is generated as O-mode radiation at a frequency $\sim \omega_{Pe}$. 
We find that this mechanism gives too low a frequency of emission.
The plasma frequency in the corona is suppressed by the large relativistic 
dispersion of the $e^\pm$, $\omega_{Pe}\sim(4\pi n_ce^2/m_e\gamma_e)^{1/2}$;
in addition, it may not be blueshifted by bulk relativistic motion as 
suggested by pulsar models. A fraction of $\sim \omega_{Pe}$ photons
may, however, be upscattered to the optical or higher bands by the 
coronal $e^\pm$ that have a broad distribution $1<\gamma_e\simlt 10^3$.

The plasma frequency is higher in the hydrostatic atmosphere, where the 
density is higher. For a hydrostatic layer of temperature $T_e$ and 
Thomson depth $\tau_T$ one finds
$\hbar\omega_{Pe}=0.65\,(\tT m_ec^2/kT_e)^{1/2}\,g_{14}^{1/2}$~eV. 
In this case, however, the brightness temperature of the O-mode photons is 
limited to a value $k_{\rm B}T_b \sim m_ec^2/\tau_T$ by induced 
downscattering in the atmosphere. The corresponding upper limit for K-band 
luminosity, $\omega L_\omega = (\kB T_b\,\omega^3/8\pi^2 c^2)\, 4\pi \RNS^2
\simlt 10^{29}\,\tT^{-1}\,(\omega/\omega_{\rm K-band})^3$~erg~s$^{-1}$,
is in conflict with observations.

\noindent 2.
Non-thermal synchrotron emission from electrons with high $\gamma_e$. 
This mechanism could work only far from the star, 
$r\simgt 200\RNS\gamma_e^{2/3}$, where the
minimum energy of the synchrotron photons, $\sim\gamma_e^2\hbar eB/m_ec$, 
is below the optical band
($B<10^8\gamma_e^{-2}$~G). The magnetic lines extending to large $r$ 
form a small cap on the stellar surface and carry a tiny fraction of the 
total magnetospheric current, 
$f\sim (r/\RNS)^{-2}\sim 3\times 10^{-5}\gamma_e^{-4/3}$.
If the voltage maintained along the extended lines is comparable to 
1~GeV, energy dissipation rate in the region is only 
$10^{30}\gamma_e^{-4/3}$~erg~s$^{-1}$, orders of magnitude smaller than 
the observed optical-IR luminosity.
Therefore this mechanism appears to be unlikely.  

\noindent 3.
Ion cyclotron emission from radii $\sim 20\RNS$, where the ion cyclotron
frequency is in the optical band. The observed 
flux then implies ion temperatures of $\sim 10^{11}$~K. Such temperatures 
are indeed achieved in the corona.
This mechanism also requires a sufficient number of ions in the 
corona, which is sensitive to
the ratio of voltage $e\Phi_e$ to the ion gravitational energy (Fig.~10).
The ions will act as a filter for the radio and microwave
radiation that is emitted by coherent processes near
the star. The radiation reaching $r\ga 500$~km 
will be partially absorbed by the ions at their cyclotron resonance,
pumping their perpendicular energy $E_\perp$. This energy further
increases as the ions
approach the star since $E_\perp/B$ is an adiabatic invariant.
Their cyclotron cooling and transit times become comparable at a radius 
$r_*\sim~200\,(v_i/c)^{-1/5}\, R_{\rm NS,6}^{6/5}\,B_{\rm pole,15}^{2/5}$~km.
Then the ions emit 
cyclotron radiation in the optical-IR range, since $\hbar eB(r_*)/m_pc\approx
B_{\rm pole,15}^{-1/5}(v_i/c)^{3/5}R_{\rm NS,6}^{-3/5}$~eV.

\noindent 4. Curvature emission by pairs in the inner magnetosphere may
extend into the optical-IR range.  Charges moving on field lines with
a curvature radius $R_C \la R_{\rm NS}$ will emit 
radiation with frequency $\nu_C\sim\gamma_e^3 c/2\pi R_C$, 
which is in the K-band ($\nu = 10^{14}$~Hz) or optical when 
$\gamma_e m_ec^2\simgt$~GeV.
As in models of pulsar radio emission, 
bunching of the radiating charges is required 
to produce the observed luminosity (e.g. Asseo, Pelletier, \& Sol 1990).
The radio luminosity of a normal pulsar can approach $\sim 10^{-2}$ of
its spindown power, and it is not implausible that the observed fraction 
$\sim 10^{-4}$ of the power dissipated in a magnetar corona
is radiated in optical-IR photons by the same coherent mechanism.  

There are indications of charge bunching on the Debye scale in our 
numerical experiment, although this detail of the experiment is 
difficult to scale to the real corona of magnetars.
Suppose a fraction $f$ of the net flow of particles through the corona 
$\dot{N}_e=I/e\sim 10^{39}L_{\rm diss\,36}\,(e\Phi_e/{\rm GeV})^{-1}$~s$^{-1}$ 
is carried by bunches of $\calN$ 
electrons (or positrons). The power emitted by one bunch is 
$\dot{E}=(2/3)({\cal N}e)^2\gamma_e^4c/R_C^2$, and the total coherent
luminosity produced by the corona is given by 
$L\approx f\calN \dot{N}_e\, \gamma_e^4\,e^2/R_C$.
Using the relation $\nu_C\sim (c/2\pi R_C)\,\gamma_e^3$ for curvature 
radiation, we find the number of charges per bunch that is required
to explain the observed luminosity $L\sim 10^{32}$~erg~s$^{-1}$, 
\be
  \calN\sim 10^4\,f^{-1}\,L_{32} \nu_{C,14}^{-4/3}.
\ee 
This is much smaller then the number of charges per Debye sphere, 
$\ND\sim (n_cr_e^3)^{-1/2}\sim 10^{11}$.
Thus, a very weak bunching of charges 
(or very small bunches, with a scale $\sim 10^{-2}\lambda_{De}$)
can generate a significant power in coherent curvature radiation.
The required bunching is small compared with what is commonly  
invoked to explain the emission of radio pulsars. 

For example, one can compare with the Crab pulsar.  The total 
flux of $e^\pm$ pairs from the pulsar, $\dot N_e \sim 10^{39}$ s$^{-1}$,
is comparable to the flux in a magnetar corona. The pulsar
$e^\pm$ flow is concentrated on the open field lines, and its density is 
enhanced by a factor $R_{\rm lc}/\RNS\sim 10^2$. The curvature radius
of the open field lines near the star is $R_C\sim 10\RNS$.
These differences are, however, modest compared with the 
much larger bunching efficiency that is implied by the observed 
brightness of the radio emission from the Crab pulsar. Supplying its 
luminosity $L \sim 10^{31}$ erg~s$^{-1}$ at $\nu\sim 10^8$~Hz would require 
$\gamma_e \sim 60$ and $\calN\sim 10^{11}f^{-1}$.

The dispersion relation for the IR-optical radiation in the corona is
dominated by plasma rather than by vacuum polarization. The curvature
radiation is, generally, a superposition of all possible polarization modes
(see Arons \& Barnard 1986 for a description of the normal modes of a
strongly magnetized plasma). Emission of the superluminal O-mode will occur
if the bunch size (measured in its rest frame) is smaller than the plasma
skin depth.
The E-mode also comprises a fraction of the curvature emission when the
magnetic field lines are twisted (Luo \& Melrose 1995). 
Curvature emission in the subluminal Alfv\'en mode is also
possible, but this mode is ducted along the magnetic field lines and has a
frequency below the ambient plasma frequency.
                                                                                
The spectrum of curvature emission is expected to cut 
off at frequencies above $\nu_C$ that corresponds to the maximum Lorentz 
factor (a few $\gres$) of $e^\pm$ in the corona.  This cutoff may be 
observable in the optical or infrared band (cf. Hulleman et al. 2004).
Note also that the K-band brightness temperature inferred from observations 
of AXPs 
is remarkably close to the effective temperature of the coronal $e^\pm$, 
$T_{\rm eff}\sim \gamma_e m_ec^2/k_{\rm B}$, 
which is bounded by $\sim 10^{13}$~K by the pair creation process.

\subsection{Evolution Following Bursts of Activity}

Bursts of magnetar activity are associated with starquakes ---
sudden crust deformations that release the internal magnetic 
helicity of the star and impart an additional twist to the 
magnetosphere. 
The increase of the twist should lead to a larger luminosity from the corona
since the luminosity is simply proportional to the net current flowing through 
the magnetosphere. Between bursts, the magnetospheric currents persistently 
dissipate and the corona disappears after the time $t_{\rm decay}$
(eqs.~[\ref{eq:tdecay}] and [\ref{eq:Ldecay}]). If there were no bursts during 
a long time $t>t_{\rm decay}$ the magnetar should enter the quiescent state 
with no coronal emission.
Such behavior has been observed in AXP~J1810-197 (Gotthelf \& Halpern 
2005). An outburst, which occured between November 2002 and January 2003,
was followed by the gradual decay of soft X-ray emission on a time-scale 
of a few years, and now the source is returning to the quiesent state.
It would be especially interesting to observe the evolution of the 
hard X-ray component, which is not detected in the relatively weak 
AXP~J1810-197.

During periods of high activity, when bursts occur frequently,
the growth of the magnetospheric twist may be faster than its decay.
The twist may grow to the point of a global instability.
When the magnetic lines achieve $\Delta\phi\simgt 1$ the magnetosphere 
becomes unstable and relaxes to a smaller twist angle. A huge release of 
energy must then occur, producing a giant flare;
a model of how this can happen is dicussed by Lyutikov (2006). 
It is possible that the 27~December~2004 giant flare from SGR~1806-20 was 
produced in this way --- a prolonged period of unusually high activity 
preceded the flare.  
If the flare indeed released the external twist energy, then the current 
through the magnetosphere should have decreased after the flare, and the 
coronal luminosity now should be smaller than it was prior to the flare.

A possible way to probe the twist evolution is to measure the history
of spindown rate $\dot{P}$ of the magnetar. The existing data confirm 
the theoretical expectations: the spin-down of SGRs was observed to 
accelerate months to years following periods of activity (Kouveliotou 
et al. 1998, 1999; Woods et al. 2002). A similar effect was also observed
following more gradual, sub-Eddington flux enhancement in AXPs
(Gavriil \& Kaspi 2004).  This ``hysteresis'' behavior of the spindown 
rate may represent the delay with which the twist is spreading to the 
outer magnetosphere.

Since the density of the plasma corona is proportional to the current that
flows through it, changes in the twist also lead to changes in the coronal 
opacity. The opacity should increase following bursts of activity and affect
the X-ray pulse profile and spectrum. Such changes, which persist for 
months to years, have been observed following large X-ray outbursts from 
two SGRs (Woods et al. 2001).

\subsection{Further Developments}

1. We focused in this paper 
on the inner magnetosphere, that is, on field lines that
do not extend beyond a radius $R_{\max}\sim 2\RNS$.  This part
of the
magnetosphere is expected to retain most of the toroidal field energy,
and to dominate the non-thermal output of the star. 
The outer corona may be qualitatively different because it is subject 
to a strong ``cyclotron drag'': the resonantly scattered X-rays can escape 
 (rather than convert to $e^\pm$), thus providing an efficient sink of energy.
This drag will
cause a local change of the electric field in the circuit (TLK; 
Thompson \& Beloborodov 2005).  At yet larger radii, 
beyond $\sim 200$~km from the star, 
ions may populate higher Landau levels $n>0$. Then a new degree of 
freedom appears in the problem: the transverse motion of the ions.

2. Like the Solar transition region, the transition layer at the base of 
a magnetar corona may respond to the coronal dissipation by increasing 
the supply of plasma. In particular, pair creation in the transition 
layer needs further investigation because pairs may feed the coronal 
current (\S~\ref{pairatm}). The state of this layer and its effect on 
the corona is determined by the level of plasma turbulence (see 
\S\S~\ref{cond} and \ref{anom}). This layer may partially insulate the 
star from the hot corona because plasma turbulence suppresses thermal 
conduction. By trapping the generated heat, the layer may overheat and 
expand to the corona.

3. The transition layer also plays a crucial role in the energy balance 
of the corona because it produces radiation that can escape the magnetar. 
A significant part of the released energy is emitted in the hard X-ray 
band. The rest of the released energy is deposited in an optically 
thick surface layer and reprocessed into blackbody radiaton with 
$k_{\rm B}T\la 1$ keV. An accurate calculation of the ratio of the 
two emission components (hard X-ray to blackbody) that are fed by 
the corona remains to be done. 

4. The role of transverse waves in the corona, which were suppressed 
in our numerical experiment, will need to be studied. The waves 
produce a three-dimensional pattern of magnetic-field fluctuations 
$\delta(\nabla\times B)\neq 0$ and their study will require 
3-D or 2-D simulations.  

5. Our plasma model of the corona may be generalized to include the 
effects of rotation. This can be done by changing the neutrality condition 
$\rho=0$ to $\rho=\rho_{\rm co}$.
Note that $\rho_{\rm co}$ may be easily maintained by a small polarization 
of the current-carrying plasma because $\rho_{\rm co}\ll en_c$ 
(see eq.~[\ref{rhoco}]).  
It can be taken into account by adding an extra source 
term $4\pi\rho_{\rm co}$ to the Poisson equation (cf. eq.~\ref{eq:Gauss1}), 
$\dd E_\parallel/\dd l=4\pi(\rho-\rho_{\rm co})$.
Then $E_\parallel$ vanishes if $\rho=\rho_{\rm co}$ along the magnetic line,
so that $\rho_{\rm co}$ is the new equilibrium charge density. 

6. It is important to understand 
the resistive redistribution of the twist in the magnetosphere,
in particular its spreading toward the open field lines, because
it is a promising source of torque variability in active magnetars (TLK).
This would require the construction of a global, time-dependent
model of the currents in the magnetosphere.

\acknowledgments
We thank A. Frolov, P. Goldreich, A. Gruzinov, M. Medvedev, and M. Ruderman 
for discussions.  AMB is supported by NASA ATP and an Alfred P. Sloan 
Fellowship. CT is supported by the Natural Sciences and Engineering Research 
Council of Canada.


\end{document}